\documentclass[aps,10pt,twocolumn,showpacs,pra,amsmath,amssymb,floatfix,superscriptaddress,dvipsnames]{revtex4-2}

\usepackage{amsmath}
\usepackage{amssymb}
\usepackage{amsfonts}
\usepackage{mathrsfs}
\usepackage{amsthm}
\usepackage{mathtools}
\usepackage{braket}
\usepackage{physics}
\usepackage{nicematrix}
\usepackage{bm} 
\usepackage{dsfont}
\usepackage{autobreak}
\usepackage{xfrac}
\usepackage{colonequals}
\newcommand{\defeq}{\colonequals}

\theoremstyle{plain}
\newtheorem{theorem}{Theorem}

\newtheorem{proposition}[theorem]{Proposition}
\newtheorem{lemma}[theorem]{Lemma}
\theoremstyle{definition}

\newtheorem{example}[theorem]{Example}
\theoremstyle{remark}

\usepackage{graphicx}
\usepackage{dcolumn}
\usepackage{bm}
\usepackage{appendix}
\usepackage{amsfonts}
\usepackage[normalem]{ulem}
\usepackage{comment}
\usepackage{lmodern}
\usepackage{stackrel}
\usepackage{bbold}
\usepackage{soul}

\usepackage[utf8]{inputenc}
\usepackage[english]{babel}
\usepackage[T1]{fontenc}

\usepackage{quantikz}
\usepackage{tikz}
\usetikzlibrary{quantikz2}

\usepackage[small,bf,justification=Justified]{caption}
\newcommand{\com}[1]{} 

\usepackage{todonotes}

\usepackage{listofitems}
\newcommand\cycle[2][\,]{%
  \readlist\thecycle{#2}%
  (\foreachitem\i\in\thecycle{\ifnum\icnt=1\else#1\fi\i})%
}

\NewDocumentEnvironment{alignb}{b}{%
  \begin{align*}
  \refstepcounter{equation} #1 \tag{\theequation}
  \end{align*}
}{\ignorespacesafterend}

\usepackage{xcolor}
\definecolor{color_scheme}{RGB}{26, 121, 199}
\usepackage[colorlinks=true,linktocpage=true]{hyperref}
\hypersetup{
    allcolors  = {color_scheme},
}

\newcommand{\CC}{\mathbb{C}}
\newcommand{\HH}{\mathcal{H}}
\newcommand{\BH}{\mathcal{B}(\HH)}
\newcommand{\permCn}{\mathtt{C}_n}
\newcommand{\permAn}{\mathtt{A}_n}
\newcommand{\permBn}{\mathtt{B}_n}
\newcommand{\permPn}{\mathtt{P}_n}
\newcommand{\permP}{\mathtt{P}}

\newcommand{\permT}{\mathtt{T}}
\newcommand{\permA}{\mathtt{A}}
\newcommand{\permB}{\mathtt{B}}

\newcommand{\SWAP}{\mathrm{SWAP}}

\begin{document}

\newcommand{\pisa}{Department of Physics ``E. Fermi'', University of Pisa, Largo B. Pontecorvo 3, 56127 Pisa, Italy} 
\newcommand{\pisashort}{Department of Physics ``E. Fermi'', University of Pisa, Pisa, Italy} 

\newcommand{\cfum}{Centro de F\'{i}sica, Universidade do Minho, Campus de Gualtar, 4710-057 Braga, Portugal} 
\newcommand{\cfumshort}{Centro de F\'{i}sica, Universidade do Minho, Braga, Portugal} 

\newcommand{\inl}{INL -- International Iberian Nanotechnology Laboratory, Av. Mestre Jos\'{e} Veiga s/n, 4715-330 Braga, Portugal} 
\newcommand{\inlshort}{INL -- International Iberian Nanotechnology Laboratory, Braga, Portugal} 
\newcommand{\uff}{Instituto de F\'{i}sica, Universidade Federal Fluminense, Av. Gal. Milton Tavares de Souza s/n, Niter\'{o}i -- RJ, 24210-340, Brazil}
\newcommand{\uffshort}{Instituto de F\'{i}sica, Universidade Federal Fluminense, Niter\'{o}i -- RJ, Brazil}
\newcommand{\ulm}{Institute of Theoretical Physics, Ulm University, Albert-Einstein-Allee 11 89081, Ulm, Germany}
\newcommand{\ulmshort}{Institute of Theoretical Physics, Ulm University, Ulm, Germany}
\newcommand{\ifsc}{Instituto de Física de São Carlos, Universidade de São Paulo, IFSC -- USP, 13566-590, São Carlos, SP, Brasil}
\newcommand{\ifscshort}{Instituto de Física de São Carlos, Universidade de São Paulo, São Carlos -- SP, Brasil}

\title{Measuring unitary invariants with the quantum switch}

\author{Pedro C. Azado}
\email{pedroazado@ifsc.usp.br}
\affiliation{\ifscshort} 
\affiliation{\inlshort}

\author{Rafael Wagner}
\email{rafael.wagner@uni-ulm.de}
\affiliation{\ulmshort}
\affiliation{\inlshort}
\affiliation{\cfumshort}

\author{Rui Soares Barbosa}
\email{rui.soaresbarbosa@inl.int}
\affiliation{\inlshort}

\author{Ernesto F. Galvão}
\email{ernesto.galvao@inl.int}
\affiliation{\inlshort}
\affiliation{\uffshort}

\begin{abstract}
Bargmann invariants, multivariate traces of states $\Tr(\rho_1 \cdots \rho_n)$, completely characterize any unitary-invariant property of a set of states. Unitary invariants enable the description of quantum resources such as basis-independent coherence and imaginarity, nonstabilizerness, and contextuality. We show that the quantum switch, a higher-order process featuring indefinite causal order, can be used to measure Bargmann invariants of arbitrary order.
We also show how simple Hadamard test circuits can deterministically simulate an arbitrary unitary quantum switch.
Our results establish a solid bridge between the theory and applications of unitary invariants and higher-order maps in quantum mechanics.
\end{abstract}

\maketitle
\section{Introduction}  

Bargmann invariants, multivariate traces of the form
 \begin{equation}\label{eq: Bargmann invariant definition}
     \Delta_n(\varrho) \defeq \Tr(\rho_1 \cdots \rho_n) , 
 \end{equation}
were introduced by Valentine Bargmann \cite{bargmann1964note} as examples of unitary-invariant functions of tuples of quantum states $\varrho = (\rho_1,\ldots,\rho_n)$.
While these were, to some extent, originally introduced almost as accidental constructions to exemplify non-trivial unitary-invariant functions in quantum theory, they proved to have deeper significance. 
In fact, it turns out that Bargmann invariants suffice to fully characterize all the unitary-invariant -- and thus physically relevant -- properties of tuples of quantum states~\cite{oszmaniec2024measuring}.

It comes therefore as no surprise that theoretical and experimental investigation of such quantities has importance across diverse subfields of quantum theory.
For example, subsequent work by Simon, Mukunda, and collaborators~\cite{simon1993Bargmann,mukunda2001Bargmann,mukunda2003Bargmann,mukunda2003Wigner} established their connection to geometric phases: for pure states, the phase of the invariant given by Eq.~\eqref{eq: Bargmann invariant definition} is equal in modulo to the Berry phase~\cite{berry1984quantal,avdoshkin2023extrinsic}.

As another example, the special case of \emph{univariate} traces of states $\Tr(\rho^n)$ -- obtained as $\Delta_n$ restricted to tuples of repeated elements $\varrho=(\rho,\ldots,\rho)$ -- has been thoroughly investigated due to their relevance in estimating the spectrum of quantum states~\cite{horodecki2003fromlimits,ekert2002direct,alves2003direct,horodecki2002method,tanaka2014determining,vanenk2012measuring,brun2004measuring,shin2024rankneedestimatingtrace,leifer2004measuring,liang2023unified,liu2025estimating}.
Such invariants also arise in connection to entanglement spectroscopy~\cite{johri2017entanglement,turkeshi2023measuring,tirrito2024quantifying} and quantum error mitigation~\cite{koczor2021exponential,huggins2021virtual}. 

Recently, more general Bargmann invariants~\cite{quek2024multivariatetrace,oszmaniec2024measuring} have attracted much attention due to newly uncovered connections with 
Kirkwood--Dirac (KD) quasiprobability~\cite{kirkwood1933quantum,dirac1945analogy,wagner2024quantumcircuits,arvidssonshukur2024properties,schmid2024kirkwood,liu2025boundarykirkwooddiracquasiprobability}, out-of-time-order correlators~\cite{yunger2018quasiprobability,gonzalez2019out,wagner2024quantumcircuits}, weak values~\cite{wagner2023anomalous,hofmann2012complex}, invariant theory~\cite{wigderson2019mathematics,chien2016characterization}, multi-photon indistinguishability \cite{menssen2017distinguishability,jones2020multiparticle,minke2021characterizing,pont2022quantifying,rodari2024experimentalobservationcounterintuitivefeatures,rodari2024semideviceindependentcharacterizationmultiphoton,seron2023boson,giordani2021witnesses,giordani2020experimental,jones2023distinguishability,jones2023distinguishability,brod2019witnessing,annoni2025incoherentbehaviorpartiallydistinguishable}, overlap uncertainty relations~\cite{bong2018strong}, and quantum thermodynamics~\cite{gherardini2024quasiprobabilities,lostaglio2022kirkwood,levy2020quasiprobability,hernandez2024projective,santini2023work,donati2024energetics}, among others~\cite{samuel1988general,fernandes2024unitary,li2024bargmann,zhang2024boundaries,wagner2024certifying,wagner2024coherence,wagner2024inequalities,galvao2020quantum,zhang2024local,giordani2023experimental,reascos2023quantum,elliott2025strictadvantagecomplexquantum}.

An apparently unrelated phenomenon is that of indefinite causal order, which in its simplest form can be probed using the so-called \emph{quantum switch}~\cite{chiribella2013quantum}. The quantum switch applies two operations $A$ and $B$ in a coherent superposition of causal orders: $A$ followed by $B$ and $B$ followed by $A$. The switch has been experimentally implemented~\cite{procopio2015experimental,goswami2018indefinite,rubino2017experimental,cao2023semideviceindependent,rozema2024experimental},
and the study of supermaps with indefinite causal order -- higher-order processes akin to the quantum switch --  is an active area of research~\cite{oreshkov2012quantum,chiribella2013quantum,chiribelly2012perfect,chiribella2021indefinite,kissinger2017picturing,wilson2025quantumsupermapscharacterizedlocality,vanderLugt2024possibilistic}.
This concept has also been generalized to post-quantum  probabilistic theories~\cite{bavaresco2024indefinitecausalorderboxworld}. Furthermore, the quantum switch and its variants have been shown to provide a query complexity advantage over conventional, causally ordered quantum circuits~\cite{renner2022computational,araujo2017quantum,simonov2024universal}.
Applications of the quantum switch include quantum metrology~\cite{yin2023experimental,zhao2020quantum}, parameter estimation \cite{procopio2023parameter}, quantum communication \cite{guerin2016exponential,ebler2018enhanced,chiribella2021indefinite,wu2024generalcommunication}, quantum thermometry \cite{mukhopadhyay2020superposition}, and quantum refrigeration~\cite{felce2020quantum,molitor2024quantum}, among others \cite{loizeau2020channel,caleffi2020quantum,procopio2020sending,simonov2022work,cao2022quantum,simonov2024universal}.
For a comprehensive review and introduction to this topic, see e.g.~Ref.~\cite{taranto2025higherorderquantumoperations}.

While significant effort has been devoted to understanding Bargmann invariants for witnessing quantum resources, the relevance of such invariants to the study of indefinite causal order remains largely unexplored. Early results have provided a few insights.
In the discrete case, Gao et al.~\cite{gao2023measuring} established a connection between outcomes of the quantum switch and fourth- or fifth-order Bargmann invariants. However, their analysis was restricted to the real parts of these invariants.
Ban~\cite{ban2021sequential} linked third-order Bargmann invariants with the quantum switch statistics, this time leveraging their connection with the real part of KD representations. A more general, albeit arguably indirect, relationship between unitary invariants and indefinite causal order has been suggested for continuous variable systems~\cite{zhao2020quantum,yin2023superHeisenberg}, where the quantum switch has been applied to estimate geometric phases in metrological contexts.

In this work, we find a systematic way of using quantum switches to measure Bargmann invariants of any order.
In doing so, we improve upon the protocols found by Gao et al.~\cite{gao2023measuring} and Ban~\cite{ban2021sequential}, which use the quantum switch to estimate the real part of \mbox{third-,} fourth-, and fifth-order Bargmann invariants.
More specifically, we construct a family of unitary quantum switches that can be used to measure the real and imaginary parts of Bargmann invariants of any-sized tuples of quantum states in arbitrary dimension. This result effectively turns all the previous applications of Bargmann invariants into potential applications for the quantum switch, opening up the possibility to use indefinite causal order in a variety of quantum information tasks.

Leveraging this connection between the quantum switch and Bargmann invariants, we also introduce a deterministic quantum circuit simulation of any unitary quantum switch, i.e.~a quantum switch where both input operations are unitary.
The quantum simulation circuit is an instance of the well-known Hadamard test~\cite{lin2022lecturenotes}. Our simulation has the peculiarity of requiring two queries of the unitary operators but also a single query of each of their inverses.
It has, therefore, worse query complexity than other simulations in the literature~\cite{renner2021reassessing,bavaresco2024can,kristjánsson2024exponentialseparationquantumquery}. It does, however, offer a space advantage: in contrast to other simulations, it requires a single auxiliary qubit as opposed to an auxiliary system of the same dimension as the Hilbert space on which the unitaries act.

\textbf{Outline.} The remainder of this document is structured as follows.
In Sec.~\ref{sec:background} we review the relevant background: Sec.~\ref{sec:unitinv} introduces Bargmann invariants and how they can be measured using cycle tests, and Sec.~\ref{ssec:qswitch} recalls the construction and basic properties of the quantum switch.
In Sec.~\ref{sec:results} we propose a measurement protocol for estimating (the real and imaginary parts of) Bargmann invariants using the quantum switch: Sec.~\ref{sec:odd order} shows how to measure odd-order invariants $\Delta_{2k+1}$ using a family of unitary quantum switches; Sec.~\ref{sec: evenBargmann} discusses why odd- and even-order invariants must be probed differently by a quantum switch and provides a protocol for estimating even-order invariants $\Delta_{2k}$.
In Sec.~\ref{sec: simulation Hadamard qs}, building upon the quantum switch protocol, we show how the Hadamard test can be used to simulate the quantum switch.
Finally, in Sec.~\ref{sec:conclusions}, we make some concluding remarks, review our results, and point out future research directions.

\section{Background}\label{sec:background}

\subsection{Bargmann invariants and the cycle test} \label{sec:unitinv}

A unitary invariant~\cite{popescu2009unitaryinvariants} is a function $f \colon \BH^n \to \mathbb{C}$, defined on tuples of bounded linear operators on a Hilbert space $\mathcal{H}$, that is invariant under the action of the same unitary on each of the operators, i.e., under transformations of the form
\[(X_1,\dots,X_n) \mapsto (U X_1 U^\dagger,  \dots, U X_n U^\dagger)\]
for $U$ a unitary operator on $\mathcal{H}$. In other words, for every tuple $(X_1,\dots,X_n)$ and every unitary $U$ one has that  $$f(U X_1 U^\dagger, \dots, U X_n U^\dagger) = f(X_1,\dots,X_n).$$

In this work, we focus on the specific family of unitary invariants defined by multivariate traces of states known as \emph{Bargmann invariants}~\cite{bargmann1964note,oszmaniec2024measuring}.
The $n$th order Bargmann invariant, $$\Delta_n \colon \mathcal{D}(\mathcal{H})^n \to \mathbb{C},$$
maps an $n$-tuple of states $\varrho = (\rho_1,\ldots,\rho_n) \in \mathcal{D}(\mathcal{H})^n$
to the expression $\Delta_n(\varrho)$ given in Eq.~\eqref{eq: Bargmann invariant definition}.
For a tuple of \emph{pure} states $(\psi_1, \ldots, \psi_n)$, where we denote $\psi_i \equiv \vert \psi_i\rangle \langle \psi_i \vert $, the Bargmann invariant becomes
\begin{equation}\label{eq: pure state Bargmann invariants}
\Delta_n(\psi_1, \ldots, \psi_n) = \braket{\psi_1}{\psi_2}\braket{\psi_2}{\psi_3}\ldots\braket{\psi_n}{\psi_1}.
\end{equation}
The simplest non-trivial Bargmann invariant is the second-order $\Delta_2(\rho_1,\rho_2) = \Tr(\rho_1\rho_2)$, the two-state overlap.
The simplest Bargmann invariant that can take non-real complex values is the third-order $\Delta_3(\rho_1,\rho_2,\rho_3) = \Tr(\rho_1\rho_2\rho_3)$~\cite{fernandes2024unitary,li2024bargmann,zhang2024boundaries}.

We now describe a circuit proposed by Oszmaniec, Galv\~ao, and Brod \cite{oszmaniec2024measuring}
for measuring the value of the Bargmann invariant $\Delta_n(\rho_1, \ldots, \rho_n)$ of an arbitrary $n$-tuple of quantum states given as input $\rho_1 \otimes \cdots \otimes \rho_n$.

The symmetric group $S_n$ acts naturally on the $n$-fold tensor product space $\mathcal{H}^{\otimes n}$ by permutations of the tensor factors, yielding a faithful unitary representation:
to each permutation $\permP$ of the set $\{1, \ldots, n\}$,
it associates the unitary $P \colon \mathcal{H}^{\otimes n} \to \mathcal{H}^{\otimes n}$, acting as
\begin{equation}
    P(\ket{\psi_1} \otimes \cdots \otimes \ket{\psi_n} ) = \ket{\psi_{\permP^{-1}(1)}} \otimes \cdots \otimes \ket{\psi_{\permP^{-1}(n)}}.
\end{equation}
We consider the cyclic permutation 
\[\permCn = (1 \;\, n \;\, n{-}1 \;\, \cdots \;\, 2) = (n \;\, n{-}1 \;\, \cdots \;\, 1),\]
which acts as
\[1 \mapsto n,\; n \mapsto n{-}1, \; \ldots, \; 2 \mapsto 1.\]
Its associated unitary $C_n$ acts on the Hilbert space $\HH^{\otimes n}$ by shifting left the tensor factors:
\[
C_n (\ket{\psi_1} \otimes \ket{\psi_2} \otimes \cdots \otimes \ket{\psi_n}) = \ket{\psi_2} \otimes \cdots \otimes \ket{\psi_{n}} \otimes \ket{\psi_1}.
\]
This permutation can be decomposed in terms of transpositions as 
\begin{align}
    \permCn &=  (n \;\, n{-}1 \;\, \cdots \;\, 2 \;\, 1) \\
    &= (n \;\, n{-}1) (n{-}1 \;\, n{-}2) \cdots (3 \;\, 2) (2 \;\, 1), \label{eq:cn-transpositions}
\end{align}
corresponding to a decomposition of the unitary $C_n$ as a sequence of nearest-neighbor SWAP gates:
\begin{equation}\label{eq:cycle_decomposition_swapcascade}
C_n = \SWAP_{n-1, n} \SWAP_{n-2, n-1} \cdots \SWAP_{2, 3} \SWAP_{1, 2}.
\end{equation}

\begin{figure}[t]
	\centering
	\includegraphics[width=\columnwidth]{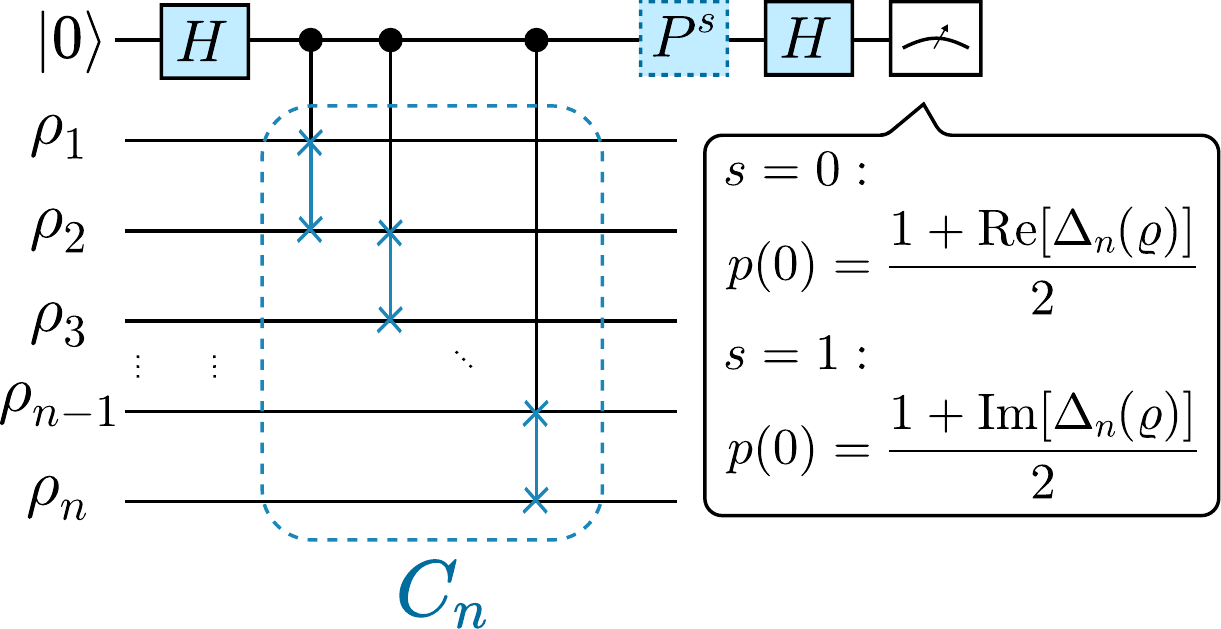}
	\caption{\textbf{Cycle test.} An instance of a Hadamard test is shown, where the unitary $C_n$ is associated to the cyclic permutation $\permCn$ of $n$ subsystems. We initialize a quantum memory of $n+1$ systems in a product state. The first wire in the circuit represents a single qubit system. The remaining wires represent systems of dimension $d \geq 2$. After the controlled cycle operation a gate $P^s=\text{diag}(1,i^s)$ is applied to the auxiliary qubit, later measured in the computational basis. When $s=0$ ($s=1)$, measurement of the auxiliary qubit gives an estimate of the real (imaginary) part of the Bargmann invariant.}    
	\label{fig: cycletest}
\end{figure}

It is well known that Bargmann invariants can be equivalently written as the expectation value of $C_n$~\cite{oszmaniec2024measuring,wagner2024quantumcircuits,quek2024multivariatetrace} as  follows:
\begin{equation}\label{eq: Bargmann expectation value}
    \Delta_n(\varrho) = \Tr\left( C_n (\rho_1 \otimes \cdots \otimes \rho_n) \right) = \langle C_n \rangle_{\rho_1 \otimes \dots \otimes \rho_n}.
\end{equation}
From this we see that Hadamard test circuits~\cite{lin2022lecturenotes} can be used to estimate such invariants, as discussed in Refs.~\cite{oszmaniec2024measuring,quek2024multivariatetrace}.
Fig.~\ref{fig: cycletest} shows the instance of the Hadamard test capable of estimating Bargmann invariants, known as a \emph{cycle test}~\cite{oszmaniec2024measuring}.
In this quantum circuit, a single-qubit auxiliary system is initialized in a coherent superposition $\vert +\rangle = \sfrac{1}{\sqrt{2}}(\ket{0} + \ket{1})$.
The remainder of the quantum memory, consisting of $n$ copies of the Hilbert space $\HH$, carries the input $n$-tuple of states in the product state $\rho_1 \otimes \cdots \otimes \rho_n$.
Then, the unitary $C_n$, controlled by the auxiliary qubit, is applied to this $n$-partite system. To conclude the cycle test, the auxiliary qubit is measured in the $\{\ket{+}, \ket{-}\}$ basis to estimate the real part of the Bargmann invariant, or in the $\{\ket{+_i}, \ket{-_i} \}$ basis to estimate its imaginary part, where $\ket{\pm} = \sfrac{1}{\sqrt{2}}(\ket{0} \pm \ket{1})$ and $\ket{\pm_i} = \sfrac{1}{\sqrt{2}}(\ket{0} \pm i\ket{1})$ are the eigenvectors of the Pauli operators $X$ and $Y$, respectively.

Depending on how one implements $C_n$ using the available gate set, one obtains different instances of the cycle test. One possible choice, shown in Fig.~\ref{fig: cycletest}, is to decompose $C_n$ as in Eq.~\eqref{eq:cycle_decomposition_swapcascade}. This decomposition  uses a single type of gate -- the $\SWAP$ gate -- applied locally on nearest-neighbor systems in the circuit, thus avoiding the need for long-range gates. The resulting circuit has linear quantum depth and uses one auxiliary qubit.

Alternatively, Refs.~\cite{oszmaniec2024measuring,quek2024multivariatetrace} propose different decompositions of the cycle operator that allow for smaller quantum circuit depth, at the cost of increasing the dimensionality of the auxiliary system. In particular, 
Quek~et~al.~\cite{quek2024multivariatetrace} found an interesting alternative decomposition of $C_n$ that will be relevant to us. They proposed a protocol for measuring $n$th-order Bargmann invariants consisting of a Hadamard test applying, in sequence, controlled unitaries
\begin{equation}\label{eq: AandB}
A_n = \prod_{\substack{i=1 \\ i \text{ odd}}}^{n-1} \SWAP_{i,i+1} \quad \text{and} \quad B_n = \prod_{\substack{i=2 \\ i \text{ even}}}^{n-1} \SWAP_{i,i+1}.
\end{equation}
Each of these unitaries consists of nearest-neighbor SWAP gates acting on disjoint pairs of qubits, and their controlled application
can be parallelized using additional entangled auxiliary qubits. The operations $A_n$ and $B_n$ are unitaries corresponding to the permutations
\begin{align*}
\permAn &= (1 \;\, 2)(3 \;\, 4) \cdots (n-2  \;\, n-1)
\\
\permBn &= (2 \;\, 3)(4 \;\, 5) \cdots (n-1 \;\, n)
\end{align*}
when $n$ is odd, and similar ones if $n$ is even.
Note that their composition $\permBn\permAn$ 
does not (quite) match the left-shift cycle permutation $\permCn$;
still, it likewise is an $n$-cycle permutation, so that it equals $\permCn$ up to relabeling.
In more formal words, $\permBn \permAn$ and $\permCn$ are in the same conjugacy class: there is a permutation $\permP$ 
such that $\permCn = \permP^{-1} \permBn \permAn \permP$.
Recall that the order of the states in a tuple generally impacts the value of its Bargmann invariants, e.g.~there are triples of states $\rho_1,\rho_2,\rho_3$ such that $\Tr(\rho_1\rho_2\rho_3) \neq \Tr(\rho_1\rho_3\rho_2)$.
Hence, in order to estimate the desired value of the Bargmann invariant using a Hadamard test with controlled $\permBn \permAn$, one first needs to permute the input states by $\permP$ by applying the corresponding unitary $P$ ahead of performing the Hadarmard test.

\subsection{Quantum switch}\label{ssec:qswitch}

\begin{figure}[t]
    \centering
    \includegraphics[width=0.35\textwidth]{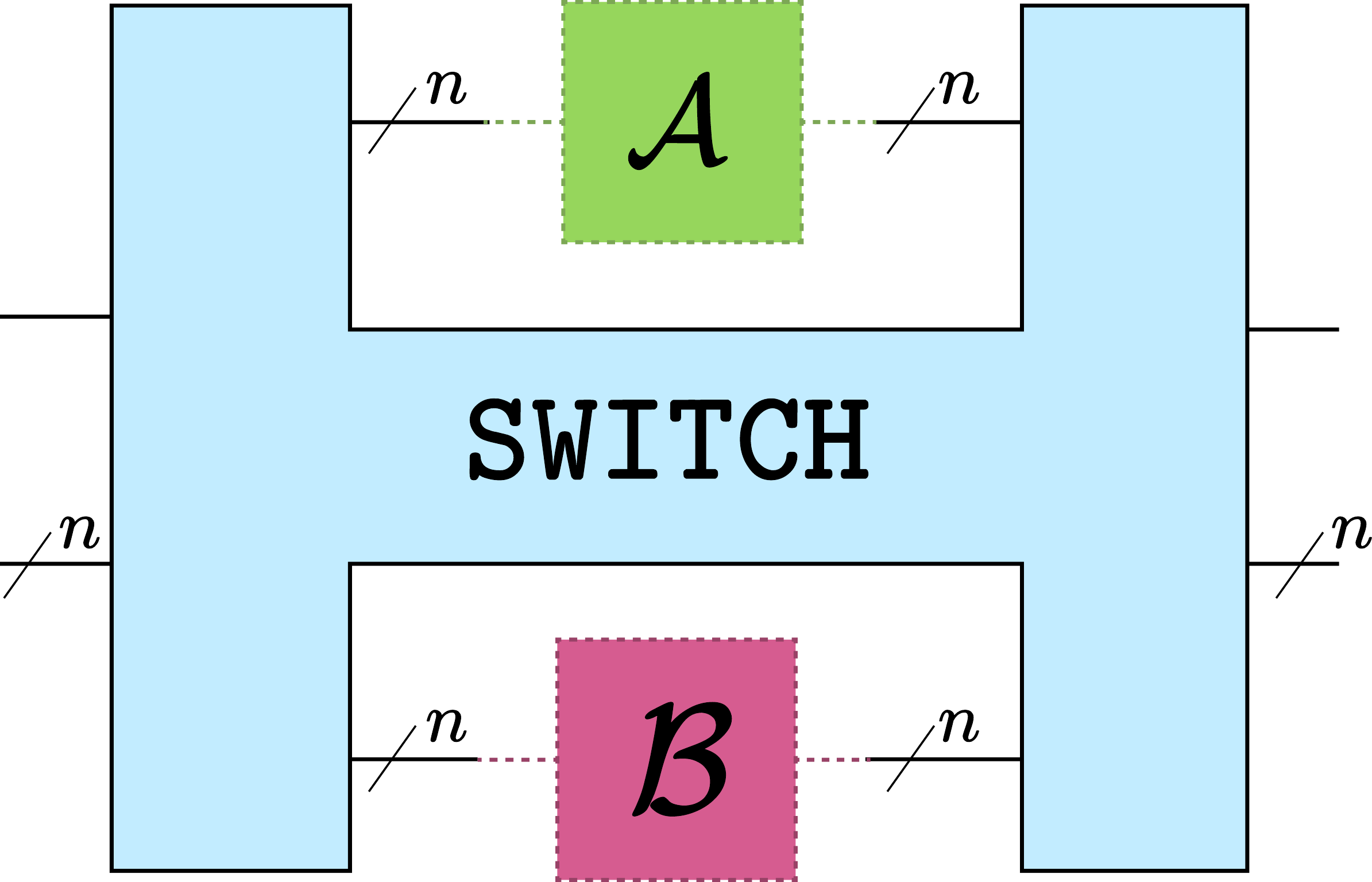}
    \caption{\textbf{Illustration of the quantum switch}. The quantum switch is a higher-order process whose inputs are a pair of quantum channels. Here we consider unitary quantum channels $\mathcal{A}$ and $\mathcal{B}$ with input and output on the composite system $\mathcal{H}^{\otimes n}$ for some fixed Hilbert space $\mathcal{H}$. The output channel of the quantum switch has input and output system $\mathbb{C}^2 \otimes \mathcal{H}^{\otimes n}$, where the first factor is the \emph{control} qubit.}
    \label{fig:qs1}
\end{figure}

The quantum switch, which we denote as $\mathcal{S}$, is the paradigmatic example of a higher-order quantum process (or quantum supermap) that is not equivalent to a process with fixed causal order (i.e., a process described by a quantum circuit with open slots for the input operations, also known as a \emph{quantum circuit board}~\cite{chiribella2008quantumcircuit}), 
nor even to a more general quantum circuit with classical control of causal order~\cite{wechs2021quantum}.

We focus on the unitary version of the quantum switch, where the input operations are taken to be unitary channels.
The switch maps a pair of unitary operators $A$ and $B$ acting on the same Hilbert space $\mathcal{H}_s$ to a unitary channel $\mathcal{S}(A, B)$ acting on the Hilbert space $\CC^2 \otimes \HH_s$, representing the original system plus an additional \emph{control} qubit. This output channel is given by $\mathcal{S}(A, B)(\,\cdot\,) = S_{A, B}(\,\cdot\,)S_{A, B}^\dagger$, where $S_{A,B}$ is the unitary operator
\begin{equation}\label{eq: switcheq}
S_{A,B} := \ketbra{0}{0} \otimes BA + \ketbra{1}{1} \otimes AB.
\end{equation}
Fig.~\ref{fig:qs1} illustrates an instance of the quantum switch, where we have taken the system of interest to be an $n$-fold tensor product $\mathcal{H}_s = \mathcal{H}^{\otimes n}$, as is required in our application to Bargmann invariants.

Let us denote the unitary channels $\mathcal{A}(\,\cdot\,) = A (\,\cdot\,)A^\dagger$ and $\mathcal{B}(\,\cdot\,) = B(\,\cdot\, )B^\dagger$. When  $\mathcal{S}(A,B)$ acts on the state $\sigma \otimes \rho$, where the control system is in an incoherent mixed state $$\sigma = p \vert 0\rangle \langle 0 \vert + (1-p)\vert 1\rangle \langle 1\vert, $$
one obtains~\cite{simonov2023indefinite}
\begin{equation*}
\mathcal{S}(A,B)(\sigma \otimes \rho) = p \vert 0\rangle \langle 0 \vert \otimes \mathcal{B}\mathcal{A}(\rho) +(1-p) \vert 1 \rangle \langle 1 \vert \otimes \mathcal{A}\mathcal{B}(\rho).
\end{equation*}
In this case, upon measuring the control system, we find that the dynamics on the system $\mathcal{H}_s$ is equivalently described by a \textit{causally separable} operation~\cite{araujo2015witnessing}, which with probability $p$  acts first with the channel $\mathcal{A}$ followed by $\mathcal{B}$, while with probability $1-p$ it applies the channels in the opposite order, acting first with $\mathcal{B}$ and then with $\mathcal{A}$.
In other words, this operation is described by a probabilistic mixture of causally ordered operations.

A more interesting situation occurs when the initial state of the control system is in a \emph{coherent} superposition state, such as $\vert +\rangle = \sfrac{1}{\sqrt{2}}(\vert 0\rangle + \vert 1 \rangle )$.
In this case, the action of $\mathcal{S}(A,B)$ is not equivalent to a convex mixture of the two causally ordered  operations just described, once we perform a measurement on the control system~\footnote{Importantly, the lack of causal order can only be inferred, e.g. by a witness of causal nonseparability~\cite{araujo2015witnessing}. Eq.~\eqref{eq: quantum switch expression} alone does not indicates the `lack of causal order', as can be seen, for example, by tracing out the control system.}.
Instead, we have that 
\begin{align}
    &\mathcal{S}(A,B)(\vert +\rangle \langle + \vert \otimes \rho) =\nonumber \\
    &= \frac{1}{2}AB \rho B^\dagger A^\dagger \otimes \vert 0\rangle \langle 0 \vert + \frac{1}{2}AB \rho A^\dagger B^\dagger \otimes \vert 1\rangle \langle 0 \vert \nonumber \\
    &+\frac{1}{2}BA\rho B^\dagger A^\dagger \otimes \vert 0\rangle \langle 1 \vert + \frac{1}{2}BA\rho A^\dagger B^\dagger \otimes \vert 1 \rangle \langle 1 \vert. \label{eq: quantum switch expression}
\end{align}
One can rewrite Eq.~\eqref{eq: quantum switch expression} as follows: 
\begin{align*}
    &\mathcal{S}(A,B)(\vert +\rangle \langle + \vert  \otimes \rho) = \\&\frac{1}{4}\Bigr ( \vert +\rangle \langle + \vert \otimes \{A,B\}\rho\{A,B\}^\dagger + \vert -\rangle \langle + \vert \otimes [A,B]\rho\{A,B\}^\dagger \\ &+ \vert +\rangle \langle - \vert  \otimes \{A,B\}\rho[A,B]^\dagger + \vert -\rangle \langle - \vert  \otimes [A,B]\rho[A,B]^\dagger \Bigr).
\end{align*}
In this form, it becomes clear how the quantum switch can discriminate between unitaries $A,B$ that either commute ($[A,B] = 0$) or anti-commute ($\{A,B\} = 0$).
This has been recognized as one of the most advantageous applications of the quantum switch in Ref.~\cite{chiribelly2012perfect}, and implemented experimentally in Ref.~\cite{procopio2015experimental}.

If we then measure the control qubit in the $X = \{\ket{+}, \ket{-} \}$ basis, the probability of observing $\ket{-}$ is 
\begin{equation}\label{eq: quantum switch prob v1}
    p_- = \frac{1}{4}\Tr(\rho\ [A,B]^\dagger [A,B]).
\end{equation}
Expanding this commutator, we find that 
\begin{align*}
    [A,B]^\dagger [A,B] = 2 \, \mathbb{1} - 2\left( \frac{D + D^\dagger}{2} \right) = 2 (\mathbb{1} - \Re[D]),
\end{align*}
where $D = D(A,B)$ is given by 
\begin{equation}\label{eq: definition of D}
    D(A,B) := A^\dagger B^\dagger AB.
\end{equation} 
We can thus rewrite $p_-$ from Eq.~\eqref{eq: quantum switch prob v1} as 
\begin{align}\label{eq: quantum switch prob}
    p_-
    &= \frac{1}{2} - \frac{1}{2}\Tr(\rho\,\Re[D]) \nonumber\\ 
    &= \frac{1}{2} - \frac{1}{2}\Re[\Tr(\rho\,D)].
\end{align}

Similarly, using  Eq.~\eqref{eq: quantum switch expression}, if we measure the control qubit in the $Y=\{\ket{+_i}, \ket{-_i} \}$ basis, the probability that we observe  $\ket{-_i}$ is
\begin{align}
    p_{-_i} = \frac{1}{2}\left( 1 + \frac{i}{2}\Tr\left( \rho [B,A]^\dagger \{B,A\} \right)\right).
\end{align}
Proceeding as before, we obtain 
\begin{alignb}\label{eq: QS Y}
    p_{-_i} &= \frac{1}{2}\left(1 +\frac{i}{2}\Tr(AB\rho A^\dagger B^\dagger) - \frac{i}{2}\Tr( BA \rho B^\dagger A^\dagger)\right)    \\
    &=\frac{1}{2}\left(1 - \frac{\Tr(\rho D)-\Tr(\rho D^\dagger)}{2i}\right) \\
    &= \frac{1}{2}\left(1 -  \Tr\left(\rho \left[\frac{D-D^\dagger}{2i}\right]\right)\right) \\
    &= \frac{1}{2} - \frac{1}{2}\Tr(\rho \, \Im[D]) \\
    &= \frac{1}{2} - \frac{1}{2}\Im[\Tr(\rho \, D)],
\end{alignb}

In summary, measuring the control qubit in the $X$ or $Y$ basis after applying the quantum switch $\mathcal{S}(A,B)$ on input state $\ket{+}\!\bra{+} \otimes \rho$ allows one to estimate the real or imaginary parts of the quantity $\Tr(\rho\, D(A,B))$.

It is through a generalization of Eq.~\eqref{eq: quantum switch prob} that connections between Bargmann invariants and the quantum switch have been previously identified in the literature.
Masashi Ban~\cite{ban2021sequential} was the first to observe -- focusing on the optical implementation of the quantum switch -- that the statistics of control measurement outcomes depend on the standard Kirkwood--Dirac quasiprobability distribution, and hence on third-order Bargmann invariants~\cite{wagner2024quantumcircuits}.

Motivated by a different task, namely that of quantifying the incompatibility of sharp measurements, Gao~et~al.~\cite{gao2023measuring} demonstrated a setting in which the real part of fourth- and fifth-order Bargmann invariants appears in Eq.~\eqref{eq: quantum switch prob}. In that work, the input channels of the quantum switch are dephasing (hence non-unitary) channels.

We show in Sec.~\ref{sec:results} that these connections are not accidental, by demonstrating that Eq.~\eqref{eq: quantum switch prob} can be used to estimate the \emph{real} part of any Bargmann invariant of arbitrary order.
We further extend these findings by showing that Eq.~\eqref{eq: QS Y} enables the estimation of the \emph{imaginary} part of these invariants.

To conclude our brief overview of the quantum switch, we recall that although the (unitary) quantum switch cannot be expressed as a quantum circuit, it can be \emph{simulated} by one, with the catch that the simulating process may need access to more than one copy of the input unitaries, as it queries them more than once.
To introduce the relevant notion of simulation, we consider a generic causally ordered higher-order process~\footnote{In our case, since we focus on quantum circuits, we are interested in situations where $\mathcal{M}$ describes a \emph{customisable} quantum circuit~\cite{taranto2025higherorderquantumoperations}. This refers to a setting where the circuit architecture is fixed, but some unitary gates are treated as variable inputs.}
$\mathcal{M}$ having inputs $(V_1,\dots,V_{k_A},W_{1},\dots,W_{k_B})$, for some $k_A,\, k_B \in \mathbb{N}$. We want to consider the situation where all the inputs labelled $V_j$ are equal, i.e.~copies of the same unitary $A$, and similarly those labelled $W_j$ are copies of the same unitary $B$. Given a unitary $U$ we write $$\mathbf{U}_{k} = (\underbrace{U,\dots,U}_{k \text{ times}})$$
for the tuple consisting of $k$ copies of $U$. 
Then, we say that $\mathcal{M}$ \emph{simulates} the quantum switch $\mathcal{S}$ if for every unitary $A$ and $B$ we have that 
\begin{equation}
\mathcal{M}(\mathbf{A}_{k_A},\mathbf{B}_{k_B}) = \mathcal{S}(A,B).
\end{equation}
We refer to $\mathcal{M}$ as a $(k_A,k_B)$-simulation. Note that our notion of simulation is restricted to unitary inputs to the quantum switch. Therefore, it is strictly weaker than what was discussed recently by Bavaresco et al.~\cite{bavaresco2024can}.

This notion is relevant to the study of query complexity advantages provided by the quantum switch. In principle, the quantum switch $\mathcal{S}(A,B)$ requires only a single use of each operation $A$ and $B$.
The switch thus offers an advantage depending on the number of \emph{additional} uses of $A$ and $B$ required to simulate $\mathcal{S}(A,B)$.
In the notation above, $k_A$ denotes the number of uses of $A$, and similarly $k_B$ denotes the number of uses of $B$. We can use such ordered pairs as a figure of merit for evaluating quantum circuits that aim to reproduce the same statistics as the quantum switch, albeit at the cost of applying the channels $A$ and $B$ more frequently, i.e., with $k_A > 1$ or $k_B > 1$.

\begin{figure}[t]
    \centering
    \includegraphics[width=\columnwidth]{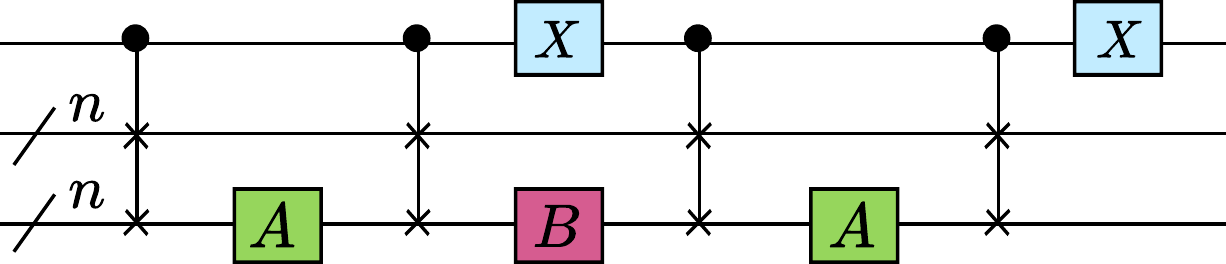}
    \caption{\textbf{Simulation of the quantum switch.} From top to bottom, the first wire represents the control system, and the second wire represents an auxiliary system of the same dimension as the system of interest. Tracing out the auxiliary system allows us to recover the same output state as the switch (Eq.~\eqref{eq: switcheq}).}
    \label{fig: sim qs}
\end{figure}

We say that a simulation of the quantum switch is \emph{deterministic} if it does not involve post-selecting the results by measuring the auxiliary system and discarding unsuccessful rounds.
There is no quantum circuit that can deterministically reproduce the statistics of the quantum switch for every possible pair of input unitary channels with
$k_A=k_B=1$~\cite{bavaresco2024can}.
Nevertheless, there exist deterministic simulations with $k_A>1$ or $k_B>1$~\cite{chiribella2013quantum,bavaresco2024can,renner2021reassessing}. Fig.~\ref{fig: sim qs} shows a deterministic $(2,1)$-simulation of the quantum switch from Ref.~\cite{chiribella2013quantum}. We refer to Ref.~\cite{bavaresco2024can} for the investigation of other $(k_A,k_B)$-simulations. 

\section{Measurement protocol}\label{sec:results}

\begin{figure*}[t]
    \centering
    \includegraphics[width=0.85\textwidth]{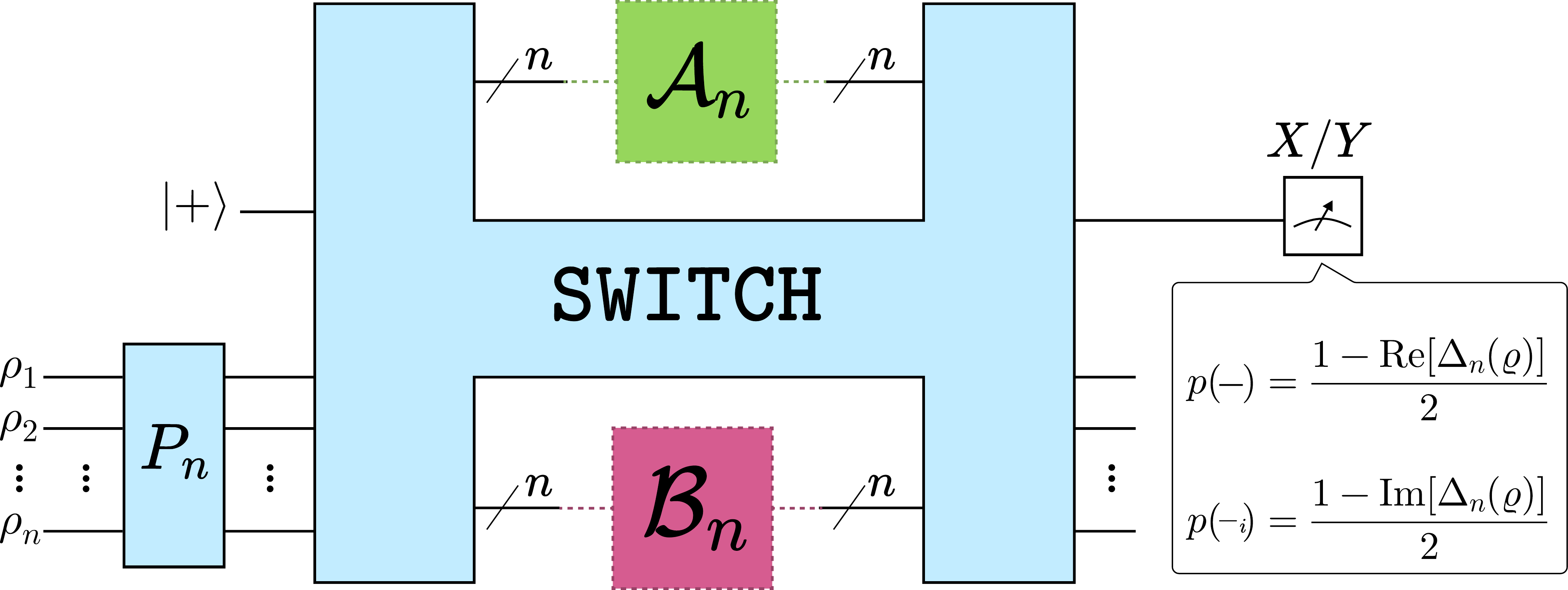}
    \caption{\textbf{Protocol for measuring Bargmann invariants $\Tr(\rho_1 \cdots \rho_n)$ with the quantum switch.} The operations $A_n$ and $B_n$ from Eq.~\eqref{eq: AandB} are inputs of the quantum switch $\mathcal{S}$. The resulting quantum channel $\mathcal{S}(A_n,B_n)$ then acts on $\vert +\rangle \langle + \vert \otimes P_n(\rho_1 \otimes \dots \otimes \rho_n) P_n^\dagger $ where $P_n$ is a permutation of $n$ symbols detailed in Eqs.~\eqref{eq: permutation P} and~\eqref{eq: inputState} (see also Appendix~\ref{app: proof of the rules}). To estimate the real (imaginary) part of the invariants we measure the auxiliary qubit in the $X$ ($Y$) basis.}
    \label{fig:protocol_figure}
\end{figure*}

We show how one can use the quantum switch to estimate generic Bargmann invariants of any order.
In summary, we use the unitaries $A_n$ and $B_n$ from Eq.~\eqref{eq: AandB} as input operations to the quantum switch after applying a pre-processing permutation to the initial states such that the probabilities of measuring the control qubit at the end of the protocol in the $X$ or $Y$ bases depend on the real or imaginary parts of the Bargmann invariant of order $n$. This protocol is shown in Fig.~\ref{fig:protocol_figure}.
More specifically, in Sec.~\ref{sec:odd order} we describe the permutation of the initial states and how measuring the control qubit in different bases leads us to Bargmann invariants.
Then, in Sec.~\ref{sec: evenBargmann} we explain why this protocol only works for odd $n$ and how repeating one of the input states allows us to access the even-order invariants.

\subsection{Odd-order invariants}\label{sec:odd order}

{

To estimate an invariant of odd order $n$, we propose the protocol depicted in Fig.~\ref{fig:protocol_figure}. The protocol applies the quantum switch with input unitaries $A_n$ and $B_n$ from Eq.~\eqref{eq: AandB} acting on a Hilbert space $\HH^{\otimes n}$, implementing the unitary channel $\mathcal{S}(A_n,B_n)$ on $\CC^2 \otimes \HH^{\otimes n}$.
The control qubit is initialized in state $\ket{+}$ and is measured subsequently in the $X$ or $Y$ basis.
As explained in Sec.~\ref{ssec:qswitch},
the outcome probabilities given in Eqs.~\eqref{eq: quantum switch prob} and~\eqref{eq: QS Y}
allow us to estimate the real or imaginary part of the quantity $\Tr({\rho_{\mathrm{sw}}} D(A_n,B_n))$,
where ${\rho_{\mathrm{sw}}}$ is the input state to the switch on the system Hilbert space $\HH^{\otimes n}$.

Since $A_n$ and $B_n$ are Hermitian (as the associated permutations $\permAn$ and $\permBn$ have order 2), we have
$$D(A_n,B_n) = A_n^\dagger B_n^\dagger A_nB_n = A_n B_n A_n B_n.$$
The corresponding permutation $\permAn\permBn\permAn\permBn$ turns out to be an $n$-cycle, hence conjugate to the cycle $\permCn$.
This means there exists a permutation $\permPn$ whose corresponding unitary $P_n$ satisfies
\begin{equation}\label{eq: condition}
    C_n = P_n^\dagger A_n B_n A_n B_n P_n.
\end{equation}
Consequently, if one would feed the state $\rho_1 \otimes \cdots \otimes \rho_n$ to the switch
directly, it would estimate the $n$th-order invariant corresponding to a permutation of the tuple $\varrho = (\rho_1,\ldots,\rho_n)$.
So, as in the protocol of Quek et al.~\cite{quek2024multivariatetrace}, in order to estimate the value of the desired Bargmann invariant $\Delta_n(\rho_1, \ldots, \rho_n)$ (with the states in increasing index order),
we must compensate by permuting the input subsystems.
This preprocessing step applies the unitary $P_n$ to the protocol's input state $\rho = \rho_1 \otimes \cdots \otimes \rho_n$, transforming it into the state
\begin{align}
    {\rho_{\mathrm{sw}}} &= P_n \rho P_n^\dagger \nonumber
    \\&= P_n(\rho_1 \otimes \cdots \otimes \rho_n)P_n^\dagger \nonumber
    \\&=\rho_{\permPn^{-1}(1)}\otimes \cdots \otimes \rho_{\permPn^{-1}(n)} , \label{eq: permutation P}
\end{align}
which is then fed as the input to the switch.
As desired, the subsequent measurement in the $X$ basis ($Y$ basis) allows us to estimate the real (imaginary) part of the quantity
\begin{align*}
\Tr({\rho_{\mathrm{sw}}} D(A_n,B_n))
   &= \Tr(P_n \rho P_n^\dagger A_nB_nA_nB_n)
 \\&= \Tr(\rho P_n^\dagger A_nB_nA_nB_n P_n)
 \\&= \Tr(\rho C_n)
 \\&= \Delta_n(\varrho).
\end{align*}

We explicitly describe the permutation $\permPn$ by the indices
$a_j \equiv \permPn^{-1}(j)$ that appear in Eq.~\eqref{eq: permutation P}.
Setting $n=2k+1$, these are given by 
\begin{alignb}\label{eq: inputState}
    a_{4s+1} &= s+1 \\
    a_{4s+2} &= k+1-s \\
    a_{4s+3} &= k+s+2 \\
    a_{4s+4} &= 2k+1-s,
\end{alignb}
for $s \in \{0, 1, 2, \ldots, \lfloor \sfrac{n}{4} \rfloor\}$.
As a concrete example, for $n=5$ we have that $k=2$, and the above expressions yield
$$a_1 = 1,\;\; a_2 = 3,\;\; a_3 = 4,\;\; a_4 = 5, \;\; a_5 = 2,$$
so that the state of the system entering the quantum switch is
\begin{equation}\label{eq:rhoin5}
    {\rho_{\mathrm{sw}}} = \rho_1\otimes \rho_3 \otimes \rho_4 \otimes \rho_5 \otimes \rho_2.
\end{equation}
We show in Appendix~\ref{app: proof of the rules} that the permutation described by the rule in Eq.~\eqref{eq: inputState} indeed satisfies the condition from Eq.~\eqref{eq: condition}.

}

Our proposed protocol generalizes early results from Refs.~\cite{ban2021sequential,gao2023measuring} which noticed the possibility of estimating the real part of Bargmann invariants of order up to $5$. With the choice of operations $A_n$ and $B_n$ (from Eq.~\eqref{eq: AandB}) and initial state ${\rho_{\mathrm{sw}}}$, we can estimate both the real and imaginary parts of any $n$th order ($n$ odd) Bargmann invariant $\Delta_n(\varrho)$ using the quantum switch.
These are, however, not the only possible choices of operations and initial state for which the statistics of the quantum switch provide the value of Bargmann invariants. We discuss some alternatives in Appendix~\ref{appendix: proof first eq: ansatz}.

We now illustrate this protocol by working through explicit calculations for small values of $n$.

\begin{example}[Third-order invariant]
Taking the case $n=3$ as an example, let us consider for simplicity the estimation of a pure state Bargmann invariant $\Delta_3(\psi_1,\psi_2,\psi_3)$.  The input state to the quantum switch is
$$\ket{{\psi_{\mathrm{sw}}}} = P_3(\ket{\psi_1\psi_2\psi_3}) =\ket{\psi_1\psi_2\psi_3},$$
so that in this case $P_3$ is the identity,
and the unitary operations $A_3$ and $B_3$ are
\begin{equation}
    A_3 = \SWAP_{1,2} \quad \text{and} \quad B_3 = \SWAP_{2,3}.
\end{equation}
The following calculation shows that the outcome probabilities $p_-$ and $p_{-i}$ depend on the real and imaginary parts of the desired third-order invariant:

\begin{align*}
    &\Tr(A_3 B_3 A_3 B_3 {\psi_{\mathrm{sw}}}) \\
    &= \Tr(A_3 B_3 A_3 B_3(\psi_1 \otimes \psi_2 \otimes \psi_3)) \\
    &= \bra{\psi_1\psi_2\psi_3} A_3 B_3 A_3 B_3\ket{\psi_1\psi_2\psi_3} \\
    &= \braket{\psi_1\psi_2\psi_3}{\psi_2\psi_3\psi_1} \\
    &= \braket{\psi_1}{\psi_2}\braket{\psi_2}{\psi_3}\braket{\psi_3}{\psi_1}\\
    &= \Delta_3(\psi_1,\psi_2,\psi_3).
\end{align*}
\end{example}

\begin{figure}[t]
    \centering
    \includegraphics[width=0.6\columnwidth]{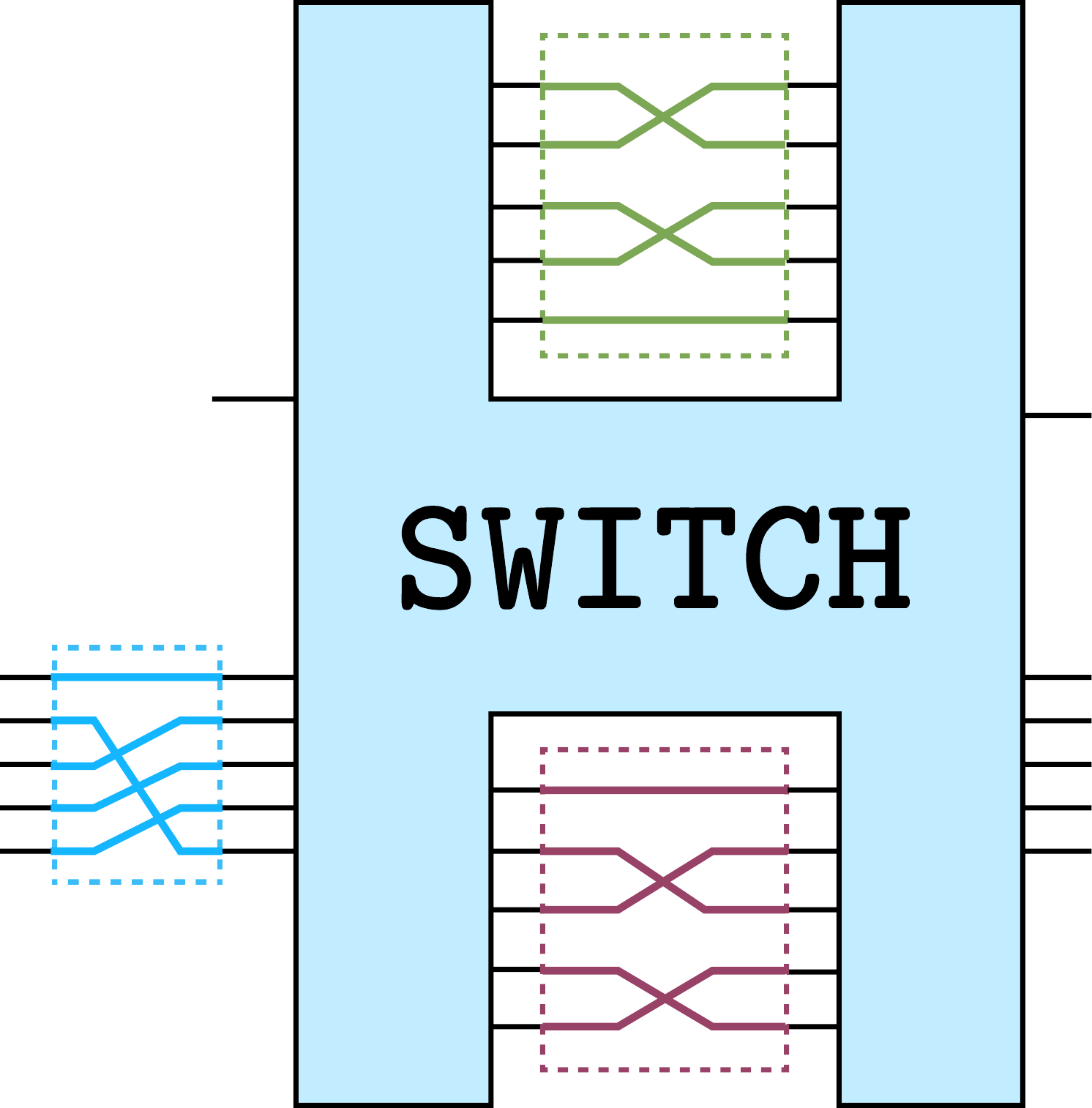}
    \caption{\textbf{Quantum switch with $A_5$ and $B_5$ as input operations.} 
    This figure illustrates the case where we estimate $\Delta_5(\varrho)$ using the operations $A_5$ and $B_5$ as input operations for the quantum switch. The preprocessing permutation $P_5$ is also shown. The input composite system to the switch is given by $\vert +\rangle \langle +\vert \otimes \rho_1 \otimes \rho_3 \otimes \rho_4 \otimes \rho_5 \otimes \rho_2$.}
    \label{fig:qs2}
\end{figure}

\begin{example}[Fifth-order invariant]
Since for $n=3$ the state $\ket{{\psi_{\mathrm{sw}}}}$ is already in increasing index order $\ket{\psi_1\psi_2\psi_3}$, we consider also the less trivial case of $n=5$. 
The input state to the quantum switch is
$\ket{{\psi_{\mathrm{sw}}}} = \ket{\psi_1\psi_3\psi_4\psi_5\psi_2}$ as per Eq.~\eqref{eq:rhoin5},
and the unitary operations $A_5$ and $B_5$ are
\begin{equation}
    A_5 = \SWAP_{1,2}\SWAP_{3,4},\,\,B_5 = \SWAP_{2,3}\SWAP_{4,5}.
\end{equation}
Just as for $n=3$, we have that
\begin{align*}
    &\Tr(A_5 B_5 A_5 B_5 {\psi_{\mathrm{sw}}})\\
    &= \Tr(A_5 B_5 A_5 B_5(\psi_1 \otimes \psi_3\otimes \psi_4 \otimes \psi_5 \otimes \psi_2)) \\
    &= \bra{\psi_1\psi_3\psi_4\psi_5\psi_2} A_5 B_5 A_5 B_5\ket{\psi_1\psi_3\psi_4\psi_5\psi_2} \\
    &= \braket{\psi_1\psi_3\psi_4\psi_5\psi_2}{\psi_2\psi_4\psi_5\psi_1\psi_3} \\
    &= \braket{\psi_1}{\psi_2}\braket{\psi_3}{\psi_4}\braket{\psi_4}{\psi_5}\braket{\psi_5}{\psi_1}\braket{\psi_2}{\psi_3} \\
    &= \braket{\psi_1}{\psi_2}\braket{\psi_2}{\psi_3}\braket{\psi_3}{\psi_4}\braket{\psi_4}{\psi_5}\braket{\psi_5}{\psi_1} \\
    &= \Delta_5(\psi_1,\psi_2,\psi_3,\psi_4,\psi_5).
\end{align*}
We show the unitary quantum switch relevant for this $n=5$ test in Fig.~\ref{fig:qs2}.
\end{example}

\subsection{Even-order invariants}\label{sec: evenBargmann}

In the previous section, we restricted the order $n$ of Bargmann invariants to be odd.
To understand why this restriction is necessary, we begin by examining the notion of \emph{parity} of permutations.

Any permutation can be written as a product of transpositions (permutations swapping only two elements).
While such decompositions are not unique, their parity is.
A permutation is said to be \emph{even} (\emph{odd}) if it can be written as a product of an even (odd) number of transpositions.
Importantly, permutation parity defines a group homomorphism from the symmetric group $S_n$ to $\mathbb{Z}_2$, the group of integers modulo $2$.
The cycle permutation $\permCn$ is \emph{odd} if $n$ is even, and it is \emph{even} if $n$ is odd, as can be seen from Eq.~\eqref{eq:cn-transpositions}.
This difference has direct consequences for the feasibility of estimating Bargmann invariants using the method discussed in Sec.~\ref{sec:odd order}.

Indeed, one may ask whether even-order Bargmann invariants can also be estimated using a similar protocol.
Specifically, we pose the following question: for $n$ even, are there permutations $\permA$, $\permB$, and $\permP$ whose corresponding unitaries $A$, $B$, and $P$ satisfy
\[
C_n = P^\dagger D(A,B) P = P^\dagger A^\dagger B^\dagger A B P ?
\]
Equivalently, at the level of permutations, this would require 
\[
\permCn = \permP^{-1}\permA^{-1}\permB^{-1}\permA\permB\permP.
\]
To see why this is impossible, note that the product of permutations on the right must necessarily be an \emph{even} permutation regardless of the parities of the three chosen permutations.
This is because parity is a homomorphism to an abelian group and each of these permutations appears paired with their inverse in the expression, with each such pair contributing trivially to the overall parity.
On the other hand, the cycle permutation $\permCn$ is \emph{odd} when $n$ is even,
so the desired equation cannot hold.

The above discussion leads naturally to a more general no-go result for estimating even-order Bargmann invariants using protocols with the same structure as our proposal for odd-order invariants shown in~Fig.~\ref{fig:protocol_figure}. Not only does it exclude any protocol of this shape where the quantum switch inputs, as well as the pre-processing step, are permutation matrices, it also excludes similar protocols where these components are arbitrary unitaries.
The key insight enabling this generalization is to regard the determinant as a complex-valued (or in the case of unitaries, unit-circle-valued) extension of permutation parity: indeed, for any permutation $\permP$ in $S_n$, the corresponding unitary $P$ has $\det P = (-1)^{\mathrm{parity}(\permP)}$.
What is more, the result also excludes similar protocols that use a variant of the quantum switch with $m$ input operations where a control qubit determines whether these $m$ operations are applied in a given sequential order or its reverse (the standard quantum switch corresponds to $m=2$).

\begin{theorem}\label{theorem: no-go results}
    For $n$ even, there exist no unitaries $P, U_1, U_2, \dots, U_m$ such that
    \begin{equation}\label{eq: theorem no go}
        C_{n} = P^\dagger U_1^\dagger U_2^\dagger \cdots U_m^\dagger U_1 U_2 \cdots U_m P.
    \end{equation}
\end{theorem}
\begin{proof}
    The result follows by taking the determinant on both sides of Eq.~\eqref{eq: theorem no go}.
    The left-hand side has determinant $-1$ since $n$ is even.
    The right-hand side, however, always has determinant $+1$ for any choice of unitaries $P,U_1,\dots,U_m$, for much the same reason as in the case of permutations (namely, $\det$ is a homomorphism from the unitary group to an abelian group, the unit circle $U(1)$, and each unitary can be paired with its adjoint/inverse in the expression on the right, so their contributions to the determinant cancel out).
\end{proof}

Theorem~\ref{theorem: no-go results} forces us to investigate \emph{alternative} ways to estimate even-order Bargmann invariants using the quantum switch. 
The no-go result does not mean that even-order Bargmann invariants are wholly inaccessible using our method, but simply that the same strategy employed for odd-order invariants cannot be directly applied to the even-order case.
One possible approach to access even-order invariants involves \emph{repeating} one of the input states.
This circumvents the no-go theorem by using the protocol for order $2k+1$ to estimate a Bargmann invariant of order $2k$,
but it also brings in some additional assumptions.

If we take, for instance, the protocol for $n=3$ and set the initial state of the system of interest to be $\ket{\psi} = \ket{\psi_1} \otimes \ket{\psi_2} \otimes \ket{\psi_2}$, then the final state (after the second Hadamard gate is applied to the control qubit in order to perform an $X$-basis measurement) will be
\begin{align*}
    \frac{1}{2} \Big[ &\ket{0} \otimes \big( \ket{\psi_2 \psi_2 \psi_1} + \ket{\psi_2 \psi_1 \psi_2} \big) \\
    + &\ket{1} \otimes \big( \ket{\psi_2 \psi_2 \psi_1} - \ket{\psi_2 \psi_1 \psi_2} \big) \Big].
\end{align*}
Since $\braket{\psi_2}{\psi_2}=1$,
the probability of obtaining outcome $1$ when measuring the control qubit in the computational basis -- which is equivalent to the probability $p_-$ given that a second Hadamard gate has already been applied --
will be $(1 - \Re[\Delta_2(\psi_1,\psi_2)])/2$, where $\Delta_2(\psi_1,\psi_2) = \braket{\psi_1}{\psi_2} \braket{\psi_2}{\psi_1} = |\langle \psi_1|\psi_2\rangle |^2$ is a second-order Bargmann invariant. 

This approach can be generalized for arbitrary even-order Bargmann invariants of \emph{pure states} as long as one has access to at least two copies of one of them.

It follows from the following simple fact about pure-state Bargmann invariants:
given an $n$-tuple of pure states
$\psi = (\psi_1,\ldots,\psi_n)$ and taking $\psi' = (\psi_1,\psi_1,\dots,\psi_n)$ to be the $(n+1)$-tuple with the first element repeated,
it holds that 
\begin{align}\label{eq: Bargmann pure trivial}
    \Delta_{n+1}(\psi') &= \Delta_{n+1}(\psi_1,\psi_1,\ldots,\psi_n) \\
    &=\Delta_n(\psi_1,\ldots,\psi_n) = \Delta_n(\psi).\nonumber 
\end{align}

Hence, to estimate pure-state even-order Bargmann invariants $\Delta_{2k}(\psi)$, one can simply use the protocol discussed before for estimating $\Delta_{2k+1}(\psi')$. Note that one could have repeated any of the other elements in $\psi$, with the two copies appearing consecutively in $\psi'$.
Of course, each run of this protocol will consume two copies of one of the states. In a situation where one has access to an ensemble of copies of each of the states, one may choose to repeat a different position of the tuple in each run.

Moving on to mixed states, the same approach works verbatim for tuples of the form $(\psi_1,\rho_2,\ldots,\rho_n)$, where at least one state in the tuple -- specifically, the one being repeated -- is pure.
It does not, however, directly extend to arbitrary tuples of mixed states.
Still, we can leverage that construction to propose a general method for estimating even-order Bargmann invariants $\Delta_n(\varrho)$ on tuples of mixed states $\varrho = (\rho_1,\rho_2\dots,\rho_{n})$.
This approach requires an additional assumption: we must have classical \emph{knowledge} of at least one of the input states. That is, we need a description of the state rather than just access to quantum systems prepared in that state.

Suppose we have a convex decomposition of the state $\rho_1$ into pure states: $\rho_1 = \sum_{j=1}^M \alpha_j \psi_{j}$.
We measure all $M$ Bargmann invariants of order $n+1$ with repeated pure states $\psi_{j}$, i.e.
\begin{equation*}
    \Delta_{n+1}(\psi_{j},\psi_{j},\rho_2,\dots,\rho_n),
\end{equation*}
using the protocol described in Sec.~\ref{sec:odd order}.
We then post-process the results using the convex weights $\alpha_j$, as follows:
\begin{align*}
    \Delta_n(\varrho)
       &=  \Delta_{n}(\textstyle\sum\nolimits_{j=1}^M \displaystyle \alpha_j \psi_{j},\rho_2,\dots,\rho_n).
    \\ &= \sum_{j=1}^M \alpha_j \Delta_{n}(\psi_{j},\rho_2,\dots,\rho_n).
    \\ &= \sum_{j=1}^M \alpha_j \Delta_{n+1}(\psi_{j},\psi_{j},\rho_2,\dots,\rho_n).
\end{align*}
While we have chosen to decompose $\rho_1$, the same argument applies to any other state in the tuple $\varrho$.
In this way, provided that we can find and prepare some convex decomposition of at least one of the states, the protocol allows us to measure even-order Bargmann invariants.

\section{Simulating the quantum switch}\label{sec: simulation Hadamard qs}

Generic protocols for simulating the unitary quantum switch operation $S_{A, B}$  typically consider the situation where one has black-box access to the unitaries $A$ and $B$~\cite{chiribella2013quantum,bavaresco2024can,bavaresco2021strict,renner2021reassessing}.
An alternative possibility is to have access not only to forward dynamics $A$ and $B$, but also to backwards-in-time dynamics $A^\dagger$ and $B^\dagger$.
Here, we show that access to such backward dynamics enables the use of a Hadamard test for the simulation of the quantum switch operation $S_{A, B}$ for any choice of unitaries $A$ and $B$. 

As noted before, 
As we have noted before, the state of the system after the quantum switch unitary $S_{A, B}$, for a generic pair of operators $A$ and $B$ acts on a state $\vert +\rangle \otimes \vert \psi \rangle $, and we apply a Hadamard to the control qubit, the state of the system becomes
\begin{equation}\label{psiQS}
    \ket{\psi_f^{\mathrm{SWITCH}}} = \frac{1}{2} \Big( \ket{0} \otimes \{B,A\} \ket{\psi} + \ket{1} \otimes [B,A] \ket{\psi} \Big).
\end{equation}

After a generic Hadamard test featuring a controlled-unitary $U$ the control qubit and the system are left in the state
\begin{equation}\label{hadamard}
    \ket{\psi_f^{\mathrm H}} = \frac{1}{2} \Big[ \ket{0} \otimes (\mathbb{1}+U) \ket{\phi} + \ket{1} \otimes (\mathbb{1}-U) \ket{\phi} \Big].
\end{equation}

\begin{figure}[t]
	\centering
	\includegraphics[width=\columnwidth]{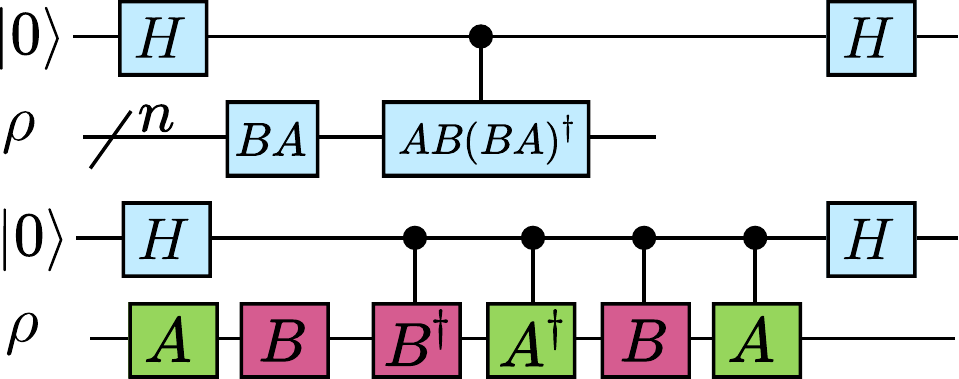}
	\caption{\textbf{Quantum switch simulation using a modified Hadamard test.} A quantum switch that superposes the order of unitary operations $A$ and $B$ can be simulated by this circuit, with $AB$ applied to the input state of a Hadamard test with a suitable controlled operation by a unitary $U= ABA^\dagger B^\dagger$. If the control state is $\ket{0}$, the lower registers undergo a transformation by the unitary $BA$, otherwise, they evolve under the unitary $AB$.}
	\label{fig: simulgeral}
\end{figure}

Now, we note that if the initial state for the Hadamard test $\ket{\phi}$ is prepared as $\ket{\phi} = BA \ket{\psi}$ and the unitary is chosen to be $U=AB(BA)^{\dagger}=ABA^\dagger B^\dagger$, the final state can be written as 
\begin{alignb}
    \ket{\psi_f^{\rm H}} = \frac{1}{2} \Big[ &\ket{0} \otimes (BA+AB) \ket{\psi} \\ 
    + &\ket{1} \otimes (BA-AB) \ket{\psi} \Big],
\end{alignb}
which is the same final state as that obtained from the quantum switch operation $S_{A,B}$.
Such modified Hadamard test is shown in Fig.~\ref{fig: simulgeral}.

This simulation of a quantum switch using a modified Hadamard test offers resource (space) efficiency by requiring only a single control qubit, as opposed to other quantum simulations such as the one from Fig.~\ref{fig: sim qs}, which require an auxiliary system of the same dimension as the system of interest.
On the other hand, it also introduces challenges related to gate complexity and experimental feasibility, as implementing controlled versions of operations like $ABA^\dagger B^\dagger$ can, in principle, increase circuit depth and lead to additional errors. 

To connect with the previous section, it is easy to see that if $ABA^\dagger B^\dagger = C_n$, we recover a structure similar to the cycle test in Fig.~\ref{fig: cycletest}, which can also be used to measure Bargmann invariants. In Appendix~\ref{appendix: proof first eq: ansatz}, we describe permutations satisfying this constraint.

Many simulations of the quantum switch have already been studied. Our simulation requires at least two calls of each operation $A$ and $B$. This would make it a $(2,2)$-simulation if not for the need for an additional call of each of their \emph{inverses}, creating an altogether different category of switch simulation. In fact, it is possible, albeit far from easy, to deterministically implement an inverse $A^\dagger$ using four calls to a black box implementing $A$~\cite{yoshida2023reversing}.
Using that protocol as a subroutine to implement both $A^\dagger$ and $B^\dagger$, the query complexity of our simulation is $(2+4,2+4)$.

\section{Conclusions}\label{sec:conclusions}

In this work, we demonstrated that by carefully choosing unitaries $A$ and $B$ as input operations to the quantum switch, a causally non-separable higher-order quantum process, we can estimate Bargmann invariants $\Delta_n(\varrho)$ of arbitrary order. We exhibit different choices of input unitaries $A$ and $B$ to the quantum switch supermap that achieve this purpose.

Furthermore, building upon the connections we explored between cycle tests and the quantum switch, we designed a universal deterministic simulation of the unitary quantum switch based on a Hadamard test. When compared to other simulation techniques, our proposal uses queries not only of the unitary inputs $A$ and $B$, but \emph{also}  of their backwards-in-time versions $A^\dagger$ and $B^\dagger$, which goes beyond what is usually considered in the literature. The Hadamard test is only one instance of the more general technique of linear combination of unitaries (LCU), which implement superpositions of different unitary dynamics, with applications to quantum simulation~\cite{childs2012} and optimization problems~\cite{Chow17}. This suggests a possible avenue for future research, exploring more systematically how LCU techniques can shed light on simulations and applications of other supermaps with indefinite causal order.  

Other proposed simulations of the quantum switch may also be used to estimate Bargmann invariants, as long as the input operations $A$ and $B$ are chosen such that $ABA^\dagger B^\dagger=C_n$ is fulfilled. As it simulates the quantum switch, any such circuit must reproduce the statistical behaviour described in Eqs.~\eqref{eq: quantum switch prob} and~\eqref{eq: QS Y}, which is key for our protocol to work. Consequently, such quantum circuits are able to estimate Bargmann invariants following the same procedure we proposed for the quantum switch. 

By measuring Bargmann invariants with the quantum switch, we are connecting two distinct research areas already rich in applications: one devoted to investigating indefinite causal order and its usefulness, with another dedicated to investigating how Bargmann invariants can be used to benchmark quantum devices and nonclassicality.
This connection, albeit simple, unlocks new opportunities. For example, it becomes possible to use the quantum switch to perform Kirkwood--Dirac quantum state tomography~\cite{wagner2024quantumcircuits,schmid2024kirkwood}.
Another possibility is to use the quantum switch to estimate out-of-time-order correlators~\cite{halpern2018quasiprobability,gonzalez2019otoc}, which quantify scrambling of quantum information, by extending our ideas to the case of more general quantum channels. Finally, just as the quantum switch was useful in Ref.~\cite{gao2023measuring} for defining a quantifier of incompatibility between quantum observables, we expect this more general connection with Bargmann invariants will lead to new uses of the quantum switch to quantify other forms of nonclassicality~\cite{budiyono2023operational,budiyono2023quantifying,debievre2021complete}. 

\begin{acknowledgments} 
\noindent We would like to thank Ismael L. Paiva for providing useful comments on an earlier version of this work. PCA acknowledges support in part by the Coordenação de Aperfeiçoamento de Pessoal de Nível Superior – Brasil (CAPES) – Finance Code 001, and also from Conselho Nacional de Desenvolvimento Científico e Tecnológico - Brazil (CNPq - Grant No. 160851/2021-1). RW acknowledges support from the European Research Council (ERC) under the European Union's Horizon 2020 research and innovation programme (grant agreement No. 856432, HyperQ). Both RW and EFG acknowledge support from FCT – Fundaç\~{a}o para a Ciência e a Tecnologia (Portugal) via PhD Grant SFRH/BD/151199/2021 and project CEECINST/00062/2018, respectively. This work was also supported by the
Digital Horizon Europe project \href{https://cordis.europa.eu/project/id/101070558}{FoQaCiA}, \emph{Foundations of Quantum Computational Advantage}, GA no. 101070558.
\end{acknowledgments}

\bibliography{bib}

\begin{thebibliography}{122}%
\makeatletter
\providecommand \@ifxundefined [1]{%
 \@ifx{#1\undefined}
}%
\providecommand \@ifnum [1]{%
 \ifnum #1\expandafter \@firstoftwo
 \else \expandafter \@secondoftwo
 \fi
}%
\providecommand \@ifx [1]{%
 \ifx #1\expandafter \@firstoftwo
 \else \expandafter \@secondoftwo
 \fi
}%
\providecommand \natexlab [1]{#1}%
\providecommand \enquote  [1]{``#1''}%
\providecommand \bibnamefont  [1]{#1}%
\providecommand \bibfnamefont [1]{#1}%
\providecommand \citenamefont [1]{#1}%
\providecommand \href@noop [0]{\@secondoftwo}%
\providecommand \href [0]{\begingroup \@sanitize@url \@href}%
\providecommand \@href[1]{\@@startlink{#1}\@@href}%
\providecommand \@@href[1]{\endgroup#1\@@endlink}%
\providecommand \@sanitize@url [0]{\catcode `\\12\catcode `\$12\catcode `\&12\catcode `\#12\catcode `\^12\catcode `\_12\catcode `\%12\relax}%
\providecommand \@@startlink[1]{}%
\providecommand \@@endlink[0]{}%
\providecommand \url  [0]{\begingroup\@sanitize@url \@url }%
\providecommand \@url [1]{\endgroup\@href {#1}{\urlprefix }}%
\providecommand \urlprefix  [0]{URL }%
\providecommand \Eprint [0]{\href }%
\providecommand \doibase [0]{https://doi.org/}%
\providecommand \selectlanguage [0]{\@gobble}%
\providecommand \bibinfo  [0]{\@secondoftwo}%
\providecommand \bibfield  [0]{\@secondoftwo}%
\providecommand \translation [1]{[#1]}%
\providecommand \BibitemOpen [0]{}%
\providecommand \bibitemStop [0]{}%
\providecommand \bibitemNoStop [0]{.\EOS\space}%
\providecommand \EOS [0]{\spacefactor3000\relax}%
\providecommand \BibitemShut  [1]{\csname bibitem#1\endcsname}%
\let\auto@bib@innerbib\@empty
\bibitem [{\citenamefont {Bargmann}(1964)}]{bargmann1964note}%
  \BibitemOpen
  \bibfield  {author} {\bibinfo {author} {\bibfnamefont {V.}~\bibnamefont {Bargmann}},\ }\bibfield  {title} {\bibinfo {title} {{Note on Wigner’s Theorem on Symmetry Operations}},\ }\href {https://doi.org/10.1063/1.1704188} {\bibfield  {journal} {\bibinfo  {journal} {Journal of Mathematical Physics}\ }\textbf {\bibinfo {volume} {5}},\ \bibinfo {pages} {862} (\bibinfo {year} {1964})}\BibitemShut {NoStop}%
\bibitem [{\citenamefont {Oszmaniec}\ \emph {et~al.}(2024)\citenamefont {Oszmaniec}, \citenamefont {Brod},\ and\ \citenamefont {Galvão}}]{oszmaniec2024measuring}%
  \BibitemOpen
  \bibfield  {author} {\bibinfo {author} {\bibfnamefont {M.}~\bibnamefont {Oszmaniec}}, \bibinfo {author} {\bibfnamefont {D.~J.}\ \bibnamefont {Brod}},\ and\ \bibinfo {author} {\bibfnamefont {E.~F.}\ \bibnamefont {Galvão}},\ }\bibfield  {title} {\bibinfo {title} {Measuring relational information between quantum states, and applications},\ }\href {https://doi.org/10.1088/1367-2630/ad1a27} {\bibfield  {journal} {\bibinfo  {journal} {New Journal of Physics}\ }\textbf {\bibinfo {volume} {26}},\ \bibinfo {pages} {013053} (\bibinfo {year} {2024})}\BibitemShut {NoStop}%
\bibitem [{\citenamefont {Simon}\ and\ \citenamefont {Mukunda}(1993)}]{simon1993Bargmann}%
  \BibitemOpen
  \bibfield  {author} {\bibinfo {author} {\bibfnamefont {R.}~\bibnamefont {Simon}}\ and\ \bibinfo {author} {\bibfnamefont {N.}~\bibnamefont {Mukunda}},\ }\bibfield  {title} {\bibinfo {title} {{Bargmann invariant and the geometry of the G\"uoy effect}},\ }\href {https://doi.org/10.1103/PhysRevLett.70.880} {\bibfield  {journal} {\bibinfo  {journal} {Phys. Rev. Lett.}\ }\textbf {\bibinfo {volume} {70}},\ \bibinfo {pages} {880} (\bibinfo {year} {1993})}\BibitemShut {NoStop}%
\bibitem [{\citenamefont {Mukunda}\ \emph {et~al.}(2001)\citenamefont {Mukunda}, \citenamefont {Arvind}, \citenamefont {Chaturvedi},\ and\ \citenamefont {Simon}}]{mukunda2001Bargmann}%
  \BibitemOpen
  \bibfield  {author} {\bibinfo {author} {\bibfnamefont {N.}~\bibnamefont {Mukunda}}, \bibinfo {author} {\bibnamefont {Arvind}}, \bibinfo {author} {\bibfnamefont {S.}~\bibnamefont {Chaturvedi}},\ and\ \bibinfo {author} {\bibfnamefont {R.}~\bibnamefont {Simon}},\ }\bibfield  {title} {\bibinfo {title} {Bargmann invariants and off-diagonal geometric phases for multilevel quantum systems: A unitary-group approach},\ }\href {https://doi.org/10.1103/PhysRevA.65.012102} {\bibfield  {journal} {\bibinfo  {journal} {Phys. Rev. A}\ }\textbf {\bibinfo {volume} {65}},\ \bibinfo {pages} {012102} (\bibinfo {year} {2001})}\BibitemShut {NoStop}%
\bibitem [{\citenamefont {Mukunda}\ \emph {et~al.}(2003{\natexlab{a}})\citenamefont {Mukunda}, \citenamefont {Arvind}, \citenamefont {Ercolessi}, \citenamefont {Marmo}, \citenamefont {Morandi},\ and\ \citenamefont {Simon}}]{mukunda2003Bargmann}%
  \BibitemOpen
  \bibfield  {author} {\bibinfo {author} {\bibfnamefont {N.}~\bibnamefont {Mukunda}}, \bibinfo {author} {\bibnamefont {Arvind}}, \bibinfo {author} {\bibfnamefont {E.}~\bibnamefont {Ercolessi}}, \bibinfo {author} {\bibfnamefont {G.}~\bibnamefont {Marmo}}, \bibinfo {author} {\bibfnamefont {G.}~\bibnamefont {Morandi}},\ and\ \bibinfo {author} {\bibfnamefont {R.}~\bibnamefont {Simon}},\ }\bibfield  {title} {\bibinfo {title} {Bargmann invariants, null phase curves, and a theory of the geometric phase},\ }\href {https://doi.org/10.1103/PhysRevA.67.042114} {\bibfield  {journal} {\bibinfo  {journal} {Phys. Rev. A}\ }\textbf {\bibinfo {volume} {67}},\ \bibinfo {pages} {042114} (\bibinfo {year} {2003}{\natexlab{a}})}\BibitemShut {NoStop}%
\bibitem [{\citenamefont {Mukunda}\ \emph {et~al.}(2003{\natexlab{b}})\citenamefont {Mukunda}, \citenamefont {Aravind},\ and\ \citenamefont {Simon}}]{mukunda2003Wigner}%
  \BibitemOpen
  \bibfield  {author} {\bibinfo {author} {\bibfnamefont {N.}~\bibnamefont {Mukunda}}, \bibinfo {author} {\bibfnamefont {P.~K.}\ \bibnamefont {Aravind}},\ and\ \bibinfo {author} {\bibfnamefont {R.}~\bibnamefont {Simon}},\ }\bibfield  {title} {\bibinfo {title} {{Wigner rotations, Bargmann invariants and geometric phases}},\ }\href {https://doi.org/10.1088/0305-4470/36/9/312} {\bibfield  {journal} {\bibinfo  {journal} {Journal of Physics A: Mathematical and General}\ }\textbf {\bibinfo {volume} {36}},\ \bibinfo {pages} {2347} (\bibinfo {year} {2003}{\natexlab{b}})}\BibitemShut {NoStop}%
\bibitem [{\citenamefont {Berry}(1984)}]{berry1984quantal}%
  \BibitemOpen
  \bibfield  {author} {\bibinfo {author} {\bibfnamefont {M.~V.}\ \bibnamefont {Berry}},\ }\bibfield  {title} {\bibinfo {title} {Quantal phase factors accompanying adiabatic changes},\ }\href {https://doi.org/https://doi.org/10.1098/rspa.1984.0023} {\bibfield  {journal} {\bibinfo  {journal} {Proceedings of the Royal Society of London. A. Mathematical and Physical Sciences}\ }\textbf {\bibinfo {volume} {392}},\ \bibinfo {pages} {45} (\bibinfo {year} {1984})}\BibitemShut {NoStop}%
\bibitem [{\citenamefont {Avdoshkin}\ and\ \citenamefont {Popov}(2023)}]{avdoshkin2023extrinsic}%
  \BibitemOpen
  \bibfield  {author} {\bibinfo {author} {\bibfnamefont {A.}~\bibnamefont {Avdoshkin}}\ and\ \bibinfo {author} {\bibfnamefont {F.~K.}\ \bibnamefont {Popov}},\ }\bibfield  {title} {\bibinfo {title} {Extrinsic geometry of quantum states},\ }\href {https://doi.org/10.1103/PhysRevB.107.245136} {\bibfield  {journal} {\bibinfo  {journal} {Phys. Rev. B}\ }\textbf {\bibinfo {volume} {107}},\ \bibinfo {pages} {245136} (\bibinfo {year} {2023})}\BibitemShut {NoStop}%
\bibitem [{\citenamefont {Horodecki}(2003)}]{horodecki2003fromlimits}%
  \BibitemOpen
  \bibfield  {author} {\bibinfo {author} {\bibfnamefont {P.}~\bibnamefont {Horodecki}},\ }\bibfield  {title} {\bibinfo {title} {From limits of quantum operations to multicopy entanglement witnesses and state-spectrum estimation},\ }\href {https://doi.org/10.1103/PhysRevA.68.052101} {\bibfield  {journal} {\bibinfo  {journal} {Phys. Rev. A}\ }\textbf {\bibinfo {volume} {68}},\ \bibinfo {pages} {052101} (\bibinfo {year} {2003})}\BibitemShut {NoStop}%
\bibitem [{\citenamefont {Ekert}\ \emph {et~al.}(2002)\citenamefont {Ekert}, \citenamefont {Alves}, \citenamefont {Oi}, \citenamefont {Horodecki}, \citenamefont {Horodecki},\ and\ \citenamefont {Kwek}}]{ekert2002direct}%
  \BibitemOpen
  \bibfield  {author} {\bibinfo {author} {\bibfnamefont {A.~K.}\ \bibnamefont {Ekert}}, \bibinfo {author} {\bibfnamefont {C.~M.}\ \bibnamefont {Alves}}, \bibinfo {author} {\bibfnamefont {D.~K.~L.}\ \bibnamefont {Oi}}, \bibinfo {author} {\bibfnamefont {M.}~\bibnamefont {Horodecki}}, \bibinfo {author} {\bibfnamefont {P.}~\bibnamefont {Horodecki}},\ and\ \bibinfo {author} {\bibfnamefont {L.~C.}\ \bibnamefont {Kwek}},\ }\bibfield  {title} {\bibinfo {title} {{Direct Estimations of Linear and Nonlinear Functionals of a Quantum State}},\ }\href {https://doi.org/10.1103/PhysRevLett.88.217901} {\bibfield  {journal} {\bibinfo  {journal} {Phys. Rev. Lett.}\ }\textbf {\bibinfo {volume} {88}},\ \bibinfo {pages} {217901} (\bibinfo {year} {2002})}\BibitemShut {NoStop}%
\bibitem [{\citenamefont {Alves}\ \emph {et~al.}(2003)\citenamefont {Alves}, \citenamefont {Horodecki}, \citenamefont {Oi}, \citenamefont {Kwek},\ and\ \citenamefont {Ekert}}]{alves2003direct}%
  \BibitemOpen
  \bibfield  {author} {\bibinfo {author} {\bibfnamefont {C.~M.}\ \bibnamefont {Alves}}, \bibinfo {author} {\bibfnamefont {P.}~\bibnamefont {Horodecki}}, \bibinfo {author} {\bibfnamefont {D.~K.~L.}\ \bibnamefont {Oi}}, \bibinfo {author} {\bibfnamefont {L.~C.}\ \bibnamefont {Kwek}},\ and\ \bibinfo {author} {\bibfnamefont {A.~K.}\ \bibnamefont {Ekert}},\ }\bibfield  {title} {\bibinfo {title} {Direct estimation of functionals of density operators by local operations and classical communication},\ }\href {https://doi.org/10.1103/PhysRevA.68.032306} {\bibfield  {journal} {\bibinfo  {journal} {Phys. Rev. A}\ }\textbf {\bibinfo {volume} {68}},\ \bibinfo {pages} {032306} (\bibinfo {year} {2003})}\BibitemShut {NoStop}%
\bibitem [{\citenamefont {Horodecki}\ and\ \citenamefont {Ekert}(2002)}]{horodecki2002method}%
  \BibitemOpen
  \bibfield  {author} {\bibinfo {author} {\bibfnamefont {P.}~\bibnamefont {Horodecki}}\ and\ \bibinfo {author} {\bibfnamefont {A.}~\bibnamefont {Ekert}},\ }\bibfield  {title} {\bibinfo {title} {{Method for Direct Detection of Quantum Entanglement}},\ }\href {https://doi.org/10.1103/PhysRevLett.89.127902} {\bibfield  {journal} {\bibinfo  {journal} {Phys. Rev. Lett.}\ }\textbf {\bibinfo {volume} {89}},\ \bibinfo {pages} {127902} (\bibinfo {year} {2002})}\BibitemShut {NoStop}%
\bibitem [{\citenamefont {Tanaka}\ \emph {et~al.}(2014)\citenamefont {Tanaka}, \citenamefont {Ota}, \citenamefont {Kanazawa}, \citenamefont {Kimura}, \citenamefont {Nakazato},\ and\ \citenamefont {Nori}}]{tanaka2014determining}%
  \BibitemOpen
  \bibfield  {author} {\bibinfo {author} {\bibfnamefont {T.}~\bibnamefont {Tanaka}}, \bibinfo {author} {\bibfnamefont {Y.}~\bibnamefont {Ota}}, \bibinfo {author} {\bibfnamefont {M.}~\bibnamefont {Kanazawa}}, \bibinfo {author} {\bibfnamefont {G.}~\bibnamefont {Kimura}}, \bibinfo {author} {\bibfnamefont {H.}~\bibnamefont {Nakazato}},\ and\ \bibinfo {author} {\bibfnamefont {F.}~\bibnamefont {Nori}},\ }\bibfield  {title} {\bibinfo {title} {Determining eigenvalues of a density matrix with minimal information in a single experimental setting},\ }\href {https://doi.org/10.1103/PhysRevA.89.012117} {\bibfield  {journal} {\bibinfo  {journal} {Phys. Rev. A}\ }\textbf {\bibinfo {volume} {89}},\ \bibinfo {pages} {012117} (\bibinfo {year} {2014})}\BibitemShut {NoStop}%
\bibitem [{\citenamefont {van Enk}\ and\ \citenamefont {Beenakker}(2012)}]{vanenk2012measuring}%
  \BibitemOpen
  \bibfield  {author} {\bibinfo {author} {\bibfnamefont {S.~J.}\ \bibnamefont {van Enk}}\ and\ \bibinfo {author} {\bibfnamefont {C.~W.~J.}\ \bibnamefont {Beenakker}},\ }\bibfield  {title} {\bibinfo {title} {Measuring $\mathrm{Tr}{\ensuremath{\rho}}^{n}$ on single copies of $\ensuremath{\rho}$ using random measurements},\ }\href {https://doi.org/10.1103/PhysRevLett.108.110503} {\bibfield  {journal} {\bibinfo  {journal} {Phys. Rev. Lett.}\ }\textbf {\bibinfo {volume} {108}},\ \bibinfo {pages} {110503} (\bibinfo {year} {2012})}\BibitemShut {NoStop}%
\bibitem [{\citenamefont {Brun}(2004)}]{brun2004measuring}%
  \BibitemOpen
  \bibfield  {author} {\bibinfo {author} {\bibfnamefont {T.}~\bibnamefont {Brun}},\ }\bibfield  {title} {\bibinfo {title} {Measuring polynomial functions of states},\ }\href {https://doi.org/10.26421/qic4.5-6} {\bibfield  {journal} {\bibinfo  {journal} {Quantum Information and Computation}\ }\textbf {\bibinfo {volume} {4}},\ \bibinfo {pages} {401–408} (\bibinfo {year} {2004})}\BibitemShut {NoStop}%
\bibitem [{\citenamefont {Shin}\ \emph {et~al.}(2025)\citenamefont {Shin}, \citenamefont {Lee}, \citenamefont {Lee},\ and\ \citenamefont {Jeong}}]{shin2024rankneedestimatingtrace}%
  \BibitemOpen
  \bibfield  {author} {\bibinfo {author} {\bibfnamefont {M.}~\bibnamefont {Shin}}, \bibinfo {author} {\bibfnamefont {J.}~\bibnamefont {Lee}}, \bibinfo {author} {\bibfnamefont {S.}~\bibnamefont {Lee}},\ and\ \bibinfo {author} {\bibfnamefont {K.}~\bibnamefont {Jeong}},\ }\href {https://arxiv.org/abs/2408.00314} {\bibinfo {title} {Resource-efficient algorithm for estimating the trace of quantum state powers}} (\bibinfo {year} {2025}),\ \Eprint {https://arxiv.org/abs/2408.00314} {arXiv:2408.00314 [quant-ph]} \BibitemShut {NoStop}%
\bibitem [{\citenamefont {Leifer}\ \emph {et~al.}(2004)\citenamefont {Leifer}, \citenamefont {Linden},\ and\ \citenamefont {Winter}}]{leifer2004measuring}%
  \BibitemOpen
  \bibfield  {author} {\bibinfo {author} {\bibfnamefont {M.~S.}\ \bibnamefont {Leifer}}, \bibinfo {author} {\bibfnamefont {N.}~\bibnamefont {Linden}},\ and\ \bibinfo {author} {\bibfnamefont {A.}~\bibnamefont {Winter}},\ }\bibfield  {title} {\bibinfo {title} {Measuring polynomial invariants of multiparty quantum states},\ }\href {https://doi.org/10.1103/PhysRevA.69.052304} {\bibfield  {journal} {\bibinfo  {journal} {Phys. Rev. A}\ }\textbf {\bibinfo {volume} {69}},\ \bibinfo {pages} {052304} (\bibinfo {year} {2004})}\BibitemShut {NoStop}%
\bibitem [{\citenamefont {Liang}\ \emph {et~al.}(2023)\citenamefont {Liang}, \citenamefont {Lv}, \citenamefont {Wang},\ and\ \citenamefont {Fei}}]{liang2023unified}%
  \BibitemOpen
  \bibfield  {author} {\bibinfo {author} {\bibfnamefont {J.-M.}\ \bibnamefont {Liang}}, \bibinfo {author} {\bibfnamefont {Q.-Q.}\ \bibnamefont {Lv}}, \bibinfo {author} {\bibfnamefont {Z.-X.}\ \bibnamefont {Wang}},\ and\ \bibinfo {author} {\bibfnamefont {S.-M.}\ \bibnamefont {Fei}},\ }\bibfield  {title} {\bibinfo {title} {Unified multivariate trace estimation and quantum error mitigation},\ }\href {https://doi.org/10.1103/PhysRevA.107.012606} {\bibfield  {journal} {\bibinfo  {journal} {Phys. Rev. A}\ }\textbf {\bibinfo {volume} {107}},\ \bibinfo {pages} {012606} (\bibinfo {year} {2023})}\BibitemShut {NoStop}%
\bibitem [{\citenamefont {Liu}\ and\ \citenamefont {Wang}(2025)}]{liu2025estimating}%
  \BibitemOpen
  \bibfield  {author} {\bibinfo {author} {\bibfnamefont {Y.}~\bibnamefont {Liu}}\ and\ \bibinfo {author} {\bibfnamefont {Q.}~\bibnamefont {Wang}},\ }\bibfield  {title} {\bibinfo {title} {On estimating the trace of quantum state powers},\ }in\ \href {https://epubs.siam.org/doi/abs/10.1137/1.9781611978322.28} {\emph {\bibinfo {booktitle} {Proceedings of the 2025 Annual ACM-SIAM Symposium on Discrete Algorithms (SODA)}}}\ (\bibinfo {organization} {SIAM},\ \bibinfo {year} {2025})\ pp.\ \bibinfo {pages} {947--993}\BibitemShut {NoStop}%
\bibitem [{\citenamefont {Johri}\ \emph {et~al.}(2017)\citenamefont {Johri}, \citenamefont {Steiger},\ and\ \citenamefont {Troyer}}]{johri2017entanglement}%
  \BibitemOpen
  \bibfield  {author} {\bibinfo {author} {\bibfnamefont {S.}~\bibnamefont {Johri}}, \bibinfo {author} {\bibfnamefont {D.~S.}\ \bibnamefont {Steiger}},\ and\ \bibinfo {author} {\bibfnamefont {M.}~\bibnamefont {Troyer}},\ }\bibfield  {title} {\bibinfo {title} {Entanglement spectroscopy on a quantum computer},\ }\href {https://doi.org/10.1103/PhysRevB.96.195136} {\bibfield  {journal} {\bibinfo  {journal} {Phys. Rev. B}\ }\textbf {\bibinfo {volume} {96}},\ \bibinfo {pages} {195136} (\bibinfo {year} {2017})}\BibitemShut {NoStop}%
\bibitem [{\citenamefont {Turkeshi}\ \emph {et~al.}(2023)\citenamefont {Turkeshi}, \citenamefont {Schir\`o},\ and\ \citenamefont {Sierant}}]{turkeshi2023measuring}%
  \BibitemOpen
  \bibfield  {author} {\bibinfo {author} {\bibfnamefont {X.}~\bibnamefont {Turkeshi}}, \bibinfo {author} {\bibfnamefont {M.}~\bibnamefont {Schir\`o}},\ and\ \bibinfo {author} {\bibfnamefont {P.}~\bibnamefont {Sierant}},\ }\bibfield  {title} {\bibinfo {title} {Measuring nonstabilizerness via multifractal flatness},\ }\href {https://doi.org/10.1103/PhysRevA.108.042408} {\bibfield  {journal} {\bibinfo  {journal} {Phys. Rev. A}\ }\textbf {\bibinfo {volume} {108}},\ \bibinfo {pages} {042408} (\bibinfo {year} {2023})}\BibitemShut {NoStop}%
\bibitem [{\citenamefont {Tirrito}\ \emph {et~al.}(2024)\citenamefont {Tirrito}, \citenamefont {Tarabunga}, \citenamefont {Lami}, \citenamefont {Chanda}, \citenamefont {Leone}, \citenamefont {Oliviero}, \citenamefont {Dalmonte}, \citenamefont {Collura},\ and\ \citenamefont {Hamma}}]{tirrito2024quantifying}%
  \BibitemOpen
  \bibfield  {author} {\bibinfo {author} {\bibfnamefont {E.}~\bibnamefont {Tirrito}}, \bibinfo {author} {\bibfnamefont {P.~S.}\ \bibnamefont {Tarabunga}}, \bibinfo {author} {\bibfnamefont {G.}~\bibnamefont {Lami}}, \bibinfo {author} {\bibfnamefont {T.}~\bibnamefont {Chanda}}, \bibinfo {author} {\bibfnamefont {L.}~\bibnamefont {Leone}}, \bibinfo {author} {\bibfnamefont {S.~F.~E.}\ \bibnamefont {Oliviero}}, \bibinfo {author} {\bibfnamefont {M.}~\bibnamefont {Dalmonte}}, \bibinfo {author} {\bibfnamefont {M.}~\bibnamefont {Collura}},\ and\ \bibinfo {author} {\bibfnamefont {A.}~\bibnamefont {Hamma}},\ }\bibfield  {title} {\bibinfo {title} {Quantifying nonstabilizerness through entanglement spectrum flatness},\ }\href {https://doi.org/10.1103/PhysRevA.109.L040401} {\bibfield  {journal} {\bibinfo  {journal} {Phys. Rev. A}\ }\textbf {\bibinfo {volume} {109}},\ \bibinfo {pages} {L040401} (\bibinfo {year} {2024})}\BibitemShut {NoStop}%
\bibitem [{\citenamefont {Koczor}(2021)}]{koczor2021exponential}%
  \BibitemOpen
  \bibfield  {author} {\bibinfo {author} {\bibfnamefont {B.}~\bibnamefont {Koczor}},\ }\bibfield  {title} {\bibinfo {title} {{Exponential Error Suppression for Near-Term Quantum Devices}},\ }\href {https://doi.org/10.1103/PhysRevX.11.031057} {\bibfield  {journal} {\bibinfo  {journal} {Phys. Rev. X}\ }\textbf {\bibinfo {volume} {11}},\ \bibinfo {pages} {031057} (\bibinfo {year} {2021})}\BibitemShut {NoStop}%
\bibitem [{\citenamefont {Huggins}\ \emph {et~al.}(2021)\citenamefont {Huggins}, \citenamefont {McArdle}, \citenamefont {O'Brien}, \citenamefont {Lee}, \citenamefont {Rubin}, \citenamefont {Boixo}, \citenamefont {Whaley}, \citenamefont {Babbush},\ and\ \citenamefont {McClean}}]{huggins2021virtual}%
  \BibitemOpen
  \bibfield  {author} {\bibinfo {author} {\bibfnamefont {W.~J.}\ \bibnamefont {Huggins}}, \bibinfo {author} {\bibfnamefont {S.}~\bibnamefont {McArdle}}, \bibinfo {author} {\bibfnamefont {T.~E.}\ \bibnamefont {O'Brien}}, \bibinfo {author} {\bibfnamefont {J.}~\bibnamefont {Lee}}, \bibinfo {author} {\bibfnamefont {N.~C.}\ \bibnamefont {Rubin}}, \bibinfo {author} {\bibfnamefont {S.}~\bibnamefont {Boixo}}, \bibinfo {author} {\bibfnamefont {K.~B.}\ \bibnamefont {Whaley}}, \bibinfo {author} {\bibfnamefont {R.}~\bibnamefont {Babbush}},\ and\ \bibinfo {author} {\bibfnamefont {J.~R.}\ \bibnamefont {McClean}},\ }\bibfield  {title} {\bibinfo {title} {{Virtual Distillation for Quantum Error Mitigation}},\ }\href {https://doi.org/10.1103/PhysRevX.11.041036} {\bibfield  {journal} {\bibinfo  {journal} {Phys. Rev. X}\ }\textbf {\bibinfo {volume} {11}},\ \bibinfo {pages} {041036} (\bibinfo {year} {2021})}\BibitemShut {NoStop}%
\bibitem [{\citenamefont {Quek}\ \emph {et~al.}(2024)\citenamefont {Quek}, \citenamefont {Kaur},\ and\ \citenamefont {Wilde}}]{quek2024multivariatetrace}%
  \BibitemOpen
  \bibfield  {author} {\bibinfo {author} {\bibfnamefont {Y.}~\bibnamefont {Quek}}, \bibinfo {author} {\bibfnamefont {E.}~\bibnamefont {Kaur}},\ and\ \bibinfo {author} {\bibfnamefont {M.~M.}\ \bibnamefont {Wilde}},\ }\bibfield  {title} {\bibinfo {title} {Multivariate trace estimation in constant quantum depth},\ }\href {https://doi.org/10.22331/q-2024-01-10-1220} {\bibfield  {journal} {\bibinfo  {journal} {{Quantum}}\ }\textbf {\bibinfo {volume} {8}},\ \bibinfo {pages} {1220} (\bibinfo {year} {2024})}\BibitemShut {NoStop}%
\bibitem [{\citenamefont {Kirkwood}(1933)}]{kirkwood1933quantum}%
  \BibitemOpen
  \bibfield  {author} {\bibinfo {author} {\bibfnamefont {J.~G.}\ \bibnamefont {Kirkwood}},\ }\bibfield  {title} {\bibinfo {title} {Quantum statistics of almost classical assemblies},\ }\href {https://doi.org/10.1103/PhysRev.44.31} {\bibfield  {journal} {\bibinfo  {journal} {Phys. Rev.}\ }\textbf {\bibinfo {volume} {44}},\ \bibinfo {pages} {31} (\bibinfo {year} {1933})}\BibitemShut {NoStop}%
\bibitem [{\citenamefont {Dirac}(1945)}]{dirac1945analogy}%
  \BibitemOpen
  \bibfield  {author} {\bibinfo {author} {\bibfnamefont {P.~A.~M.}\ \bibnamefont {Dirac}},\ }\bibfield  {title} {\bibinfo {title} {On the analogy between classical and quantum mechanics},\ }\href {https://doi.org/10.1103/RevModPhys.17.195} {\bibfield  {journal} {\bibinfo  {journal} {Rev. Mod. Phys.}\ }\textbf {\bibinfo {volume} {17}},\ \bibinfo {pages} {195} (\bibinfo {year} {1945})}\BibitemShut {NoStop}%
\bibitem [{\citenamefont {Wagner}\ \emph {et~al.}(2024{\natexlab{a}})\citenamefont {Wagner}, \citenamefont {Schwartzman-Nowik}, \citenamefont {Paiva}, \citenamefont {Te’eni}, \citenamefont {Ruiz-Molero}, \citenamefont {Barbosa}, \citenamefont {Cohen},\ and\ \citenamefont {Galvão}}]{wagner2024quantumcircuits}%
  \BibitemOpen
  \bibfield  {author} {\bibinfo {author} {\bibfnamefont {R.}~\bibnamefont {Wagner}}, \bibinfo {author} {\bibfnamefont {Z.}~\bibnamefont {Schwartzman-Nowik}}, \bibinfo {author} {\bibfnamefont {I.~L.}\ \bibnamefont {Paiva}}, \bibinfo {author} {\bibfnamefont {A.}~\bibnamefont {Te’eni}}, \bibinfo {author} {\bibfnamefont {A.}~\bibnamefont {Ruiz-Molero}}, \bibinfo {author} {\bibfnamefont {R.~S.}\ \bibnamefont {Barbosa}}, \bibinfo {author} {\bibfnamefont {E.}~\bibnamefont {Cohen}},\ and\ \bibinfo {author} {\bibfnamefont {E.~F.}\ \bibnamefont {Galvão}},\ }\bibfield  {title} {\bibinfo {title} {{Quantum circuits for measuring weak values, Kirkwood–Dirac quasiprobability distributions, and state spectra}},\ }\href {https://doi.org/10.1088/2058-9565/ad124c} {\bibfield  {journal} {\bibinfo  {journal} {Quantum Science and Technology}\ }\textbf {\bibinfo {volume} {9}},\ \bibinfo {pages} {015030} (\bibinfo {year} {2024}{\natexlab{a}})}\BibitemShut {NoStop}%
\bibitem [{\citenamefont {Arvidsson-Shukur}\ \emph {et~al.}(2024)\citenamefont {Arvidsson-Shukur}, \citenamefont {Braasch~Jr}, \citenamefont {De~Bièvre}, \citenamefont {Dressel}, \citenamefont {Jordan}, \citenamefont {Langrenez}, \citenamefont {Lostaglio}, \citenamefont {Lundeen},\ and\ \citenamefont {Halpern}}]{arvidssonshukur2024properties}%
  \BibitemOpen
  \bibfield  {author} {\bibinfo {author} {\bibfnamefont {D.~R.~M.}\ \bibnamefont {Arvidsson-Shukur}}, \bibinfo {author} {\bibfnamefont {W.~F.}\ \bibnamefont {Braasch~Jr}}, \bibinfo {author} {\bibfnamefont {S.}~\bibnamefont {De~Bièvre}}, \bibinfo {author} {\bibfnamefont {J.}~\bibnamefont {Dressel}}, \bibinfo {author} {\bibfnamefont {A.~N.}\ \bibnamefont {Jordan}}, \bibinfo {author} {\bibfnamefont {C.}~\bibnamefont {Langrenez}}, \bibinfo {author} {\bibfnamefont {M.}~\bibnamefont {Lostaglio}}, \bibinfo {author} {\bibfnamefont {J.~S.}\ \bibnamefont {Lundeen}},\ and\ \bibinfo {author} {\bibfnamefont {N.~Y.}\ \bibnamefont {Halpern}},\ }\bibfield  {title} {\bibinfo {title} {{Properties and applications of the Kirkwood–Dirac distribution}},\ }\href {https://doi.org/10.1088/1367-2630/ada05d} {\bibfield  {journal} {\bibinfo  {journal} {New Journal of Physics}\ }\textbf {\bibinfo {volume} {26}},\ \bibinfo {pages} {121201} (\bibinfo {year} {2024})}\BibitemShut {NoStop}%
\bibitem [{\citenamefont {Schmid}\ \emph {et~al.}(2024)\citenamefont {Schmid}, \citenamefont {Baldij\~ao}, \citenamefont {Yīng}, \citenamefont {Wagner},\ and\ \citenamefont {Selby}}]{schmid2024kirkwood}%
  \BibitemOpen
  \bibfield  {author} {\bibinfo {author} {\bibfnamefont {D.}~\bibnamefont {Schmid}}, \bibinfo {author} {\bibfnamefont {R.~D.}\ \bibnamefont {Baldij\~ao}}, \bibinfo {author} {\bibfnamefont {Y.}~\bibnamefont {Yīng}}, \bibinfo {author} {\bibfnamefont {R.}~\bibnamefont {Wagner}},\ and\ \bibinfo {author} {\bibfnamefont {J.~H.}\ \bibnamefont {Selby}},\ }\bibfield  {title} {\bibinfo {title} {{Kirkwood-Dirac representations beyond quantum states and their relation to noncontextuality}},\ }\href {https://doi.org/10.1103/PhysRevA.110.052206} {\bibfield  {journal} {\bibinfo  {journal} {Phys. Rev. A}\ }\textbf {\bibinfo {volume} {110}},\ \bibinfo {pages} {052206} (\bibinfo {year} {2024})}\BibitemShut {NoStop}%
\bibitem [{\citenamefont {Liu}\ and\ \citenamefont {Cheng}(2025)}]{liu2025boundarykirkwooddiracquasiprobability}%
  \BibitemOpen
  \bibfield  {author} {\bibinfo {author} {\bibfnamefont {L.}~\bibnamefont {Liu}}\ and\ \bibinfo {author} {\bibfnamefont {S.}~\bibnamefont {Cheng}},\ }\href {https://arxiv.org/abs/2504.09238} {\bibinfo {title} {The boundary of {K}irkwood--{D}irac quasiprobability}} (\bibinfo {year} {2025}),\ \Eprint {https://arxiv.org/abs/2504.09238} {arXiv:2504.09238 [quant-ph]} \BibitemShut {NoStop}%
\bibitem [{\citenamefont {Yunger~Halpern}\ \emph {et~al.}(2018)\citenamefont {Yunger~Halpern}, \citenamefont {Swingle},\ and\ \citenamefont {Dressel}}]{yunger2018quasiprobability}%
  \BibitemOpen
  \bibfield  {author} {\bibinfo {author} {\bibfnamefont {N.}~\bibnamefont {Yunger~Halpern}}, \bibinfo {author} {\bibfnamefont {B.}~\bibnamefont {Swingle}},\ and\ \bibinfo {author} {\bibfnamefont {J.}~\bibnamefont {Dressel}},\ }\bibfield  {title} {\bibinfo {title} {Quasiprobability behind the out-of-time-ordered correlator},\ }\href {https://doi.org/https://doi.org/10.1103/PhysRevA.97.042105} {\bibfield  {journal} {\bibinfo  {journal} {Physical Review A}\ }\textbf {\bibinfo {volume} {97}},\ \bibinfo {pages} {042105} (\bibinfo {year} {2018})}\BibitemShut {NoStop}%
\bibitem [{\citenamefont {Gonz{\'a}lez~Alonso}\ \emph {et~al.}(2019)\citenamefont {Gonz{\'a}lez~Alonso}, \citenamefont {Yunger~Halpern},\ and\ \citenamefont {Dressel}}]{gonzalez2019out}%
  \BibitemOpen
  \bibfield  {author} {\bibinfo {author} {\bibfnamefont {J.~R.}\ \bibnamefont {Gonz{\'a}lez~Alonso}}, \bibinfo {author} {\bibfnamefont {N.}~\bibnamefont {Yunger~Halpern}},\ and\ \bibinfo {author} {\bibfnamefont {J.}~\bibnamefont {Dressel}},\ }\bibfield  {title} {\bibinfo {title} {Out-of-time-ordered-correlator quasiprobabilities robustly witness scrambling},\ }\href {https://doi.org/https://doi.org/10.1103/PhysRevLett.122.040404} {\bibfield  {journal} {\bibinfo  {journal} {Physical Review Letters}\ }\textbf {\bibinfo {volume} {122}},\ \bibinfo {pages} {040404} (\bibinfo {year} {2019})}\BibitemShut {NoStop}%
\bibitem [{\citenamefont {Wagner}\ and\ \citenamefont {Galv\~ao}(2023)}]{wagner2023anomalous}%
  \BibitemOpen
  \bibfield  {author} {\bibinfo {author} {\bibfnamefont {R.}~\bibnamefont {Wagner}}\ and\ \bibinfo {author} {\bibfnamefont {E.~F.}\ \bibnamefont {Galv\~ao}},\ }\bibfield  {title} {\bibinfo {title} {Simple proof that anomalous weak values require coherence},\ }\href {https://doi.org/10.1103/PhysRevA.108.L040202} {\bibfield  {journal} {\bibinfo  {journal} {Phys. Rev. A}\ }\textbf {\bibinfo {volume} {108}},\ \bibinfo {pages} {L040202} (\bibinfo {year} {2023})}\BibitemShut {NoStop}%
\bibitem [{\citenamefont {Hofmann}(2012)}]{hofmann2012complex}%
  \BibitemOpen
  \bibfield  {author} {\bibinfo {author} {\bibfnamefont {H.~F.}\ \bibnamefont {Hofmann}},\ }\bibfield  {title} {\bibinfo {title} {{Complex Joint Probabilities as Expressions of Reversible Transformations in Quantum Mechanics}},\ }\href {https://doi.org/10.1088/1367-2630/14/4/043031} {\bibfield  {journal} {\bibinfo  {journal} {New Journal of Physics}\ }\textbf {\bibinfo {volume} {14}},\ \bibinfo {pages} {043031} (\bibinfo {year} {2012})}\BibitemShut {NoStop}%
\bibitem [{\citenamefont {Wigderson}(2019)}]{wigderson2019mathematics}%
  \BibitemOpen
  \bibfield  {author} {\bibinfo {author} {\bibfnamefont {A.}~\bibnamefont {Wigderson}},\ }\bibfield  {title} {\bibinfo {title} {Mathematics and computation},\ }in\ \href {https://press.princeton.edu/books/hardcover/9780691189130/mathematics-and-computation} {\emph {\bibinfo {booktitle} {Mathematics and Computation}}}\ (\bibinfo  {publisher} {Princeton University Press},\ \bibinfo {address} {Princeton, NJ},\ \bibinfo {year} {2019})\BibitemShut {NoStop}%
\bibitem [{\citenamefont {Chien}\ and\ \citenamefont {Waldron}(2016)}]{chien2016characterization}%
  \BibitemOpen
  \bibfield  {author} {\bibinfo {author} {\bibfnamefont {T.-Y.}\ \bibnamefont {Chien}}\ and\ \bibinfo {author} {\bibfnamefont {S.}~\bibnamefont {Waldron}},\ }\bibfield  {title} {\bibinfo {title} {{A Characterization of Projective Unitary Equivalence of Finite Frames and Applications}},\ }\href {https://doi.org/10.1137/15m1042140} {\bibfield  {journal} {\bibinfo  {journal} {SIAM Journal on Discrete Mathematics}\ }\textbf {\bibinfo {volume} {30}},\ \bibinfo {pages} {976} (\bibinfo {year} {2016})}\BibitemShut {NoStop}%
\bibitem [{\citenamefont {Menssen}\ \emph {et~al.}(2017)\citenamefont {Menssen}, \citenamefont {Jones}, \citenamefont {Metcalf}, \citenamefont {Tichy}, \citenamefont {Barz}, \citenamefont {Kolthammer},\ and\ \citenamefont {Walmsley}}]{menssen2017distinguishability}%
  \BibitemOpen
  \bibfield  {author} {\bibinfo {author} {\bibfnamefont {A.~J.}\ \bibnamefont {Menssen}}, \bibinfo {author} {\bibfnamefont {A.~E.}\ \bibnamefont {Jones}}, \bibinfo {author} {\bibfnamefont {B.~J.}\ \bibnamefont {Metcalf}}, \bibinfo {author} {\bibfnamefont {M.~C.}\ \bibnamefont {Tichy}}, \bibinfo {author} {\bibfnamefont {S.}~\bibnamefont {Barz}}, \bibinfo {author} {\bibfnamefont {W.~S.}\ \bibnamefont {Kolthammer}},\ and\ \bibinfo {author} {\bibfnamefont {I.~A.}\ \bibnamefont {Walmsley}},\ }\bibfield  {title} {\bibinfo {title} {Distinguishability and many-particle interference},\ }\href {https://doi.org/10.1103/PhysRevLett.118.153603} {\bibfield  {journal} {\bibinfo  {journal} {Phys. Rev. Lett.}\ }\textbf {\bibinfo {volume} {118}},\ \bibinfo {pages} {153603} (\bibinfo {year} {2017})}\BibitemShut {NoStop}%
\bibitem [{\citenamefont {Jones}\ \emph {et~al.}(2020)\citenamefont {Jones}, \citenamefont {Menssen}, \citenamefont {Chrzanowski}, \citenamefont {Wolterink}, \citenamefont {Shchesnovich},\ and\ \citenamefont {Walmsley}}]{jones2020multiparticle}%
  \BibitemOpen
  \bibfield  {author} {\bibinfo {author} {\bibfnamefont {A.~E.}\ \bibnamefont {Jones}}, \bibinfo {author} {\bibfnamefont {A.~J.}\ \bibnamefont {Menssen}}, \bibinfo {author} {\bibfnamefont {H.~M.}\ \bibnamefont {Chrzanowski}}, \bibinfo {author} {\bibfnamefont {T.~A.~W.}\ \bibnamefont {Wolterink}}, \bibinfo {author} {\bibfnamefont {V.~S.}\ \bibnamefont {Shchesnovich}},\ and\ \bibinfo {author} {\bibfnamefont {I.~A.}\ \bibnamefont {Walmsley}},\ }\bibfield  {title} {\bibinfo {title} {{Multiparticle Interference of Pairwise Distinguishable Photons}},\ }\href {https://doi.org/10.1103/PhysRevLett.125.123603} {\bibfield  {journal} {\bibinfo  {journal} {Phys. Rev. Lett.}\ }\textbf {\bibinfo {volume} {125}},\ \bibinfo {pages} {123603} (\bibinfo {year} {2020})}\BibitemShut {NoStop}%
\bibitem [{\citenamefont {Minke}\ \emph {et~al.}(2021)\citenamefont {Minke}, \citenamefont {Buchleitner},\ and\ \citenamefont {Dittel}}]{minke2021characterizing}%
  \BibitemOpen
  \bibfield  {author} {\bibinfo {author} {\bibfnamefont {A.~M.}\ \bibnamefont {Minke}}, \bibinfo {author} {\bibfnamefont {A.}~\bibnamefont {Buchleitner}},\ and\ \bibinfo {author} {\bibfnamefont {C.}~\bibnamefont {Dittel}},\ }\bibfield  {title} {\bibinfo {title} {{Characterizing four-body indistinguishability via symmetries}},\ }\href {https://doi.org/10.1088/1367-2630/ac0fb1} {\bibfield  {journal} {\bibinfo  {journal} {New Journal of Physics}\ }\textbf {\bibinfo {volume} {23}},\ \bibinfo {pages} {073028} (\bibinfo {year} {2021})}\BibitemShut {NoStop}%
\bibitem [{\citenamefont {Pont}\ \emph {et~al.}(2022)\citenamefont {Pont}, \citenamefont {Albiero}, \citenamefont {Thomas}, \citenamefont {Spagnolo}, \citenamefont {Ceccarelli}, \citenamefont {Corrielli}, \citenamefont {Brieussel}, \citenamefont {Somaschi}, \citenamefont {Huet}, \citenamefont {Harouri}, \citenamefont {Lema\^{\i}tre}, \citenamefont {Sagnes}, \citenamefont {Belabas}, \citenamefont {Sciarrino}, \citenamefont {Osellame}, \citenamefont {Senellart},\ and\ \citenamefont {Crespi}}]{pont2022quantifying}%
  \BibitemOpen
  \bibfield  {author} {\bibinfo {author} {\bibfnamefont {M.}~\bibnamefont {Pont}}, \bibinfo {author} {\bibfnamefont {R.}~\bibnamefont {Albiero}}, \bibinfo {author} {\bibfnamefont {S.~E.}\ \bibnamefont {Thomas}}, \bibinfo {author} {\bibfnamefont {N.}~\bibnamefont {Spagnolo}}, \bibinfo {author} {\bibfnamefont {F.}~\bibnamefont {Ceccarelli}}, \bibinfo {author} {\bibfnamefont {G.}~\bibnamefont {Corrielli}}, \bibinfo {author} {\bibfnamefont {A.}~\bibnamefont {Brieussel}}, \bibinfo {author} {\bibfnamefont {N.}~\bibnamefont {Somaschi}}, \bibinfo {author} {\bibfnamefont {H.}~\bibnamefont {Huet}}, \bibinfo {author} {\bibfnamefont {A.}~\bibnamefont {Harouri}}, \bibinfo {author} {\bibfnamefont {A.}~\bibnamefont {Lema\^{\i}tre}}, \bibinfo {author} {\bibfnamefont {I.}~\bibnamefont {Sagnes}}, \bibinfo {author} {\bibfnamefont {N.}~\bibnamefont {Belabas}}, \bibinfo {author} {\bibfnamefont {F.}~\bibnamefont {Sciarrino}}, \bibinfo {author} {\bibfnamefont {R.}~\bibnamefont {Osellame}}, \bibinfo {author} {\bibfnamefont
  {P.}~\bibnamefont {Senellart}},\ and\ \bibinfo {author} {\bibfnamefont {A.}~\bibnamefont {Crespi}},\ }\bibfield  {title} {\bibinfo {title} {{Quantifying $n$-Photon Indistinguishability with a Cyclic Integrated Interferometer}},\ }\href {https://doi.org/10.1103/PhysRevX.12.031033} {\bibfield  {journal} {\bibinfo  {journal} {Phys. Rev. X}\ }\textbf {\bibinfo {volume} {12}},\ \bibinfo {pages} {031033} (\bibinfo {year} {2022})}\BibitemShut {NoStop}%
\bibitem [{\citenamefont {Rodari}\ \emph {et~al.}(2024)\citenamefont {Rodari}, \citenamefont {Fernandes}, \citenamefont {Caruccio}, \citenamefont {Suprano}, \citenamefont {Hoch}, \citenamefont {Giordani}, \citenamefont {Carvacho}, \citenamefont {Albiero}, \citenamefont {Giano}, \citenamefont {Corrielli}, \citenamefont {Ceccarelli}, \citenamefont {Osellame}, \citenamefont {Brod}, \citenamefont {Novo}, \citenamefont {Spagnolo}, \citenamefont {Galvão},\ and\ \citenamefont {Sciarrino}}]{rodari2024experimentalobservationcounterintuitivefeatures}%
  \BibitemOpen
  \bibfield  {author} {\bibinfo {author} {\bibfnamefont {G.}~\bibnamefont {Rodari}}, \bibinfo {author} {\bibfnamefont {C.}~\bibnamefont {Fernandes}}, \bibinfo {author} {\bibfnamefont {E.}~\bibnamefont {Caruccio}}, \bibinfo {author} {\bibfnamefont {A.}~\bibnamefont {Suprano}}, \bibinfo {author} {\bibfnamefont {F.}~\bibnamefont {Hoch}}, \bibinfo {author} {\bibfnamefont {T.}~\bibnamefont {Giordani}}, \bibinfo {author} {\bibfnamefont {G.}~\bibnamefont {Carvacho}}, \bibinfo {author} {\bibfnamefont {R.}~\bibnamefont {Albiero}}, \bibinfo {author} {\bibfnamefont {N.~D.}\ \bibnamefont {Giano}}, \bibinfo {author} {\bibfnamefont {G.}~\bibnamefont {Corrielli}}, \bibinfo {author} {\bibfnamefont {F.}~\bibnamefont {Ceccarelli}}, \bibinfo {author} {\bibfnamefont {R.}~\bibnamefont {Osellame}}, \bibinfo {author} {\bibfnamefont {D.~J.}\ \bibnamefont {Brod}}, \bibinfo {author} {\bibfnamefont {L.}~\bibnamefont {Novo}}, \bibinfo {author} {\bibfnamefont {N.}~\bibnamefont {Spagnolo}}, \bibinfo {author} {\bibfnamefont {E.~F.}\
  \bibnamefont {Galvão}},\ and\ \bibinfo {author} {\bibfnamefont {F.}~\bibnamefont {Sciarrino}},\ }\href {https://doi.org/https://doi.org/10.48550/arXiv.2410.15883} {\bibinfo {title} {Experimental observation of counter-intuitive features of photonic bunching}} (\bibinfo {year} {2024}),\ \Eprint {https://arxiv.org/abs/2410.15883} {arXiv:2410.15883 [quant-ph]} \BibitemShut {NoStop}%
\bibitem [{\citenamefont {Rodari}\ \emph {et~al.}(2025)\citenamefont {Rodari}, \citenamefont {Novo}, \citenamefont {Albiero}, \citenamefont {Suprano}, \citenamefont {Tavares}, \citenamefont {Caruccio}, \citenamefont {Hoch}, \citenamefont {Giordani}, \citenamefont {Carvacho}, \citenamefont {Gardina}, \citenamefont {Di~Giano}, \citenamefont {Di~Giorgio}, \citenamefont {Corrielli}, \citenamefont {Ceccarelli}, \citenamefont {Osellame}, \citenamefont {Spagnolo}, \citenamefont {Galv\~ao},\ and\ \citenamefont {Sciarrino}}]{rodari2024semideviceindependentcharacterizationmultiphoton}%
  \BibitemOpen
  \bibfield  {author} {\bibinfo {author} {\bibfnamefont {G.}~\bibnamefont {Rodari}}, \bibinfo {author} {\bibfnamefont {L.}~\bibnamefont {Novo}}, \bibinfo {author} {\bibfnamefont {R.}~\bibnamefont {Albiero}}, \bibinfo {author} {\bibfnamefont {A.}~\bibnamefont {Suprano}}, \bibinfo {author} {\bibfnamefont {C.~T.}\ \bibnamefont {Tavares}}, \bibinfo {author} {\bibfnamefont {E.}~\bibnamefont {Caruccio}}, \bibinfo {author} {\bibfnamefont {F.}~\bibnamefont {Hoch}}, \bibinfo {author} {\bibfnamefont {T.}~\bibnamefont {Giordani}}, \bibinfo {author} {\bibfnamefont {G.}~\bibnamefont {Carvacho}}, \bibinfo {author} {\bibfnamefont {M.}~\bibnamefont {Gardina}}, \bibinfo {author} {\bibfnamefont {N.}~\bibnamefont {Di~Giano}}, \bibinfo {author} {\bibfnamefont {S.}~\bibnamefont {Di~Giorgio}}, \bibinfo {author} {\bibfnamefont {G.}~\bibnamefont {Corrielli}}, \bibinfo {author} {\bibfnamefont {F.}~\bibnamefont {Ceccarelli}}, \bibinfo {author} {\bibfnamefont {R.}~\bibnamefont {Osellame}}, \bibinfo {author} {\bibfnamefont
  {N.}~\bibnamefont {Spagnolo}}, \bibinfo {author} {\bibfnamefont {E.~F.}\ \bibnamefont {Galv\~ao}},\ and\ \bibinfo {author} {\bibfnamefont {F.}~\bibnamefont {Sciarrino}},\ }\bibfield  {title} {\bibinfo {title} {Semi-device-independent characterization of multiphoton indistinguishability},\ }\href {https://doi.org/10.1103/PRXQuantum.6.020340} {\bibfield  {journal} {\bibinfo  {journal} {PRX Quantum}\ }\textbf {\bibinfo {volume} {6}},\ \bibinfo {pages} {020340} (\bibinfo {year} {2025})}\BibitemShut {NoStop}%
\bibitem [{\citenamefont {Seron}\ \emph {et~al.}(2023)\citenamefont {Seron}, \citenamefont {Novo},\ and\ \citenamefont {Cerf}}]{seron2023boson}%
  \BibitemOpen
  \bibfield  {author} {\bibinfo {author} {\bibfnamefont {B.}~\bibnamefont {Seron}}, \bibinfo {author} {\bibfnamefont {L.}~\bibnamefont {Novo}},\ and\ \bibinfo {author} {\bibfnamefont {N.~J.}\ \bibnamefont {Cerf}},\ }\bibfield  {title} {\bibinfo {title} {{Boson bunching is not maximized by indistinguishable particles}},\ }\href {https://doi.org/10.1038/s41566-023-01213-0} {\bibfield  {journal} {\bibinfo  {journal} {Nature Photonics}\ }\textbf {\bibinfo {volume} {17}},\ \bibinfo {pages} {702} (\bibinfo {year} {2023})}\BibitemShut {NoStop}%
\bibitem [{\citenamefont {Giordani}\ \emph {et~al.}(2021)\citenamefont {Giordani}, \citenamefont {Esposito}, \citenamefont {Hoch}, \citenamefont {Carvacho}, \citenamefont {Brod}, \citenamefont {Galv\~ao}, \citenamefont {Spagnolo},\ and\ \citenamefont {Sciarrino}}]{giordani2021witnesses}%
  \BibitemOpen
  \bibfield  {author} {\bibinfo {author} {\bibfnamefont {T.}~\bibnamefont {Giordani}}, \bibinfo {author} {\bibfnamefont {C.}~\bibnamefont {Esposito}}, \bibinfo {author} {\bibfnamefont {F.}~\bibnamefont {Hoch}}, \bibinfo {author} {\bibfnamefont {G.}~\bibnamefont {Carvacho}}, \bibinfo {author} {\bibfnamefont {D.~J.}\ \bibnamefont {Brod}}, \bibinfo {author} {\bibfnamefont {E.~F.}\ \bibnamefont {Galv\~ao}}, \bibinfo {author} {\bibfnamefont {N.}~\bibnamefont {Spagnolo}},\ and\ \bibinfo {author} {\bibfnamefont {F.}~\bibnamefont {Sciarrino}},\ }\bibfield  {title} {\bibinfo {title} {Witnesses of coherence and dimension from multiphoton indistinguishability tests},\ }\href {https://doi.org/10.1103/PhysRevResearch.3.023031} {\bibfield  {journal} {\bibinfo  {journal} {Phys. Rev. Res.}\ }\textbf {\bibinfo {volume} {3}},\ \bibinfo {pages} {023031} (\bibinfo {year} {2021})}\BibitemShut {NoStop}%
\bibitem [{\citenamefont {Giordani}\ \emph {et~al.}(2020)\citenamefont {Giordani}, \citenamefont {Brod}, \citenamefont {Esposito}, \citenamefont {Viggianiello}, \citenamefont {Romano}, \citenamefont {Flamini}, \citenamefont {Carvacho}, \citenamefont {Spagnolo}, \citenamefont {Galv{\~{a}}o},\ and\ \citenamefont {Sciarrino}}]{giordani2020experimental}%
  \BibitemOpen
  \bibfield  {author} {\bibinfo {author} {\bibfnamefont {T.}~\bibnamefont {Giordani}}, \bibinfo {author} {\bibfnamefont {D.~J.}\ \bibnamefont {Brod}}, \bibinfo {author} {\bibfnamefont {C.}~\bibnamefont {Esposito}}, \bibinfo {author} {\bibfnamefont {N.}~\bibnamefont {Viggianiello}}, \bibinfo {author} {\bibfnamefont {M.}~\bibnamefont {Romano}}, \bibinfo {author} {\bibfnamefont {F.}~\bibnamefont {Flamini}}, \bibinfo {author} {\bibfnamefont {G.}~\bibnamefont {Carvacho}}, \bibinfo {author} {\bibfnamefont {N.}~\bibnamefont {Spagnolo}}, \bibinfo {author} {\bibfnamefont {E.~F.}\ \bibnamefont {Galv{\~{a}}o}},\ and\ \bibinfo {author} {\bibfnamefont {F.}~\bibnamefont {Sciarrino}},\ }\bibfield  {title} {\bibinfo {title} {Experimental quantification of four-photon indistinguishability},\ }\href {https://doi.org/10.1088/1367-2630/ab7a30} {\bibfield  {journal} {\bibinfo  {journal} {New Journal of Physics}\ }\textbf {\bibinfo {volume} {22}},\ \bibinfo {pages} {043001} (\bibinfo {year} {2020})}\BibitemShut {NoStop}%
\bibitem [{\citenamefont {Jones}\ \emph {et~al.}(2023)\citenamefont {Jones}, \citenamefont {Kumar}, \citenamefont {D'Aurelio}, \citenamefont {Bayerbach}, \citenamefont {Menssen},\ and\ \citenamefont {Barz}}]{jones2023distinguishability}%
  \BibitemOpen
  \bibfield  {author} {\bibinfo {author} {\bibfnamefont {A.~E.}\ \bibnamefont {Jones}}, \bibinfo {author} {\bibfnamefont {S.}~\bibnamefont {Kumar}}, \bibinfo {author} {\bibfnamefont {S.}~\bibnamefont {D'Aurelio}}, \bibinfo {author} {\bibfnamefont {M.}~\bibnamefont {Bayerbach}}, \bibinfo {author} {\bibfnamefont {A.~J.}\ \bibnamefont {Menssen}},\ and\ \bibinfo {author} {\bibfnamefont {S.}~\bibnamefont {Barz}},\ }\bibfield  {title} {\bibinfo {title} {Distinguishability and mixedness in quantum interference},\ }\href {https://doi.org/10.1103/PhysRevA.108.053701} {\bibfield  {journal} {\bibinfo  {journal} {Phys. Rev. A}\ }\textbf {\bibinfo {volume} {108}},\ \bibinfo {pages} {053701} (\bibinfo {year} {2023})}\BibitemShut {NoStop}%
\bibitem [{\citenamefont {Brod}\ \emph {et~al.}(2019)\citenamefont {Brod}, \citenamefont {Galv\~ao}, \citenamefont {Viggianiello}, \citenamefont {Flamini}, \citenamefont {Spagnolo},\ and\ \citenamefont {Sciarrino}}]{brod2019witnessing}%
  \BibitemOpen
  \bibfield  {author} {\bibinfo {author} {\bibfnamefont {D.~J.}\ \bibnamefont {Brod}}, \bibinfo {author} {\bibfnamefont {E.~F.}\ \bibnamefont {Galv\~ao}}, \bibinfo {author} {\bibfnamefont {N.}~\bibnamefont {Viggianiello}}, \bibinfo {author} {\bibfnamefont {F.}~\bibnamefont {Flamini}}, \bibinfo {author} {\bibfnamefont {N.}~\bibnamefont {Spagnolo}},\ and\ \bibinfo {author} {\bibfnamefont {F.}~\bibnamefont {Sciarrino}},\ }\bibfield  {title} {\bibinfo {title} {{Witnessing Genuine Multiphoton Indistinguishability}},\ }\href {https://doi.org/10.1103/PhysRevLett.122.063602} {\bibfield  {journal} {\bibinfo  {journal} {Phys. Rev. Lett.}\ }\textbf {\bibinfo {volume} {122}},\ \bibinfo {pages} {063602} (\bibinfo {year} {2019})}\BibitemShut {NoStop}%
\bibitem [{\citenamefont {Annoni}\ and\ \citenamefont {Wein}(2025)}]{annoni2025incoherentbehaviorpartiallydistinguishable}%
  \BibitemOpen
  \bibfield  {author} {\bibinfo {author} {\bibfnamefont {E.}~\bibnamefont {Annoni}}\ and\ \bibinfo {author} {\bibfnamefont {S.~C.}\ \bibnamefont {Wein}},\ }\href {https://doi.org/https://doi.org/10.48550/arXiv.2502.05047} {\bibinfo {title} {Incoherent behavior of partially distinguishable photons}},\ \bibinfo {howpublished} {arXiv:2502.05047 [quant-ph]} (\bibinfo {year} {2025})\BibitemShut {NoStop}%
\bibitem [{\citenamefont {Bong}\ \emph {et~al.}(2018)\citenamefont {Bong}, \citenamefont {Tischler}, \citenamefont {Patel}, \citenamefont {Wollmann}, \citenamefont {Pryde},\ and\ \citenamefont {Hall}}]{bong2018strong}%
  \BibitemOpen
  \bibfield  {author} {\bibinfo {author} {\bibfnamefont {K.-W.}\ \bibnamefont {Bong}}, \bibinfo {author} {\bibfnamefont {N.}~\bibnamefont {Tischler}}, \bibinfo {author} {\bibfnamefont {R.~B.}\ \bibnamefont {Patel}}, \bibinfo {author} {\bibfnamefont {S.}~\bibnamefont {Wollmann}}, \bibinfo {author} {\bibfnamefont {G.~J.}\ \bibnamefont {Pryde}},\ and\ \bibinfo {author} {\bibfnamefont {M.~J.~W.}\ \bibnamefont {Hall}},\ }\bibfield  {title} {\bibinfo {title} {{Strong Unitary and Overlap Uncertainty Relations: Theory and Experiment}},\ }\href {https://doi.org/10.1103/PhysRevLett.120.230402} {\bibfield  {journal} {\bibinfo  {journal} {Phys. Rev. Lett.}\ }\textbf {\bibinfo {volume} {120}},\ \bibinfo {pages} {230402} (\bibinfo {year} {2018})}\BibitemShut {NoStop}%
\bibitem [{\citenamefont {Gherardini}\ and\ \citenamefont {De~Chiara}(2024)}]{gherardini2024quasiprobabilities}%
  \BibitemOpen
  \bibfield  {author} {\bibinfo {author} {\bibfnamefont {S.}~\bibnamefont {Gherardini}}\ and\ \bibinfo {author} {\bibfnamefont {G.}~\bibnamefont {De~Chiara}},\ }\bibfield  {title} {\bibinfo {title} {{Quasiprobabilities in Quantum Thermodynamics and Many-Body Systems}},\ }\href {https://doi.org/10.1103/PRXQuantum.5.030201} {\bibfield  {journal} {\bibinfo  {journal} {PRX Quantum}\ }\textbf {\bibinfo {volume} {5}},\ \bibinfo {pages} {030201} (\bibinfo {year} {2024})}\BibitemShut {NoStop}%
\bibitem [{\citenamefont {Lostaglio}\ \emph {et~al.}(2023)\citenamefont {Lostaglio}, \citenamefont {Belenchia}, \citenamefont {Levy}, \citenamefont {Hern\'{a}ndez-G\'{o}mez}, \citenamefont {Fabbri},\ and\ \citenamefont {Gherardini}}]{lostaglio2022kirkwood}%
  \BibitemOpen
  \bibfield  {author} {\bibinfo {author} {\bibfnamefont {M.}~\bibnamefont {Lostaglio}}, \bibinfo {author} {\bibfnamefont {A.}~\bibnamefont {Belenchia}}, \bibinfo {author} {\bibfnamefont {A.}~\bibnamefont {Levy}}, \bibinfo {author} {\bibfnamefont {S.}~\bibnamefont {Hern\'{a}ndez-G\'{o}mez}}, \bibinfo {author} {\bibfnamefont {N.}~\bibnamefont {Fabbri}},\ and\ \bibinfo {author} {\bibfnamefont {S.}~\bibnamefont {Gherardini}},\ }\bibfield  {title} {\bibinfo {title} {Kirkwood-{D}irac quasiprobability approach to the statistics of incompatible observables},\ }\href {https://doi.org/10.22331/q-2023-10-09-1128} {\bibfield  {journal} {\bibinfo  {journal} {Quantum}\ }\textbf {\bibinfo {volume} {7}},\ \bibinfo {pages} {1128} (\bibinfo {year} {2023})}\BibitemShut {NoStop}%
\bibitem [{\citenamefont {Levy}\ and\ \citenamefont {Lostaglio}(2020)}]{levy2020quasiprobability}%
  \BibitemOpen
  \bibfield  {author} {\bibinfo {author} {\bibfnamefont {A.}~\bibnamefont {Levy}}\ and\ \bibinfo {author} {\bibfnamefont {M.}~\bibnamefont {Lostaglio}},\ }\bibfield  {title} {\bibinfo {title} {Quasiprobability distribution for heat fluctuations in the quantum regime},\ }\href {https://doi.org/10.1103/PRXQuantum.1.010309} {\bibfield  {journal} {\bibinfo  {journal} {PRX Quantum}\ }\textbf {\bibinfo {volume} {1}},\ \bibinfo {pages} {010309} (\bibinfo {year} {2020})}\BibitemShut {NoStop}%
\bibitem [{\citenamefont {Hern\'andez-G\'omez}\ \emph {et~al.}(2024)\citenamefont {Hern\'andez-G\'omez}, \citenamefont {Gherardini}, \citenamefont {Belenchia}, \citenamefont {Lostaglio}, \citenamefont {Levy},\ and\ \citenamefont {Fabbri}}]{hernandez2024projective}%
  \BibitemOpen
  \bibfield  {author} {\bibinfo {author} {\bibfnamefont {S.}~\bibnamefont {Hern\'andez-G\'omez}}, \bibinfo {author} {\bibfnamefont {S.}~\bibnamefont {Gherardini}}, \bibinfo {author} {\bibfnamefont {A.}~\bibnamefont {Belenchia}}, \bibinfo {author} {\bibfnamefont {M.}~\bibnamefont {Lostaglio}}, \bibinfo {author} {\bibfnamefont {A.}~\bibnamefont {Levy}},\ and\ \bibinfo {author} {\bibfnamefont {N.}~\bibnamefont {Fabbri}},\ }\bibfield  {title} {\bibinfo {title} {Projective measurements can probe nonclassical work extraction and time correlations},\ }\href {https://doi.org/10.1103/PhysRevResearch.6.023280} {\bibfield  {journal} {\bibinfo  {journal} {Phys. Rev. Res.}\ }\textbf {\bibinfo {volume} {6}},\ \bibinfo {pages} {023280} (\bibinfo {year} {2024})}\BibitemShut {NoStop}%
\bibitem [{\citenamefont {Santini}\ \emph {et~al.}(2023)\citenamefont {Santini}, \citenamefont {Solfanelli}, \citenamefont {Gherardini},\ and\ \citenamefont {Collura}}]{santini2023work}%
  \BibitemOpen
  \bibfield  {author} {\bibinfo {author} {\bibfnamefont {A.}~\bibnamefont {Santini}}, \bibinfo {author} {\bibfnamefont {A.}~\bibnamefont {Solfanelli}}, \bibinfo {author} {\bibfnamefont {S.}~\bibnamefont {Gherardini}},\ and\ \bibinfo {author} {\bibfnamefont {M.}~\bibnamefont {Collura}},\ }\bibfield  {title} {\bibinfo {title} {Work statistics, quantum signatures, and enhanced work extraction in quadratic fermionic models},\ }\href {https://doi.org/10.1103/PhysRevB.108.104308} {\bibfield  {journal} {\bibinfo  {journal} {Phys. Rev. B}\ }\textbf {\bibinfo {volume} {108}},\ \bibinfo {pages} {104308} (\bibinfo {year} {2023})}\BibitemShut {NoStop}%
\bibitem [{\citenamefont {Donati}\ \emph {et~al.}(2024)\citenamefont {Donati}, \citenamefont {Cataliotti},\ and\ \citenamefont {Gherardini}}]{donati2024energetics}%
  \BibitemOpen
  \bibfield  {author} {\bibinfo {author} {\bibfnamefont {L.}~\bibnamefont {Donati}}, \bibinfo {author} {\bibfnamefont {F.~S.}\ \bibnamefont {Cataliotti}},\ and\ \bibinfo {author} {\bibfnamefont {S.}~\bibnamefont {Gherardini}},\ }\bibfield  {title} {\bibinfo {title} {Energetics and quantumness of {F}ano coherence generation},\ }\href {https://doi.org/10.1038/s41598-024-67037-2} {\bibfield  {journal} {\bibinfo  {journal} {Scientific Reports}\ }\textbf {\bibinfo {volume} {14}},\ \bibinfo {pages} {20145} (\bibinfo {year} {2024})}\BibitemShut {NoStop}%
\bibitem [{\citenamefont {Samuel}\ and\ \citenamefont {Bhandari}(1988)}]{samuel1988general}%
  \BibitemOpen
  \bibfield  {author} {\bibinfo {author} {\bibfnamefont {J.}~\bibnamefont {Samuel}}\ and\ \bibinfo {author} {\bibfnamefont {R.}~\bibnamefont {Bhandari}},\ }\bibfield  {title} {\bibinfo {title} {{General Setting for Berry's Phase}},\ }\href {https://doi.org/10.1103/PhysRevLett.60.2339} {\bibfield  {journal} {\bibinfo  {journal} {Phys. Rev. Lett.}\ }\textbf {\bibinfo {volume} {60}},\ \bibinfo {pages} {2339} (\bibinfo {year} {1988})}\BibitemShut {NoStop}%
\bibitem [{\citenamefont {Fernandes}\ \emph {et~al.}(2024)\citenamefont {Fernandes}, \citenamefont {Wagner}, \citenamefont {Novo},\ and\ \citenamefont {Galv\~ao}}]{fernandes2024unitary}%
  \BibitemOpen
  \bibfield  {author} {\bibinfo {author} {\bibfnamefont {C.}~\bibnamefont {Fernandes}}, \bibinfo {author} {\bibfnamefont {R.}~\bibnamefont {Wagner}}, \bibinfo {author} {\bibfnamefont {L.}~\bibnamefont {Novo}},\ and\ \bibinfo {author} {\bibfnamefont {E.~F.}\ \bibnamefont {Galv\~ao}},\ }\bibfield  {title} {\bibinfo {title} {Unitary-invariant witnesses of quantum imaginarity},\ }\href {https://doi.org/10.1103/PhysRevLett.133.190201} {\bibfield  {journal} {\bibinfo  {journal} {Phys. Rev. Lett.}\ }\textbf {\bibinfo {volume} {133}},\ \bibinfo {pages} {190201} (\bibinfo {year} {2024})}\BibitemShut {NoStop}%
\bibitem [{\citenamefont {Li}\ and\ \citenamefont {Tan}(2025)}]{li2024bargmann}%
  \BibitemOpen
  \bibfield  {author} {\bibinfo {author} {\bibfnamefont {M.-S.}\ \bibnamefont {Li}}\ and\ \bibinfo {author} {\bibfnamefont {Y.-X.}\ \bibnamefont {Tan}},\ }\bibfield  {title} {\bibinfo {title} {Bargmann invariants for quantum imaginarity},\ }\href {https://doi.org/10.1103/PhysRevA.111.022409} {\bibfield  {journal} {\bibinfo  {journal} {Phys. Rev. A}\ }\textbf {\bibinfo {volume} {111}},\ \bibinfo {pages} {022409} (\bibinfo {year} {2025})}\BibitemShut {NoStop}%
\bibitem [{\citenamefont {Zhang}\ \emph {et~al.}(2025)\citenamefont {Zhang}, \citenamefont {Xie},\ and\ \citenamefont {Li}}]{zhang2024boundaries}%
  \BibitemOpen
  \bibfield  {author} {\bibinfo {author} {\bibfnamefont {L.}~\bibnamefont {Zhang}}, \bibinfo {author} {\bibfnamefont {B.}~\bibnamefont {Xie}},\ and\ \bibinfo {author} {\bibfnamefont {B.}~\bibnamefont {Li}},\ }\bibfield  {title} {\bibinfo {title} {Geometry of sets of {B}argmann invariants},\ }\href {https://doi.org/10.1103/PhysRevA.111.042417} {\bibfield  {journal} {\bibinfo  {journal} {Phys. Rev. A}\ }\textbf {\bibinfo {volume} {111}},\ \bibinfo {pages} {042417} (\bibinfo {year} {2025})}\BibitemShut {NoStop}%
\bibitem [{\citenamefont {Wagner}\ \emph {et~al.}(2025)\citenamefont {Wagner}, \citenamefont {Peres}, \citenamefont {Zambrini~Cruzeiro},\ and\ \citenamefont {Galv\~{a}o}}]{wagner2024certifying}%
  \BibitemOpen
  \bibfield  {author} {\bibinfo {author} {\bibfnamefont {R.}~\bibnamefont {Wagner}}, \bibinfo {author} {\bibfnamefont {F.~C.~R.}\ \bibnamefont {Peres}}, \bibinfo {author} {\bibfnamefont {E.}~\bibnamefont {Zambrini~Cruzeiro}},\ and\ \bibinfo {author} {\bibfnamefont {E.~F.}\ \bibnamefont {Galv\~{a}o}},\ }\bibfield  {title} {\bibinfo {title} {{Unitary‐invariant method for witnessing nonstabilizerness in quantum processors}},\ }\href {https://doi.org/10.1088/1751-8121/ade9ff} {\bibfield  {journal} {\bibinfo  {journal} {Journal of Physics A: Mathematical and Theoretical}\ }\textbf {\bibinfo {volume} {58}},\ \bibinfo {pages} {285302} (\bibinfo {year} {2025})}\BibitemShut {NoStop}%
\bibitem [{\citenamefont {Wagner}\ \emph {et~al.}(2024{\natexlab{b}})\citenamefont {Wagner}, \citenamefont {Camillini},\ and\ \citenamefont {Galv{\~{a}}o}}]{wagner2024coherence}%
  \BibitemOpen
  \bibfield  {author} {\bibinfo {author} {\bibfnamefont {R.}~\bibnamefont {Wagner}}, \bibinfo {author} {\bibfnamefont {A.}~\bibnamefont {Camillini}},\ and\ \bibinfo {author} {\bibfnamefont {E.~F.}\ \bibnamefont {Galv{\~{a}}o}},\ }\bibfield  {title} {\bibinfo {title} {Coherence and contextuality in a {M}ach-{Z}ehnder interferometer},\ }\href {https://doi.org/10.22331/q-2024-02-05-1240} {\bibfield  {journal} {\bibinfo  {journal} {{Quantum}}\ }\textbf {\bibinfo {volume} {8}},\ \bibinfo {pages} {1240} (\bibinfo {year} {2024}{\natexlab{b}})}\BibitemShut {NoStop}%
\bibitem [{\citenamefont {Wagner}\ \emph {et~al.}(2024{\natexlab{c}})\citenamefont {Wagner}, \citenamefont {Barbosa},\ and\ \citenamefont {Galv\~ao}}]{wagner2024inequalities}%
  \BibitemOpen
  \bibfield  {author} {\bibinfo {author} {\bibfnamefont {R.}~\bibnamefont {Wagner}}, \bibinfo {author} {\bibfnamefont {R.~S.}\ \bibnamefont {Barbosa}},\ and\ \bibinfo {author} {\bibfnamefont {E.~F.}\ \bibnamefont {Galv\~ao}},\ }\bibfield  {title} {\bibinfo {title} {Inequalities witnessing coherence, nonlocality, and contextuality},\ }\href {https://doi.org/10.1103/PhysRevA.109.032220} {\bibfield  {journal} {\bibinfo  {journal} {Phys. Rev. A}\ }\textbf {\bibinfo {volume} {109}},\ \bibinfo {pages} {032220} (\bibinfo {year} {2024}{\natexlab{c}})}\BibitemShut {NoStop}%
\bibitem [{\citenamefont {Galvão}\ and\ \citenamefont {Brod}(2020)}]{galvao2020quantum}%
  \BibitemOpen
  \bibfield  {author} {\bibinfo {author} {\bibfnamefont {E.~F.}\ \bibnamefont {Galvão}}\ and\ \bibinfo {author} {\bibfnamefont {D.~J.}\ \bibnamefont {Brod}},\ }\bibfield  {title} {\bibinfo {title} {Quantum and classical bounds for two-state overlaps},\ }\href {https://doi.org/10.1103/PhysRevA.101.062110} {\bibfield  {journal} {\bibinfo  {journal} {Phys. Rev. A}\ }\textbf {\bibinfo {volume} {101}},\ \bibinfo {pages} {062110} (\bibinfo {year} {2020})}\BibitemShut {NoStop}%
\bibitem [{\citenamefont {Zhang}\ \emph {et~al.}(2024)\citenamefont {Zhang}, \citenamefont {Xie},\ and\ \citenamefont {Tao}}]{zhang2024local}%
  \BibitemOpen
  \bibfield  {author} {\bibinfo {author} {\bibfnamefont {L.}~\bibnamefont {Zhang}}, \bibinfo {author} {\bibfnamefont {B.}~\bibnamefont {Xie}},\ and\ \bibinfo {author} {\bibfnamefont {Y.}~\bibnamefont {Tao}},\ }\href {https://arxiv.org/abs/2412.17237} {\bibinfo {title} {Local unitarty equivalence and entanglement by bargmann invariants}} (\bibinfo {year} {2024}),\ \Eprint {https://arxiv.org/abs/2412.17237} {arXiv:2412.17237 [quant-ph]} \BibitemShut {NoStop}%
\bibitem [{\citenamefont {Giordani}\ \emph {et~al.}(2023)\citenamefont {Giordani}, \citenamefont {Wagner}, \citenamefont {Esposito}, \citenamefont {Camillini}, \citenamefont {Hoch}, \citenamefont {Carvacho}, \citenamefont {Pentangelo}, \citenamefont {Ceccarelli}, \citenamefont {Piacentini}, \citenamefont {Crespi}, \citenamefont {Spagnolo}, \citenamefont {Osellame}, \citenamefont {Galv{\~{a}}o},\ and\ \citenamefont {Sciarrino}}]{giordani2023experimental}%
  \BibitemOpen
  \bibfield  {author} {\bibinfo {author} {\bibfnamefont {T.}~\bibnamefont {Giordani}}, \bibinfo {author} {\bibfnamefont {R.}~\bibnamefont {Wagner}}, \bibinfo {author} {\bibfnamefont {C.}~\bibnamefont {Esposito}}, \bibinfo {author} {\bibfnamefont {A.}~\bibnamefont {Camillini}}, \bibinfo {author} {\bibfnamefont {F.}~\bibnamefont {Hoch}}, \bibinfo {author} {\bibfnamefont {G.}~\bibnamefont {Carvacho}}, \bibinfo {author} {\bibfnamefont {C.}~\bibnamefont {Pentangelo}}, \bibinfo {author} {\bibfnamefont {F.}~\bibnamefont {Ceccarelli}}, \bibinfo {author} {\bibfnamefont {S.}~\bibnamefont {Piacentini}}, \bibinfo {author} {\bibfnamefont {A.}~\bibnamefont {Crespi}}, \bibinfo {author} {\bibfnamefont {N.}~\bibnamefont {Spagnolo}}, \bibinfo {author} {\bibfnamefont {R.}~\bibnamefont {Osellame}}, \bibinfo {author} {\bibfnamefont {E.~F.}\ \bibnamefont {Galv{\~{a}}o}},\ and\ \bibinfo {author} {\bibfnamefont {F.}~\bibnamefont {Sciarrino}},\ }\bibfield  {title} {\bibinfo {title} {Experimental certification of contextuality,
  coherence, and dimension in a programmable universal photonic processor},\ }\href {https://doi.org/10.1126/sciadv.adj4249} {\bibfield  {journal} {\bibinfo  {journal} {Science Advances}\ }\textbf {\bibinfo {volume} {9}},\ \bibinfo {pages} {eadj4249} (\bibinfo {year} {2023})}\BibitemShut {NoStop}%
\bibitem [{\citenamefont {Reascos}\ \emph {et~al.}(2023)\citenamefont {Reascos}, \citenamefont {Murta}, \citenamefont {Galv\~ao},\ and\ \citenamefont {Fern\'andez-Rossier}}]{reascos2023quantum}%
  \BibitemOpen
  \bibfield  {author} {\bibinfo {author} {\bibfnamefont {L.~I.}\ \bibnamefont {Reascos}}, \bibinfo {author} {\bibfnamefont {B.}~\bibnamefont {Murta}}, \bibinfo {author} {\bibfnamefont {E.~F.}\ \bibnamefont {Galv\~ao}},\ and\ \bibinfo {author} {\bibfnamefont {J.}~\bibnamefont {Fern\'andez-Rossier}},\ }\bibfield  {title} {\bibinfo {title} {Quantum circuits to measure scalar spin chirality},\ }\href {https://doi.org/10.1103/PhysRevResearch.5.043087} {\bibfield  {journal} {\bibinfo  {journal} {Phys. Rev. Res.}\ }\textbf {\bibinfo {volume} {5}},\ \bibinfo {pages} {043087} (\bibinfo {year} {2023})}\BibitemShut {NoStop}%
\bibitem [{\citenamefont {Elliott}(2025)}]{elliott2025strictadvantagecomplexquantum}%
  \BibitemOpen
  \bibfield  {author} {\bibinfo {author} {\bibfnamefont {T.~J.}\ \bibnamefont {Elliott}},\ }\bibfield  {title} {\bibinfo {title} {Strict advantage of complex quantum theory in a communication task},\ }\href {https://doi.org/10.1103/PhysRevA.111.062401} {\bibfield  {journal} {\bibinfo  {journal} {Phys. Rev. A}\ }\textbf {\bibinfo {volume} {111}},\ \bibinfo {pages} {062401} (\bibinfo {year} {2025})}\BibitemShut {NoStop}%
\bibitem [{\citenamefont {Chiribella}\ \emph {et~al.}(2013)\citenamefont {Chiribella}, \citenamefont {D'Ariano}, \citenamefont {Perinotti},\ and\ \citenamefont {Valiron}}]{chiribella2013quantum}%
  \BibitemOpen
  \bibfield  {author} {\bibinfo {author} {\bibfnamefont {G.}~\bibnamefont {Chiribella}}, \bibinfo {author} {\bibfnamefont {G.~M.}\ \bibnamefont {D'Ariano}}, \bibinfo {author} {\bibfnamefont {P.}~\bibnamefont {Perinotti}},\ and\ \bibinfo {author} {\bibfnamefont {B.}~\bibnamefont {Valiron}},\ }\bibfield  {title} {\bibinfo {title} {Quantum computations without definite causal structure},\ }\href {https://doi.org/10.1103/PhysRevA.88.022318} {\bibfield  {journal} {\bibinfo  {journal} {Phys. Rev. A}\ }\textbf {\bibinfo {volume} {88}},\ \bibinfo {pages} {022318} (\bibinfo {year} {2013})}\BibitemShut {NoStop}%
\bibitem [{\citenamefont {Procopio}\ \emph {et~al.}(2015)\citenamefont {Procopio}, \citenamefont {Moqanaki}, \citenamefont {Araújo}, \citenamefont {Costa}, \citenamefont {Alonso~Calafell}, \citenamefont {Dowd}, \citenamefont {Hamel}, \citenamefont {Rozema}, \citenamefont {Brukner},\ and\ \citenamefont {Walther}}]{procopio2015experimental}%
  \BibitemOpen
  \bibfield  {author} {\bibinfo {author} {\bibfnamefont {L.~M.}\ \bibnamefont {Procopio}}, \bibinfo {author} {\bibfnamefont {A.}~\bibnamefont {Moqanaki}}, \bibinfo {author} {\bibfnamefont {M.}~\bibnamefont {Araújo}}, \bibinfo {author} {\bibfnamefont {F.}~\bibnamefont {Costa}}, \bibinfo {author} {\bibfnamefont {I.}~\bibnamefont {Alonso~Calafell}}, \bibinfo {author} {\bibfnamefont {E.~G.}\ \bibnamefont {Dowd}}, \bibinfo {author} {\bibfnamefont {D.~R.}\ \bibnamefont {Hamel}}, \bibinfo {author} {\bibfnamefont {L.~A.}\ \bibnamefont {Rozema}}, \bibinfo {author} {\bibfnamefont {{\v{C}}.}~\bibnamefont {Brukner}},\ and\ \bibinfo {author} {\bibfnamefont {P.}~\bibnamefont {Walther}},\ }\bibfield  {title} {\bibinfo {title} {Experimental superposition of orders of quantum gates},\ }\href {https://doi.org/10.1038/ncomms8913} {\bibfield  {journal} {\bibinfo  {journal} {Nat. Commun.}\ }\textbf {\bibinfo {volume} {6}},\ \bibinfo {pages} {7913} (\bibinfo {year} {2015})}\BibitemShut {NoStop}%
\bibitem [{\citenamefont {Goswami}\ \emph {et~al.}(2018)\citenamefont {Goswami}, \citenamefont {Giarmatzi}, \citenamefont {Kewming}, \citenamefont {Costa}, \citenamefont {Branciard}, \citenamefont {Romero},\ and\ \citenamefont {White}}]{goswami2018indefinite}%
  \BibitemOpen
  \bibfield  {author} {\bibinfo {author} {\bibfnamefont {K.}~\bibnamefont {Goswami}}, \bibinfo {author} {\bibfnamefont {C.}~\bibnamefont {Giarmatzi}}, \bibinfo {author} {\bibfnamefont {M.}~\bibnamefont {Kewming}}, \bibinfo {author} {\bibfnamefont {F.}~\bibnamefont {Costa}}, \bibinfo {author} {\bibfnamefont {C.}~\bibnamefont {Branciard}}, \bibinfo {author} {\bibfnamefont {J.}~\bibnamefont {Romero}},\ and\ \bibinfo {author} {\bibfnamefont {A.~G.}\ \bibnamefont {White}},\ }\bibfield  {title} {\bibinfo {title} {Indefinite causal order in a quantum switch},\ }\href {https://doi.org/10.1103/PhysRevLett.121.090503} {\bibfield  {journal} {\bibinfo  {journal} {Phys. Rev. Lett.}\ }\textbf {\bibinfo {volume} {121}},\ \bibinfo {pages} {090503} (\bibinfo {year} {2018})}\BibitemShut {NoStop}%
\bibitem [{\citenamefont {Rubino}\ \emph {et~al.}(2017)\citenamefont {Rubino}, \citenamefont {Rozema}, \citenamefont {Feix}, \citenamefont {Ara{\'u}jo}, \citenamefont {Zeuner}, \citenamefont {Procopio}, \citenamefont {Brukner},\ and\ \citenamefont {Walther}}]{rubino2017experimental}%
  \BibitemOpen
  \bibfield  {author} {\bibinfo {author} {\bibfnamefont {G.}~\bibnamefont {Rubino}}, \bibinfo {author} {\bibfnamefont {L.~A.}\ \bibnamefont {Rozema}}, \bibinfo {author} {\bibfnamefont {A.}~\bibnamefont {Feix}}, \bibinfo {author} {\bibfnamefont {M.}~\bibnamefont {Ara{\'u}jo}}, \bibinfo {author} {\bibfnamefont {J.~M.}\ \bibnamefont {Zeuner}}, \bibinfo {author} {\bibfnamefont {L.~M.}\ \bibnamefont {Procopio}}, \bibinfo {author} {\bibfnamefont {{\v{C}}.}~\bibnamefont {Brukner}},\ and\ \bibinfo {author} {\bibfnamefont {P.}~\bibnamefont {Walther}},\ }\bibfield  {title} {\bibinfo {title} {Experimental verification of an indefinite causal order},\ }\href {https://doi.org/10.1126/sciadv.1602589} {\bibfield  {journal} {\bibinfo  {journal} {Sci. Adv.}\ }\textbf {\bibinfo {volume} {3}},\ \bibinfo {pages} {e1602589} (\bibinfo {year} {2017})}\BibitemShut {NoStop}%
\bibitem [{\citenamefont {Cao}\ \emph {et~al.}(2023)\citenamefont {Cao}, \citenamefont {Bavaresco}, \citenamefont {Wang}, \citenamefont {Rozema}, \citenamefont {Zhang}, \citenamefont {Huang}, \citenamefont {Liu}, \citenamefont {Li}, \citenamefont {Guo},\ and\ \citenamefont {Walther}}]{cao2023semideviceindependent}%
  \BibitemOpen
  \bibfield  {author} {\bibinfo {author} {\bibfnamefont {H.}~\bibnamefont {Cao}}, \bibinfo {author} {\bibfnamefont {J.}~\bibnamefont {Bavaresco}}, \bibinfo {author} {\bibfnamefont {N.-N.}\ \bibnamefont {Wang}}, \bibinfo {author} {\bibfnamefont {L.~A.}\ \bibnamefont {Rozema}}, \bibinfo {author} {\bibfnamefont {C.}~\bibnamefont {Zhang}}, \bibinfo {author} {\bibfnamefont {Y.-F.}\ \bibnamefont {Huang}}, \bibinfo {author} {\bibfnamefont {B.-H.}\ \bibnamefont {Liu}}, \bibinfo {author} {\bibfnamefont {C.-F.}\ \bibnamefont {Li}}, \bibinfo {author} {\bibfnamefont {G.-C.}\ \bibnamefont {Guo}},\ and\ \bibinfo {author} {\bibfnamefont {P.}~\bibnamefont {Walther}},\ }\bibfield  {title} {\bibinfo {title} {Semi-device-independent certification of indefinite causal order in a photonic quantum switch},\ }\href {https://doi.org/10.1364/optica.483876} {\bibfield  {journal} {\bibinfo  {journal} {Optica}\ }\textbf {\bibinfo {volume} {10}},\ \bibinfo {pages} {561} (\bibinfo {year} {2023})}\BibitemShut {NoStop}%
\bibitem [{\citenamefont {Rozema}\ \emph {et~al.}(2024)\citenamefont {Rozema}, \citenamefont {Str{\"o}mberg}, \citenamefont {Cao}, \citenamefont {Guo}, \citenamefont {Liu},\ and\ \citenamefont {Walther}}]{rozema2024experimental}%
  \BibitemOpen
  \bibfield  {author} {\bibinfo {author} {\bibfnamefont {L.~A.}\ \bibnamefont {Rozema}}, \bibinfo {author} {\bibfnamefont {T.}~\bibnamefont {Str{\"o}mberg}}, \bibinfo {author} {\bibfnamefont {H.}~\bibnamefont {Cao}}, \bibinfo {author} {\bibfnamefont {Y.}~\bibnamefont {Guo}}, \bibinfo {author} {\bibfnamefont {B.-H.}\ \bibnamefont {Liu}},\ and\ \bibinfo {author} {\bibfnamefont {P.}~\bibnamefont {Walther}},\ }\bibfield  {title} {\bibinfo {title} {Experimental aspects of indefinite causal order in quantum mechanics},\ }\href {https://doi.org/10.1038/s42254-024-00739-8} {\bibfield  {journal} {\bibinfo  {journal} {Nature Reviews Physics}\ }\textbf {\bibinfo {volume} {6}},\ \bibinfo {pages} {483} (\bibinfo {year} {2024})}\BibitemShut {NoStop}%
\bibitem [{\citenamefont {Oreshkov}\ \emph {et~al.}(2012)\citenamefont {Oreshkov}, \citenamefont {Costa},\ and\ \citenamefont {Brukner}}]{oreshkov2012quantum}%
  \BibitemOpen
  \bibfield  {author} {\bibinfo {author} {\bibfnamefont {O.}~\bibnamefont {Oreshkov}}, \bibinfo {author} {\bibfnamefont {F.}~\bibnamefont {Costa}},\ and\ \bibinfo {author} {\bibfnamefont {{\v{C}}.}~\bibnamefont {Brukner}},\ }\bibfield  {title} {\bibinfo {title} {Quantum correlations with no causal order},\ }\href {https://doi.org/10.1038/ncomms2076} {\bibfield  {journal} {\bibinfo  {journal} {Nature Communications}\ }\textbf {\bibinfo {volume} {3}},\ \bibinfo {pages} {1092} (\bibinfo {year} {2012})}\BibitemShut {NoStop}%
\bibitem [{\citenamefont {Chiribella}(2012)}]{chiribelly2012perfect}%
  \BibitemOpen
  \bibfield  {author} {\bibinfo {author} {\bibfnamefont {G.}~\bibnamefont {Chiribella}},\ }\bibfield  {title} {\bibinfo {title} {Perfect discrimination of no-signalling channels via quantum superposition of causal structures},\ }\href {https://doi.org/10.1103/PhysRevA.86.040301} {\bibfield  {journal} {\bibinfo  {journal} {Phys. Rev. A}\ }\textbf {\bibinfo {volume} {86}},\ \bibinfo {pages} {040301} (\bibinfo {year} {2012})}\BibitemShut {NoStop}%
\bibitem [{\citenamefont {Chiribella}\ \emph {et~al.}(2021)\citenamefont {Chiribella}, \citenamefont {Banik}, \citenamefont {Bhattacharya}, \citenamefont {Guha}, \citenamefont {Alimuddin}, \citenamefont {Roy}, \citenamefont {Saha}, \citenamefont {Agrawal},\ and\ \citenamefont {Kar}}]{chiribella2021indefinite}%
  \BibitemOpen
  \bibfield  {author} {\bibinfo {author} {\bibfnamefont {G.}~\bibnamefont {Chiribella}}, \bibinfo {author} {\bibfnamefont {M.}~\bibnamefont {Banik}}, \bibinfo {author} {\bibfnamefont {S.~S.}\ \bibnamefont {Bhattacharya}}, \bibinfo {author} {\bibfnamefont {T.}~\bibnamefont {Guha}}, \bibinfo {author} {\bibfnamefont {M.}~\bibnamefont {Alimuddin}}, \bibinfo {author} {\bibfnamefont {A.}~\bibnamefont {Roy}}, \bibinfo {author} {\bibfnamefont {S.}~\bibnamefont {Saha}}, \bibinfo {author} {\bibfnamefont {S.}~\bibnamefont {Agrawal}},\ and\ \bibinfo {author} {\bibfnamefont {G.}~\bibnamefont {Kar}},\ }\bibfield  {title} {\bibinfo {title} {Indefinite causal order enables perfect quantum communication with zero capacity channels},\ }\href {https://doi.org/0.1088/1367-2630/abe7a0} {\bibfield  {journal} {\bibinfo  {journal} {New Journal of Physics}\ }\textbf {\bibinfo {volume} {23}},\ \bibinfo {pages} {033039} (\bibinfo {year} {2021})}\BibitemShut {NoStop}%
\bibitem [{\citenamefont {Kissinger}\ and\ \citenamefont {Uijlen}(2017)}]{kissinger2017picturing}%
  \BibitemOpen
  \bibfield  {author} {\bibinfo {author} {\bibfnamefont {A.}~\bibnamefont {Kissinger}}\ and\ \bibinfo {author} {\bibfnamefont {S.}~\bibnamefont {Uijlen}},\ }\bibfield  {title} {\bibinfo {title} {Picturing indefinite causal structure},\ }\href {https://doi.org/10.4204/eptcs.236.6} {\bibfield  {journal} {\bibinfo  {journal} {Electronic Proceedings in Theoretical Computer Science}\ }\textbf {\bibinfo {volume} {236}},\ \bibinfo {pages} {87–94} (\bibinfo {year} {2017})}\BibitemShut {NoStop}%
\bibitem [{\citenamefont {Wilson}\ \emph {et~al.}(2025)\citenamefont {Wilson}, \citenamefont {Chiribella},\ and\ \citenamefont {Kissinger}}]{wilson2025quantumsupermapscharacterizedlocality}%
  \BibitemOpen
  \bibfield  {author} {\bibinfo {author} {\bibfnamefont {M.}~\bibnamefont {Wilson}}, \bibinfo {author} {\bibfnamefont {G.}~\bibnamefont {Chiribella}},\ and\ \bibinfo {author} {\bibfnamefont {A.}~\bibnamefont {Kissinger}},\ }\href {https://arxiv.org/abs/2205.09844} {\bibinfo {title} {Quantum supermaps are characterized by locality}} (\bibinfo {year} {2025}),\ \Eprint {https://arxiv.org/abs/2205.09844} {arXiv:2205.09844 [quant-ph]} \BibitemShut {NoStop}%
\bibitem [{\citenamefont {van~der Lugt}\ and\ \citenamefont {Ormrod}(2024)}]{vanderLugt2024possibilistic}%
  \BibitemOpen
  \bibfield  {author} {\bibinfo {author} {\bibfnamefont {T.}~\bibnamefont {van~der Lugt}}\ and\ \bibinfo {author} {\bibfnamefont {N.}~\bibnamefont {Ormrod}},\ }\bibfield  {title} {\bibinfo {title} {Possibilistic and maximal indefinite causal order in the quantum switch},\ }\href {https://doi.org/10.22331/q-2024-12-03-1543} {\bibfield  {journal} {\bibinfo  {journal} {{Quantum}}\ }\textbf {\bibinfo {volume} {8}},\ \bibinfo {pages} {1543} (\bibinfo {year} {2024})}\BibitemShut {NoStop}%
\bibitem [{\citenamefont {Bavaresco}\ \emph {et~al.}(2024{\natexlab{a}})\citenamefont {Bavaresco}, \citenamefont {Ämin Baumeler}, \citenamefont {Guryanova},\ and\ \citenamefont {Budroni}}]{bavaresco2024indefinitecausalorderboxworld}%
  \BibitemOpen
  \bibfield  {author} {\bibinfo {author} {\bibfnamefont {J.}~\bibnamefont {Bavaresco}}, \bibinfo {author} {\bibnamefont {Ämin Baumeler}}, \bibinfo {author} {\bibfnamefont {Y.}~\bibnamefont {Guryanova}},\ and\ \bibinfo {author} {\bibfnamefont {C.}~\bibnamefont {Budroni}},\ }\href {https://arxiv.org/abs/2411.00951} {\bibinfo {title} {Indefinite causal order in boxworld theories}} (\bibinfo {year} {2024}{\natexlab{a}}),\ \Eprint {https://arxiv.org/abs/2411.00951} {arXiv:2411.00951 [quant-ph]} \BibitemShut {NoStop}%
\bibitem [{\citenamefont {Renner}\ and\ \citenamefont {Brukner}(2022)}]{renner2022computational}%
  \BibitemOpen
  \bibfield  {author} {\bibinfo {author} {\bibfnamefont {M.~J.}\ \bibnamefont {Renner}}\ and\ \bibinfo {author} {\bibfnamefont {{\v{C}}.}~\bibnamefont {Brukner}},\ }\bibfield  {title} {\bibinfo {title} {Computational advantage from a quantum superposition of qubit gate orders},\ }\href {https://doi.org/https://doi.org/10.1103/PhysRevLett.128.230503} {\bibfield  {journal} {\bibinfo  {journal} {Physical Review Letters}\ }\textbf {\bibinfo {volume} {128}},\ \bibinfo {pages} {230503} (\bibinfo {year} {2022})}\BibitemShut {NoStop}%
\bibitem [{\citenamefont {Ara{\'u}jo}\ \emph {et~al.}(2017)\citenamefont {Ara{\'u}jo}, \citenamefont {Gu{\'e}rin},\ and\ \citenamefont {Baumeler}}]{araujo2017quantum}%
  \BibitemOpen
  \bibfield  {author} {\bibinfo {author} {\bibfnamefont {M.}~\bibnamefont {Ara{\'u}jo}}, \bibinfo {author} {\bibfnamefont {P.~A.}\ \bibnamefont {Gu{\'e}rin}},\ and\ \bibinfo {author} {\bibfnamefont {{\"A}.}~\bibnamefont {Baumeler}},\ }\bibfield  {title} {\bibinfo {title} {Quantum computation with indefinite causal structures},\ }\href {https://doi.org/https://doi.org/10.1103/PhysRevA.96.052315} {\bibfield  {journal} {\bibinfo  {journal} {Physical Review A}\ }\textbf {\bibinfo {volume} {96}},\ \bibinfo {pages} {052315} (\bibinfo {year} {2017})}\BibitemShut {NoStop}%
\bibitem [{\citenamefont {Simonov}\ \emph {et~al.}(2024)\citenamefont {Simonov}, \citenamefont {Caleffi}, \citenamefont {Illiano}, \citenamefont {Romero},\ and\ \citenamefont {Cacciapuoti}}]{simonov2024universal}%
  \BibitemOpen
  \bibfield  {author} {\bibinfo {author} {\bibfnamefont {K.}~\bibnamefont {Simonov}}, \bibinfo {author} {\bibfnamefont {M.}~\bibnamefont {Caleffi}}, \bibinfo {author} {\bibfnamefont {J.}~\bibnamefont {Illiano}}, \bibinfo {author} {\bibfnamefont {J.}~\bibnamefont {Romero}},\ and\ \bibinfo {author} {\bibfnamefont {A.~S.}\ \bibnamefont {Cacciapuoti}},\ }\href {https://arxiv.org/abs/2311.13654} {\bibinfo {title} {{Universal Quantum Computation via Superposed Orders of Single-Qubit Gates}}} (\bibinfo {year} {2024}),\ \Eprint {https://arxiv.org/abs/2311.13654} {arXiv:2311.13654 [quant-ph]} \BibitemShut {NoStop}%
\bibitem [{\citenamefont {Yin}\ \emph {et~al.}(2023{\natexlab{a}})\citenamefont {Yin}, \citenamefont {Zhao}, \citenamefont {Yang}, \citenamefont {Guo}, \citenamefont {Zhang}, \citenamefont {Li}, \citenamefont {Han}, \citenamefont {Liu}, \citenamefont {Xu}, \citenamefont {Chiribella} \emph {et~al.}}]{yin2023experimental}%
  \BibitemOpen
  \bibfield  {author} {\bibinfo {author} {\bibfnamefont {P.}~\bibnamefont {Yin}}, \bibinfo {author} {\bibfnamefont {X.}~\bibnamefont {Zhao}}, \bibinfo {author} {\bibfnamefont {Y.}~\bibnamefont {Yang}}, \bibinfo {author} {\bibfnamefont {Y.}~\bibnamefont {Guo}}, \bibinfo {author} {\bibfnamefont {W.-H.}\ \bibnamefont {Zhang}}, \bibinfo {author} {\bibfnamefont {G.-C.}\ \bibnamefont {Li}}, \bibinfo {author} {\bibfnamefont {Y.-J.}\ \bibnamefont {Han}}, \bibinfo {author} {\bibfnamefont {B.-H.}\ \bibnamefont {Liu}}, \bibinfo {author} {\bibfnamefont {J.-S.}\ \bibnamefont {Xu}}, \bibinfo {author} {\bibfnamefont {G.}~\bibnamefont {Chiribella}}, \emph {et~al.},\ }\bibfield  {title} {\bibinfo {title} {Experimental super-heisenberg quantum metrology with indefinite gate order},\ }\href {https://doi.org/https://doi.org/10.1038/s41567-023-02046-y} {\bibfield  {journal} {\bibinfo  {journal} {Nature Physics}\ }\textbf {\bibinfo {volume} {19}},\ \bibinfo {pages} {1122} (\bibinfo {year} {2023}{\natexlab{a}})}\BibitemShut
  {NoStop}%
\bibitem [{\citenamefont {Zhao}\ \emph {et~al.}(2020)\citenamefont {Zhao}, \citenamefont {Yang},\ and\ \citenamefont {Chiribella}}]{zhao2020quantum}%
  \BibitemOpen
  \bibfield  {author} {\bibinfo {author} {\bibfnamefont {X.}~\bibnamefont {Zhao}}, \bibinfo {author} {\bibfnamefont {Y.}~\bibnamefont {Yang}},\ and\ \bibinfo {author} {\bibfnamefont {G.}~\bibnamefont {Chiribella}},\ }\bibfield  {title} {\bibinfo {title} {{Quantum Metrology with Indefinite Causal Order}},\ }\href {https://doi.org/10.1103/PhysRevLett.124.190503} {\bibfield  {journal} {\bibinfo  {journal} {Phys. Rev. Lett.}\ }\textbf {\bibinfo {volume} {124}},\ \bibinfo {pages} {190503} (\bibinfo {year} {2020})}\BibitemShut {NoStop}%
\bibitem [{\citenamefont {Procopio}(2023)}]{procopio2023parameter}%
  \BibitemOpen
  \bibfield  {author} {\bibinfo {author} {\bibfnamefont {L.~M.}\ \bibnamefont {Procopio}},\ }\bibfield  {title} {\bibinfo {title} {Parameter estimation via indefinite causal structures},\ }in\ \href {https://doi.org/10.1088/1742-6596/2448/1/012007} {\emph {\bibinfo {booktitle} {Journal of Physics: Conference Series}}},\ Vol.\ \bibinfo {volume} {2448}\ (\bibinfo {organization} {IOP Publishing},\ \bibinfo {year} {2023})\ p.\ \bibinfo {pages} {012007}\BibitemShut {NoStop}%
\bibitem [{\citenamefont {Gu\'erin}\ \emph {et~al.}(2016)\citenamefont {Gu\'erin}, \citenamefont {Feix}, \citenamefont {Ara\'ujo},\ and\ \citenamefont {Brukner}}]{guerin2016exponential}%
  \BibitemOpen
  \bibfield  {author} {\bibinfo {author} {\bibfnamefont {P.~A.}\ \bibnamefont {Gu\'erin}}, \bibinfo {author} {\bibfnamefont {A.}~\bibnamefont {Feix}}, \bibinfo {author} {\bibfnamefont {M.}~\bibnamefont {Ara\'ujo}},\ and\ \bibinfo {author} {\bibfnamefont {{\v{C}}.}~\bibnamefont {Brukner}},\ }\bibfield  {title} {\bibinfo {title} {{Exponential Communication Complexity Advantage from Quantum Superposition of the Direction of Communication}},\ }\href {https://doi.org/10.1103/PhysRevLett.117.100502} {\bibfield  {journal} {\bibinfo  {journal} {Phys. Rev. Lett.}\ }\textbf {\bibinfo {volume} {117}},\ \bibinfo {pages} {100502} (\bibinfo {year} {2016})}\BibitemShut {NoStop}%
\bibitem [{\citenamefont {Ebler}\ \emph {et~al.}(2018)\citenamefont {Ebler}, \citenamefont {Salek},\ and\ \citenamefont {Chiribella}}]{ebler2018enhanced}%
  \BibitemOpen
  \bibfield  {author} {\bibinfo {author} {\bibfnamefont {D.}~\bibnamefont {Ebler}}, \bibinfo {author} {\bibfnamefont {S.}~\bibnamefont {Salek}},\ and\ \bibinfo {author} {\bibfnamefont {G.}~\bibnamefont {Chiribella}},\ }\bibfield  {title} {\bibinfo {title} {{Enhanced Communication with the Assistance of Indefinite Causal Order}},\ }\href {https://doi.org/10.1103/PhysRevLett.120.120502} {\bibfield  {journal} {\bibinfo  {journal} {Phys. Rev. Lett.}\ }\textbf {\bibinfo {volume} {120}},\ \bibinfo {pages} {120502} (\bibinfo {year} {2018})}\BibitemShut {NoStop}%
\bibitem [{\citenamefont {Wu}\ \emph {et~al.}(2024)\citenamefont {Wu}, \citenamefont {Fullwood}, \citenamefont {Ma}, \citenamefont {Zhou}, \citenamefont {Zhao},\ and\ \citenamefont {Chiribella}}]{wu2024generalcommunication}%
  \BibitemOpen
  \bibfield  {author} {\bibinfo {author} {\bibfnamefont {Z.}~\bibnamefont {Wu}}, \bibinfo {author} {\bibfnamefont {J.}~\bibnamefont {Fullwood}}, \bibinfo {author} {\bibfnamefont {Z.}~\bibnamefont {Ma}}, \bibinfo {author} {\bibfnamefont {S.}~\bibnamefont {Zhou}}, \bibinfo {author} {\bibfnamefont {Q.}~\bibnamefont {Zhao}},\ and\ \bibinfo {author} {\bibfnamefont {G.}~\bibnamefont {Chiribella}},\ }\href {https://arxiv.org/abs/2407.02726} {\bibinfo {title} {{General Communication Enhancement via the Quantum Switch}}} (\bibinfo {year} {2024}),\ \Eprint {https://arxiv.org/abs/2407.02726} {arXiv:2407.02726 [quant-ph]} \BibitemShut {NoStop}%
\bibitem [{\citenamefont {Mukhopadhyay}\ and\ \citenamefont {Pati}(2020)}]{mukhopadhyay2020superposition}%
  \BibitemOpen
  \bibfield  {author} {\bibinfo {author} {\bibfnamefont {C.}~\bibnamefont {Mukhopadhyay}}\ and\ \bibinfo {author} {\bibfnamefont {A.~K.}\ \bibnamefont {Pati}},\ }\bibfield  {title} {\bibinfo {title} {Superposition of causal order enables quantum advantage in teleportation under very noisy channels},\ }\href {https://doi.org/0.1088/2399-6528/abbd77} {\bibfield  {journal} {\bibinfo  {journal} {Journal of Physics Communications}\ }\textbf {\bibinfo {volume} {4}},\ \bibinfo {pages} {105003} (\bibinfo {year} {2020})}\BibitemShut {NoStop}%
\bibitem [{\citenamefont {Felce}\ and\ \citenamefont {Vedral}(2020)}]{felce2020quantum}%
  \BibitemOpen
  \bibfield  {author} {\bibinfo {author} {\bibfnamefont {D.}~\bibnamefont {Felce}}\ and\ \bibinfo {author} {\bibfnamefont {V.}~\bibnamefont {Vedral}},\ }\bibfield  {title} {\bibinfo {title} {Quantum refrigeration with indefinite causal order},\ }\href {https://doi.org/https://doi.org/10.1103/PhysRevLett.125.070603} {\bibfield  {journal} {\bibinfo  {journal} {Physical Review Letters}\ }\textbf {\bibinfo {volume} {125}},\ \bibinfo {pages} {070603} (\bibinfo {year} {2020})}\BibitemShut {NoStop}%
\bibitem [{\citenamefont {Molitor}\ \emph {et~al.}(2024)\citenamefont {Molitor}, \citenamefont {Malavazi}, \citenamefont {Baldij{\~a}o}, \citenamefont {Orthey}, \citenamefont {Paiva},\ and\ \citenamefont {Dieguez}}]{molitor2024quantum}%
  \BibitemOpen
  \bibfield  {author} {\bibinfo {author} {\bibfnamefont {O.~A.~D.}\ \bibnamefont {Molitor}}, \bibinfo {author} {\bibfnamefont {A.~H.~A.}\ \bibnamefont {Malavazi}}, \bibinfo {author} {\bibfnamefont {R.~D.}\ \bibnamefont {Baldij{\~a}o}}, \bibinfo {author} {\bibfnamefont {A.~C.}\ \bibnamefont {Orthey}}, \bibinfo {author} {\bibfnamefont {I.~L.}\ \bibnamefont {Paiva}},\ and\ \bibinfo {author} {\bibfnamefont {P.~R.}\ \bibnamefont {Dieguez}},\ }\bibfield  {title} {\bibinfo {title} {Quantum switch instabilities with an open control},\ }\href {https://doi.org/10.1038/s42005-024-01843-y} {\bibfield  {journal} {\bibinfo  {journal} {Commun. Phys.}\ }\textbf {\bibinfo {volume} {7}},\ \bibinfo {pages} {373} (\bibinfo {year} {2024})}\BibitemShut {NoStop}%
\bibitem [{\citenamefont {Loizeau}\ and\ \citenamefont {Grinbaum}(2020)}]{loizeau2020channel}%
  \BibitemOpen
  \bibfield  {author} {\bibinfo {author} {\bibfnamefont {N.}~\bibnamefont {Loizeau}}\ and\ \bibinfo {author} {\bibfnamefont {A.}~\bibnamefont {Grinbaum}},\ }\bibfield  {title} {\bibinfo {title} {Channel capacity enhancement with indefinite causal order},\ }\href {https://doi.org/https://doi.org/10.1103/PhysRevA.101.012340} {\bibfield  {journal} {\bibinfo  {journal} {Physical Review A}\ }\textbf {\bibinfo {volume} {101}},\ \bibinfo {pages} {012340} (\bibinfo {year} {2020})}\BibitemShut {NoStop}%
\bibitem [{\citenamefont {Caleffi}\ and\ \citenamefont {Cacciapuoti}(2020)}]{caleffi2020quantum}%
  \BibitemOpen
  \bibfield  {author} {\bibinfo {author} {\bibfnamefont {M.}~\bibnamefont {Caleffi}}\ and\ \bibinfo {author} {\bibfnamefont {A.~S.}\ \bibnamefont {Cacciapuoti}},\ }\bibfield  {title} {\bibinfo {title} {Quantum switch for the quantum internet: Noiseless communications through noisy channels},\ }\href {https://doi.org/10.1109/JSAC.2020.2969035} {\bibfield  {journal} {\bibinfo  {journal} {IEEE Journal on Selected Areas in Communications}\ }\textbf {\bibinfo {volume} {38}},\ \bibinfo {pages} {575} (\bibinfo {year} {2020})}\BibitemShut {NoStop}%
\bibitem [{\citenamefont {Procopio}\ \emph {et~al.}(2020)\citenamefont {Procopio}, \citenamefont {Delgado}, \citenamefont {Enr{\'\i}quez}, \citenamefont {Belabas},\ and\ \citenamefont {Levenson}}]{procopio2020sending}%
  \BibitemOpen
  \bibfield  {author} {\bibinfo {author} {\bibfnamefont {L.~M.}\ \bibnamefont {Procopio}}, \bibinfo {author} {\bibfnamefont {F.}~\bibnamefont {Delgado}}, \bibinfo {author} {\bibfnamefont {M.}~\bibnamefont {Enr{\'\i}quez}}, \bibinfo {author} {\bibfnamefont {N.}~\bibnamefont {Belabas}},\ and\ \bibinfo {author} {\bibfnamefont {J.~A.}\ \bibnamefont {Levenson}},\ }\bibfield  {title} {\bibinfo {title} {Sending classical information via three noisy channels in superposition of causal orders},\ }\href {https://doi.org/https://doi.org/10.1103/PhysRevA.101.012346} {\bibfield  {journal} {\bibinfo  {journal} {Physical Review A}\ }\textbf {\bibinfo {volume} {101}},\ \bibinfo {pages} {012346} (\bibinfo {year} {2020})}\BibitemShut {NoStop}%
\bibitem [{\citenamefont {Simonov}\ \emph {et~al.}(2022)\citenamefont {Simonov}, \citenamefont {Francica}, \citenamefont {Guarnieri},\ and\ \citenamefont {Paternostro}}]{simonov2022work}%
  \BibitemOpen
  \bibfield  {author} {\bibinfo {author} {\bibfnamefont {K.}~\bibnamefont {Simonov}}, \bibinfo {author} {\bibfnamefont {G.}~\bibnamefont {Francica}}, \bibinfo {author} {\bibfnamefont {G.}~\bibnamefont {Guarnieri}},\ and\ \bibinfo {author} {\bibfnamefont {M.}~\bibnamefont {Paternostro}},\ }\bibfield  {title} {\bibinfo {title} {Work extraction from coherently activated maps via quantum switch},\ }\href {https://doi.org/https://doi.org/10.1103/PhysRevA.105.032217} {\bibfield  {journal} {\bibinfo  {journal} {Physical Review A}\ }\textbf {\bibinfo {volume} {105}},\ \bibinfo {pages} {032217} (\bibinfo {year} {2022})}\BibitemShut {NoStop}%
\bibitem [{\citenamefont {Cao}\ \emph {et~al.}(2022)\citenamefont {Cao}, \citenamefont {Wang}, \citenamefont {Jia}, \citenamefont {Zhang}, \citenamefont {Guo}, \citenamefont {Liu}, \citenamefont {Huang}, \citenamefont {Li},\ and\ \citenamefont {Guo}}]{cao2022quantum}%
  \BibitemOpen
  \bibfield  {author} {\bibinfo {author} {\bibfnamefont {H.}~\bibnamefont {Cao}}, \bibinfo {author} {\bibfnamefont {N.-N.}\ \bibnamefont {Wang}}, \bibinfo {author} {\bibfnamefont {Z.}~\bibnamefont {Jia}}, \bibinfo {author} {\bibfnamefont {C.}~\bibnamefont {Zhang}}, \bibinfo {author} {\bibfnamefont {Y.}~\bibnamefont {Guo}}, \bibinfo {author} {\bibfnamefont {B.-H.}\ \bibnamefont {Liu}}, \bibinfo {author} {\bibfnamefont {Y.-F.}\ \bibnamefont {Huang}}, \bibinfo {author} {\bibfnamefont {C.-F.}\ \bibnamefont {Li}},\ and\ \bibinfo {author} {\bibfnamefont {G.-C.}\ \bibnamefont {Guo}},\ }\bibfield  {title} {\bibinfo {title} {Quantum simulation of indefinite causal order induced quantum refrigeration},\ }\href {https://doi.org/https://doi.org/10.1103/PhysRevResearch.4.L032029} {\bibfield  {journal} {\bibinfo  {journal} {Physical Review Research}\ }\textbf {\bibinfo {volume} {4}},\ \bibinfo {pages} {L032029} (\bibinfo {year} {2022})}\BibitemShut {NoStop}%
\bibitem [{\citenamefont {Taranto}\ \emph {et~al.}(2025)\citenamefont {Taranto}, \citenamefont {Milz}, \citenamefont {Murao}, \citenamefont {Quintino},\ and\ \citenamefont {Modi}}]{taranto2025higherorderquantumoperations}%
  \BibitemOpen
  \bibfield  {author} {\bibinfo {author} {\bibfnamefont {P.}~\bibnamefont {Taranto}}, \bibinfo {author} {\bibfnamefont {S.}~\bibnamefont {Milz}}, \bibinfo {author} {\bibfnamefont {M.}~\bibnamefont {Murao}}, \bibinfo {author} {\bibfnamefont {M.~T.}\ \bibnamefont {Quintino}},\ and\ \bibinfo {author} {\bibfnamefont {K.}~\bibnamefont {Modi}},\ }\href {https://arxiv.org/abs/2503.09693} {\bibinfo {title} {Higher-order quantum operations}} (\bibinfo {year} {2025}),\ \Eprint {https://arxiv.org/abs/2503.09693} {arXiv:2503.09693 [quant-ph]} \BibitemShut {NoStop}%
\bibitem [{\citenamefont {Gao}\ \emph {et~al.}(2023)\citenamefont {Gao}, \citenamefont {Li}, \citenamefont {Mishra}, \citenamefont {Yan}, \citenamefont {Simonov},\ and\ \citenamefont {Chiribella}}]{gao2023measuring}%
  \BibitemOpen
  \bibfield  {author} {\bibinfo {author} {\bibfnamefont {N.}~\bibnamefont {Gao}}, \bibinfo {author} {\bibfnamefont {D.}~\bibnamefont {Li}}, \bibinfo {author} {\bibfnamefont {A.}~\bibnamefont {Mishra}}, \bibinfo {author} {\bibfnamefont {J.}~\bibnamefont {Yan}}, \bibinfo {author} {\bibfnamefont {K.}~\bibnamefont {Simonov}},\ and\ \bibinfo {author} {\bibfnamefont {G.}~\bibnamefont {Chiribella}},\ }\bibfield  {title} {\bibinfo {title} {{Measuring Incompatibility and Clustering Quantum Observables with a Quantum Switch}},\ }\href {https://doi.org/10.1103/PhysRevLett.130.170201} {\bibfield  {journal} {\bibinfo  {journal} {Phys. Rev. Lett.}\ }\textbf {\bibinfo {volume} {130}},\ \bibinfo {pages} {170201} (\bibinfo {year} {2023})}\BibitemShut {NoStop}%
\bibitem [{\citenamefont {Ban}(2021)}]{ban2021sequential}%
  \BibitemOpen
  \bibfield  {author} {\bibinfo {author} {\bibfnamefont {M.}~\bibnamefont {Ban}},\ }\bibfield  {title} {\bibinfo {title} {On sequential measurements with indefinite causal order},\ }\href {https://doi.org/10.1016/j.physleta.2021.127383} {\bibfield  {journal} {\bibinfo  {journal} {Physics Letters A}\ }\textbf {\bibinfo {volume} {403}},\ \bibinfo {pages} {127383} (\bibinfo {year} {2021})}\BibitemShut {NoStop}%
\bibitem [{\citenamefont {Yin}\ \emph {et~al.}(2023{\natexlab{b}})\citenamefont {Yin}, \citenamefont {Zhao}, \citenamefont {Yang}, \citenamefont {Guo}, \citenamefont {Zhang}, \citenamefont {Li}, \citenamefont {Han}, \citenamefont {Liu}, \citenamefont {Xu}, \citenamefont {Chiribella}, \citenamefont {Chen}, \citenamefont {Li},\ and\ \citenamefont {Guo}}]{yin2023superHeisenberg}%
  \BibitemOpen
  \bibfield  {author} {\bibinfo {author} {\bibfnamefont {P.}~\bibnamefont {Yin}}, \bibinfo {author} {\bibfnamefont {X.}~\bibnamefont {Zhao}}, \bibinfo {author} {\bibfnamefont {Y.}~\bibnamefont {Yang}}, \bibinfo {author} {\bibfnamefont {Y.}~\bibnamefont {Guo}}, \bibinfo {author} {\bibfnamefont {W.-H.}\ \bibnamefont {Zhang}}, \bibinfo {author} {\bibfnamefont {G.-C.}\ \bibnamefont {Li}}, \bibinfo {author} {\bibfnamefont {Y.-J.}\ \bibnamefont {Han}}, \bibinfo {author} {\bibfnamefont {B.-H.}\ \bibnamefont {Liu}}, \bibinfo {author} {\bibfnamefont {J.-S.}\ \bibnamefont {Xu}}, \bibinfo {author} {\bibfnamefont {G.}~\bibnamefont {Chiribella}}, \bibinfo {author} {\bibfnamefont {G.}~\bibnamefont {Chen}}, \bibinfo {author} {\bibfnamefont {C.-F.}\ \bibnamefont {Li}},\ and\ \bibinfo {author} {\bibfnamefont {G.-C.}\ \bibnamefont {Guo}},\ }\bibfield  {title} {\bibinfo {title} {{Experimental super-Heisenberg quantum metrology with indefinite gate order}},\ }\href {https://doi.org/10.1038/s41567-023-02046-y} {\bibfield
  {journal} {\bibinfo  {journal} {Nat. Phys.}\ }\textbf {\bibinfo {volume} {19}},\ \bibinfo {pages} {1122} (\bibinfo {year} {2023}{\natexlab{b}})}\BibitemShut {NoStop}%
\bibitem [{\citenamefont {Lin}(2022)}]{lin2022lecturenotes}%
  \BibitemOpen
  \bibfield  {author} {\bibinfo {author} {\bibfnamefont {L.}~\bibnamefont {Lin}},\ }\href {https://arxiv.org/abs/2201.08309} {\bibinfo {title} {{Lecture Notes on Quantum Algorithms for Scientific Computation}}} (\bibinfo {year} {2022}),\ \Eprint {https://arxiv.org/abs/2201.08309} {arXiv:2201.08309 [quant-ph]} \BibitemShut {NoStop}%
\bibitem [{\citenamefont {Renner}\ and\ \citenamefont {Brukner}(2021)}]{renner2021reassessing}%
  \BibitemOpen
  \bibfield  {author} {\bibinfo {author} {\bibfnamefont {M.~J.}\ \bibnamefont {Renner}}\ and\ \bibinfo {author} {\bibfnamefont {{\v{C}}.}~\bibnamefont {Brukner}},\ }\bibfield  {title} {\bibinfo {title} {Reassessing the computational advantage of quantum-controlled ordering of gates},\ }\href {https://doi.org/10.1103/PhysRevResearch.3.043012} {\bibfield  {journal} {\bibinfo  {journal} {Phys. Rev. Res.}\ }\textbf {\bibinfo {volume} {3}},\ \bibinfo {pages} {043012} (\bibinfo {year} {2021})}\BibitemShut {NoStop}%
\bibitem [{\citenamefont {Bavaresco}\ \emph {et~al.}(2024{\natexlab{b}})\citenamefont {Bavaresco}, \citenamefont {Yoshida}, \citenamefont {Odake}, \citenamefont {Kristjánsson}, \citenamefont {Taranto}, \citenamefont {Murao},\ and\ \citenamefont {Quintino}}]{bavaresco2024can}%
  \BibitemOpen
  \bibfield  {author} {\bibinfo {author} {\bibfnamefont {J.}~\bibnamefont {Bavaresco}}, \bibinfo {author} {\bibfnamefont {S.}~\bibnamefont {Yoshida}}, \bibinfo {author} {\bibfnamefont {T.}~\bibnamefont {Odake}}, \bibinfo {author} {\bibfnamefont {H.}~\bibnamefont {Kristjánsson}}, \bibinfo {author} {\bibfnamefont {P.}~\bibnamefont {Taranto}}, \bibinfo {author} {\bibfnamefont {M.}~\bibnamefont {Murao}},\ and\ \bibinfo {author} {\bibfnamefont {M.~T.}\ \bibnamefont {Quintino}},\ }\href {https://arxiv.org/abs/2409.18202} {\bibinfo {title} {Can the quantum switch be deterministically simulated?}} (\bibinfo {year} {2024}{\natexlab{b}}),\ \Eprint {https://arxiv.org/abs/2409.18202} {arXiv:2409.18202 [quant-ph]} \BibitemShut {NoStop}%
\bibitem [{\citenamefont {Kristjánsson}\ \emph {et~al.}(2024)\citenamefont {Kristjánsson}, \citenamefont {Odake}, \citenamefont {Yoshida}, \citenamefont {Taranto}, \citenamefont {Bavaresco}, \citenamefont {Quintino},\ and\ \citenamefont {Murao}}]{kristjánsson2024exponentialseparationquantumquery}%
  \BibitemOpen
  \bibfield  {author} {\bibinfo {author} {\bibfnamefont {H.}~\bibnamefont {Kristjánsson}}, \bibinfo {author} {\bibfnamefont {T.}~\bibnamefont {Odake}}, \bibinfo {author} {\bibfnamefont {S.}~\bibnamefont {Yoshida}}, \bibinfo {author} {\bibfnamefont {P.}~\bibnamefont {Taranto}}, \bibinfo {author} {\bibfnamefont {J.}~\bibnamefont {Bavaresco}}, \bibinfo {author} {\bibfnamefont {M.~T.}\ \bibnamefont {Quintino}},\ and\ \bibinfo {author} {\bibfnamefont {M.}~\bibnamefont {Murao}},\ }\href {https://arxiv.org/abs/2409.18420} {\bibinfo {title} {Exponential separation in quantum query complexity of the quantum switch with respect to simulations with standard quantum circuits}} (\bibinfo {year} {2024}),\ \Eprint {https://arxiv.org/abs/2409.18420} {arXiv:2409.18420 [quant-ph]} \BibitemShut {NoStop}%
\bibitem [{\citenamefont {Popescu}(2009)}]{popescu2009unitaryinvariants}%
  \BibitemOpen
  \bibfield  {author} {\bibinfo {author} {\bibfnamefont {G.}~\bibnamefont {Popescu}},\ }\href@noop {} {{\selectlanguage {English}\emph {\bibinfo {title} {Unitary Invariants in Multivariable Operator Theory}}}}\ (\bibinfo  {publisher} {American Mathematical Society},\ \bibinfo {year} {2009})\ p.~\bibinfo {pages} {91}\BibitemShut {NoStop}%
\bibitem [{\citenamefont {Chiribella}\ \emph {et~al.}(2008)\citenamefont {Chiribella}, \citenamefont {D'Ariano},\ and\ \citenamefont {Perinotti}}]{chiribella2008quantumcircuit}%
  \BibitemOpen
  \bibfield  {author} {\bibinfo {author} {\bibfnamefont {G.}~\bibnamefont {Chiribella}}, \bibinfo {author} {\bibfnamefont {G.~M.}\ \bibnamefont {D'Ariano}},\ and\ \bibinfo {author} {\bibfnamefont {P.}~\bibnamefont {Perinotti}},\ }\bibfield  {title} {\bibinfo {title} {Quantum circuit architecture},\ }\href {https://doi.org/10.1103/PhysRevLett.101.060401} {\bibfield  {journal} {\bibinfo  {journal} {Phys. Rev. Lett.}\ }\textbf {\bibinfo {volume} {101}},\ \bibinfo {pages} {060401} (\bibinfo {year} {2008})}\BibitemShut {NoStop}%
\bibitem [{\citenamefont {Wechs}\ \emph {et~al.}(2021)\citenamefont {Wechs}, \citenamefont {Dourdent}, \citenamefont {Abbott},\ and\ \citenamefont {Branciard}}]{wechs2021quantum}%
  \BibitemOpen
  \bibfield  {author} {\bibinfo {author} {\bibfnamefont {J.}~\bibnamefont {Wechs}}, \bibinfo {author} {\bibfnamefont {H.}~\bibnamefont {Dourdent}}, \bibinfo {author} {\bibfnamefont {A.~A.}\ \bibnamefont {Abbott}},\ and\ \bibinfo {author} {\bibfnamefont {C.}~\bibnamefont {Branciard}},\ }\bibfield  {title} {\bibinfo {title} {{Quantum Circuits with Classical Versus Quantum Control of Causal Order}},\ }\href {https://doi.org/10.1103/PRXQuantum.2.030335} {\bibfield  {journal} {\bibinfo  {journal} {Phys. Rev. X Quantum}\ }\textbf {\bibinfo {volume} {2}},\ \bibinfo {pages} {030335} (\bibinfo {year} {2021})}\BibitemShut {NoStop}%
\bibitem [{\citenamefont {Simonov}(2023)}]{simonov2023indefinite}%
  \BibitemOpen
  \bibfield  {author} {\bibinfo {author} {\bibfnamefont {K.}~\bibnamefont {Simonov}},\ }\bibfield  {title} {\bibinfo {title} {Indefinite causality in quantum mechanics and its thermodynamic applications},\ }\href {https://doi.org/10.1088/1742-6596/2533/1/012039} {\bibfield  {journal} {\bibinfo  {journal} {Journal of Physics: Conference Series}\ }\textbf {\bibinfo {volume} {2533}},\ \bibinfo {pages} {012039} (\bibinfo {year} {2023})}\BibitemShut {NoStop}%
\bibitem [{\citenamefont {Araújo}\ \emph {et~al.}(2015)\citenamefont {Araújo}, \citenamefont {Branciard}, \citenamefont {Costa}, \citenamefont {Feix}, \citenamefont {Giarmatzi},\ and\ \citenamefont {Brukner}}]{araujo2015witnessing}%
  \BibitemOpen
  \bibfield  {author} {\bibinfo {author} {\bibfnamefont {M.}~\bibnamefont {Araújo}}, \bibinfo {author} {\bibfnamefont {C.}~\bibnamefont {Branciard}}, \bibinfo {author} {\bibfnamefont {F.}~\bibnamefont {Costa}}, \bibinfo {author} {\bibfnamefont {A.}~\bibnamefont {Feix}}, \bibinfo {author} {\bibfnamefont {C.}~\bibnamefont {Giarmatzi}},\ and\ \bibinfo {author} {\bibfnamefont {{\v{C}}.}~\bibnamefont {Brukner}},\ }\bibfield  {title} {\bibinfo {title} {Witnessing causal nonseparability},\ }\href {https://doi.org/10.1088/1367-2630/17/10/102001} {\bibfield  {journal} {\bibinfo  {journal} {New Journal of Physics}\ }\textbf {\bibinfo {volume} {17}},\ \bibinfo {pages} {102001} (\bibinfo {year} {2015})}\BibitemShut {NoStop}%
\bibitem [{Note1()}]{Note1}%
  \BibitemOpen
  \bibinfo {note} {Importantly, the lack of causal order can only be inferred, e.g. by a witness of causal nonseparability~\cite {araujo2015witnessing}. Eq.~\protect \eqref {eq: quantum switch expression} alone does not indicates the `lack of causal order', as can be seen, for example, by tracing out the control system.}\BibitemShut {Stop}%
\bibitem [{Note2()}]{Note2}%
  \BibitemOpen
  \bibinfo {note} {In our case, since we focus on quantum circuits, we are interested in situations where $\protect \mathcal {M}$ describes a \protect \emph {customisable} quantum circuit~\cite {taranto2025higherorderquantumoperations}. This refers to a setting where the circuit architecture is fixed, but some unitary gates are treated as variable inputs.}\BibitemShut {Stop}%
\bibitem [{\citenamefont {Bavaresco}\ \emph {et~al.}(2021)\citenamefont {Bavaresco}, \citenamefont {Murao},\ and\ \citenamefont {Quintino}}]{bavaresco2021strict}%
  \BibitemOpen
  \bibfield  {author} {\bibinfo {author} {\bibfnamefont {J.}~\bibnamefont {Bavaresco}}, \bibinfo {author} {\bibfnamefont {M.}~\bibnamefont {Murao}},\ and\ \bibinfo {author} {\bibfnamefont {M.~T.}\ \bibnamefont {Quintino}},\ }\bibfield  {title} {\bibinfo {title} {{Strict Hierarchy between Parallel, Sequential, and Indefinite-Causal-Order Strategies for Channel Discrimination}},\ }\href {https://doi.org/10.1103/PhysRevLett.127.200504} {\bibfield  {journal} {\bibinfo  {journal} {Phys. Rev. Lett.}\ }\textbf {\bibinfo {volume} {127}},\ \bibinfo {pages} {200504} (\bibinfo {year} {2021})}\BibitemShut {NoStop}%
\bibitem [{\citenamefont {Yoshida}\ \emph {et~al.}(2023)\citenamefont {Yoshida}, \citenamefont {Soeda},\ and\ \citenamefont {Murao}}]{yoshida2023reversing}%
  \BibitemOpen
  \bibfield  {author} {\bibinfo {author} {\bibfnamefont {S.}~\bibnamefont {Yoshida}}, \bibinfo {author} {\bibfnamefont {A.}~\bibnamefont {Soeda}},\ and\ \bibinfo {author} {\bibfnamefont {M.}~\bibnamefont {Murao}},\ }\bibfield  {title} {\bibinfo {title} {Reversing unknown qubit-unitary operation, deterministically and exactly},\ }\href {https://doi.org/https://doi.org/10.1103/PhysRevLett.131.120602} {\bibfield  {journal} {\bibinfo  {journal} {Physical Review Letters}\ }\textbf {\bibinfo {volume} {131}},\ \bibinfo {pages} {120602} (\bibinfo {year} {2023})}\BibitemShut {NoStop}%
\bibitem [{\citenamefont {Childs}\ and\ \citenamefont {Wiebe}(2012)}]{childs2012}%
  \BibitemOpen
  \bibfield  {author} {\bibinfo {author} {\bibfnamefont {A.~M.}\ \bibnamefont {Childs}}\ and\ \bibinfo {author} {\bibfnamefont {N.}~\bibnamefont {Wiebe}},\ }\bibfield  {title} {\bibinfo {title} {Hamiltonian simulation using linear combinations of unitary operations},\ }\href {https://doi.org/10.26421/qic12.11-12-1} {\bibfield  {journal} {\bibinfo  {journal} {Quantum Information and Computation}\ }\textbf {\bibinfo {volume} {12}},\ \bibinfo {pages} {901–924} (\bibinfo {year} {2012})}\BibitemShut {NoStop}%
\bibitem [{\citenamefont {Chowdhury}\ and\ \citenamefont {Somma}(2017)}]{Chow17}%
  \BibitemOpen
  \bibfield  {author} {\bibinfo {author} {\bibfnamefont {A.~N.}\ \bibnamefont {Chowdhury}}\ and\ \bibinfo {author} {\bibfnamefont {R.~D.}\ \bibnamefont {Somma}},\ }\bibfield  {title} {\bibinfo {title} {Quantum algorithms for gibbs sampling and hitting-time estimation},\ }\href {https://doi.org/https://dl.acm.org/doi/10.5555/3179483.3179486} {\bibfield  {journal} {\bibinfo  {journal} {Quantum Information and Computation}\ }\textbf {\bibinfo {volume} {17}},\ \bibinfo {pages} {41–64} (\bibinfo {year} {2017})}\BibitemShut {NoStop}%
\bibitem [{\citenamefont {Halpern}\ \emph {et~al.}(2018)\citenamefont {Halpern}, \citenamefont {Swingle},\ and\ \citenamefont {Dressel}}]{halpern2018quasiprobability}%
  \BibitemOpen
  \bibfield  {author} {\bibinfo {author} {\bibfnamefont {N.~Y.}\ \bibnamefont {Halpern}}, \bibinfo {author} {\bibfnamefont {B.}~\bibnamefont {Swingle}},\ and\ \bibinfo {author} {\bibfnamefont {J.}~\bibnamefont {Dressel}},\ }\bibfield  {title} {\bibinfo {title} {Quasiprobability behind the out-of-time-ordered correlator},\ }\href {https://doi.org/10.1103/physreva.97.042105} {\bibfield  {journal} {\bibinfo  {journal} {Phys. Rev. A}\ }\textbf {\bibinfo {volume} {97}},\ \bibinfo {pages} {042105} (\bibinfo {year} {2018})}\BibitemShut {NoStop}%
\bibitem [{\citenamefont {Gonz\'alez~Alonso}\ \emph {et~al.}(2019)\citenamefont {Gonz\'alez~Alonso}, \citenamefont {Yunger~Halpern},\ and\ \citenamefont {Dressel}}]{gonzalez2019otoc}%
  \BibitemOpen
  \bibfield  {author} {\bibinfo {author} {\bibfnamefont {J.~R.}\ \bibnamefont {Gonz\'alez~Alonso}}, \bibinfo {author} {\bibfnamefont {N.}~\bibnamefont {Yunger~Halpern}},\ and\ \bibinfo {author} {\bibfnamefont {J.}~\bibnamefont {Dressel}},\ }\bibfield  {title} {\bibinfo {title} {Out-of-time-ordered-correlator quasiprobabilities robustly witness scrambling},\ }\href {https://doi.org/10.1103/PhysRevLett.122.040404} {\bibfield  {journal} {\bibinfo  {journal} {Phys. Rev. Lett.}\ }\textbf {\bibinfo {volume} {122}},\ \bibinfo {pages} {040404} (\bibinfo {year} {2019})}\BibitemShut {NoStop}%
\bibitem [{\citenamefont {Budiyono}(2023)}]{budiyono2023operational}%
  \BibitemOpen
  \bibfield  {author} {\bibinfo {author} {\bibfnamefont {A.}~\bibnamefont {Budiyono}},\ }\bibfield  {title} {\bibinfo {title} {Operational interpretation and estimation of quantum trace-norm asymmetry based on weak-value measurement and some bounds},\ }\href {https://doi.org/10.1103/PhysRevA.108.012431} {\bibfield  {journal} {\bibinfo  {journal} {Phys. Rev. A}\ }\textbf {\bibinfo {volume} {108}},\ \bibinfo {pages} {012431} (\bibinfo {year} {2023})}\BibitemShut {NoStop}%
\bibitem [{\citenamefont {Budiyono}\ and\ \citenamefont {Dipojono}(2023)}]{budiyono2023quantifying}%
  \BibitemOpen
  \bibfield  {author} {\bibinfo {author} {\bibfnamefont {A.}~\bibnamefont {Budiyono}}\ and\ \bibinfo {author} {\bibfnamefont {H.~K.}\ \bibnamefont {Dipojono}},\ }\bibfield  {title} {\bibinfo {title} {Quantifying quantum coherence via {K}irkwood-{D}irac quasiprobability},\ }\href {https://doi.org/10.1103/PhysRevA.107.022408} {\bibfield  {journal} {\bibinfo  {journal} {Phys. Rev. A}\ }\textbf {\bibinfo {volume} {107}},\ \bibinfo {pages} {022408} (\bibinfo {year} {2023})}\BibitemShut {NoStop}%
\bibitem [{\citenamefont {De~Bi\`evre}(2021)}]{debievre2021complete}%
  \BibitemOpen
  \bibfield  {author} {\bibinfo {author} {\bibfnamefont {S.}~\bibnamefont {De~Bi\`evre}},\ }\bibfield  {title} {\bibinfo {title} {{Complete Incompatibility, Support Uncertainty, and Kirkwood-Dirac Nonclassicality}},\ }\href {https://doi.org/10.1103/PhysRevLett.127.190404} {\bibfield  {journal} {\bibinfo  {journal} {Phys. Rev. Lett.}\ }\textbf {\bibinfo {volume} {127}},\ \bibinfo {pages} {190404} (\bibinfo {year} {2021})}\BibitemShut {NoStop}%
\end{thebibliography}%

\onecolumngrid
\appendix

\section{Input states permutation}\label{app: proof of the rules}

In this Appendix we show that if we input the states $\rho_{a_1} \otimes \dots \otimes \rho_{a_n}$, for labels $a_1,\ldots,a_n \in \{1,\ldots,n\}$ chosen via Eq.~\eqref{eq: inputState}, we estimate the Bargmann invariants $\Tr(\rho_1 \cdots \rho_n)$ using the protocol discussed in Sec.~\ref{sec:results}. In what follows, $n$ is always an odd integer satisfying $n \geq 3$. Recall that our choice of unitaries $A_n$ and $B_n$ is given by
\begin{equation}\label{eq: def of A and B}
    A_n = \prod_{\substack{i=1\\ i\text{ odd}}}^{n-1} \mathrm{SWAP}_{i,i+1}, \quad B_n = \prod_{\substack{i=2\\ i\text{ even}}}^{n-1} \mathrm{SWAP}_{i,i+1}.
\end{equation}
In the main text, we have presented a rule to find the coefficients $a_1,\ldots,a_n$, which is as follows:
\begin{align}
a_{4s+1} &= s+1, \nonumber\\
a_{4s+2} &= k+1-s, \label{eq: blockruleAPP}\\
a_{4s+3} &= k+s+2, \nonumber\\
a_{4s+4} &= 2k+1-s, \nonumber
\end{align}
where $s = 0, 1, 2, \dots, \lfloor n/4 \rfloor$, i.e. until we reach the component $a_n$, and we have written $n=2k+1$. 

In summary, we want to show the following theorem. 
\begin{theorem}\label{theorem: main theorem appendix}
For every odd integer $n \geq 3$  odd and unitaries $A_n$ and $ B_n$ given by Eq.~\eqref{eq: def of A and B}, it holds that
\begin{equation}\label{eq: equation Tn}
    \Tr(T_n {\psi_{\mathrm{sw}}}^{(n)}) = \Tr(T_n(\psi_{a_1}  \otimes \dots \otimes \psi_{a_n})) = \Tr(\psi_1 \cdots \psi_n),
\end{equation}
where $\psi \equiv \vert \psi\rangle \langle \psi \vert$, $T_n := A_nB_nA_nB_n$,  and ${\psi_{\mathrm{sw}}}^{(n)} := \psi_{a_1} \otimes \dots \otimes \psi_{a_n}$ with $a_1,\dots,a_n$ given by Eqs.~\eqref{eq: blockruleAPP}.
\end{theorem}
If it is clear from the context, we simply write ${\psi_{\mathrm{sw}}}^{(n)} \equiv {\psi_{\mathrm{sw}}}$. Our strategy will be to use induction. The coefficients $\{a_i\}_i$ from Eqs.~\eqref{eq: blockruleAPP} do not immediately provide intuition on how to apply the inductive step. To do so, we first recognize a pattern of the inner products $\langle \psi_{a_i}|\psi_{a_j}\rangle$ present in $\Tr(T_n{\psi_{\mathrm{sw}}}^{(n)})$. Let us also denote the action of $T_n$ as 
    $$\vert \psi_{a_1}\dots \psi_{a_n}\rangle \mapsto \vert \psi_{\tilde{a}_1}\dots \psi_{\tilde{a}_n}\rangle = \vert \psi_{a_{\permT^{-1}(1)}}\dots \psi_{a_{\permT^{-1}(n)}}\rangle,$$ where $\permT$ is the associated permutation. This implies that we can equivalently write  
    \begin{equation*}
        \Tr(T_n {\psi_{\mathrm{sw}}}^{(n)}) = \prod_{i=1}^n\langle \psi_{a_i}|\psi_{a_{\permT^{-1}(i)}}\rangle = \prod_{(a_i,a_j) \in I_n} \langle \psi_{a_i}|\psi_{a_j}\rangle, 
    \end{equation*}
where we have defined $I_n$ as the set of all pairs $(a_i,a_j) \equiv a_{ij}$ for which $\langle \psi_{a_i}|\psi_{a_j}\rangle$ is an inner-product entering $\Tr(T_n {\psi_{\mathrm{sw}}}^{(n)})$. We now show a Lemma on how to construct $I_{n+2}$ from $I_n$.

\begin{lemma}\label{lemma: update rule}
Given $I_n$, we can algorithmically construct the set $I_{n+2}$, for every $n \geq 5$ as follows: 
    \begin{enumerate}
        \item If $a \in I_{n}$, then include $a \in I_{n+2}.$
        \item Transform $a_{n-2,n-1} \to a_{n-2,n+2}.$
        \item Transform $a_{n,n-3} \to a_{n,n+1}$.
        \item Include $a_{n+1,n-3} \in I_{n+2}$.
        \item Include $a_{n+2,n-1}\in I_{n+2}$.
    \end{enumerate}
\end{lemma}

\begin{proof}
    If we show that  $\prod_{(a_i,a_j) \in I_{n+2}} \langle \psi_{a_i}|\psi_{a_j}\rangle $ is equal to $\Tr(T_{n+2}{\psi_{\mathrm{sw}}}^{(n+2)})$, for the set $I_{n+2}$ constructed from $I_n$ as described above, we conclude the proof. 
    \begin{figure}
        \centering
        \includegraphics[width=0.75\linewidth]{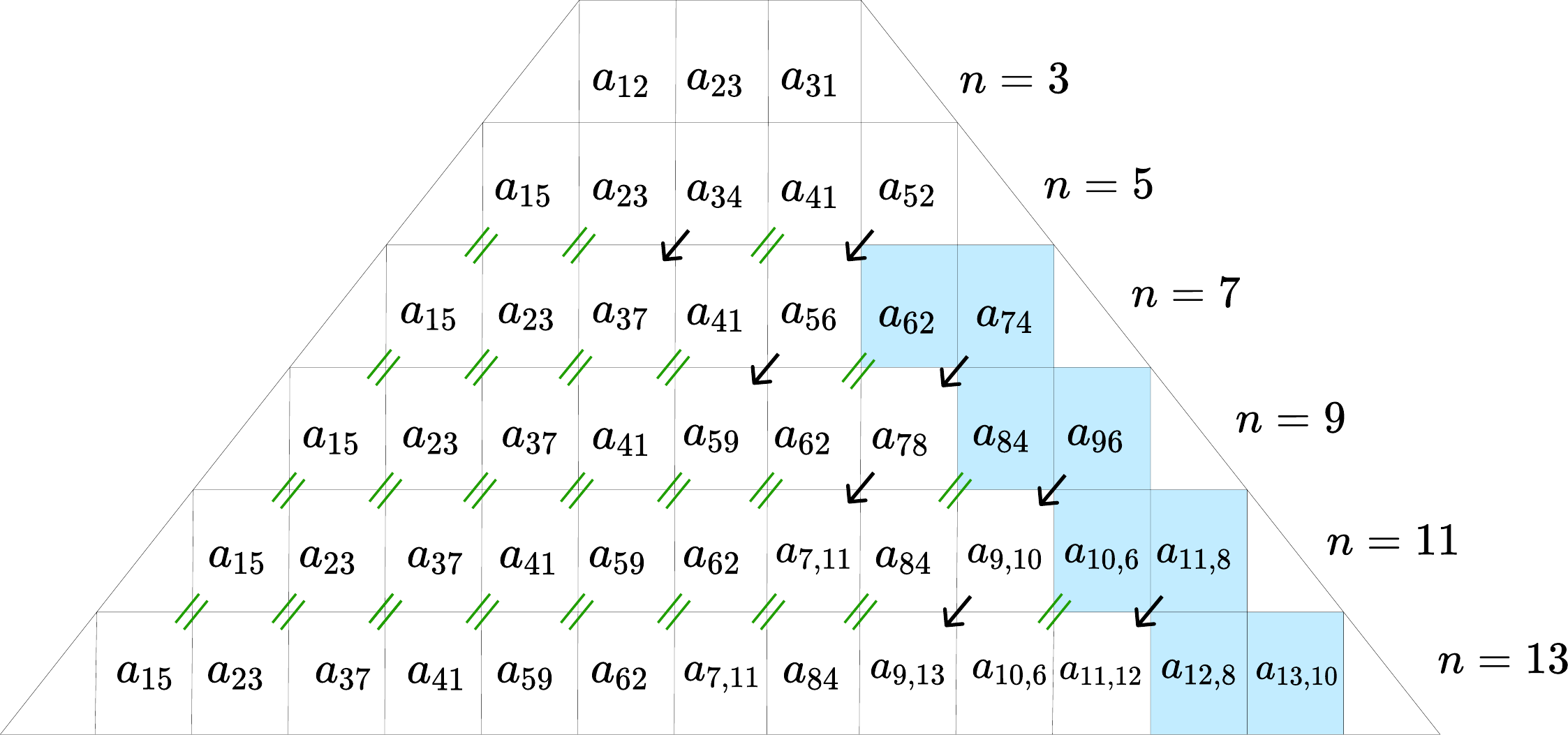}
        \caption{\textbf{Algorithm from Lemma~\ref{lemma: update rule}.} We can find out which labels for the inner products $a_{ij} = (a_i,a_j)$ that appear in $\Tr(T_{n+2} {\psi_{\mathrm{sw}}}^{(n+2)})$ by updating them from those that appear in $\Tr(T_n {\psi_{\mathrm{sw}}}^{(n)})$ according to Lemma~\ref{lemma: update rule}. The equality signs illustrate step 1, the arrows illustrate steps 2 and 3, while the blue squares the steps 4 and 5 from Lemma~\ref{lemma: update rule}. }
        \label{fig: rule}
    \end{figure}
    Let us start by noting that it is simple to see that we only need to update an inner product $\langle \psi_{a_i}|\psi_{a_j}\rangle $ labeled by $(a_i,a_j)$ appearing in $\Tr(T_{n+2}{\psi_{\mathrm{sw}}}^{(n+2)})$ that had not appeared in $\Tr(T_{n}{\psi_{\mathrm{sw}}}^{(n)})$ if the action of $T_n$ to produce $\langle \psi_{a_i}|\psi_{a_j}\rangle $ interacted with the system $n$ (apart from the label $a_n$). This only happens to the labels $a_{n-1}$ and $a_{n-3}$, due to the structure of $A_n$ and $B_n$. This can be seen in Fig.~\ref{fig: circuit inductive step}. This gives us the rules $1,2$ and $3$ from the algorithm. To conclude, we notice that we need to include $\langle \psi_{a_{n+1}}|\psi_{a_{\permT^{-1}(n+1)}}\rangle$ and $\langle \psi_{a_{n+2}}|\psi_{a_{\permT^{-1}(n+2)}}\rangle$. It is simple to see, again from Fig.~\ref{fig: circuit inductive step}, that $\langle \psi_{a_{n+1}}|\psi_{a_{\permT^{-1}(n+1)}}\rangle  = \langle \psi_{a_{n+1}}|\psi_{a_{n-3}}\rangle$ and $\langle \psi_{a_{n+2}}|\psi_{a_{n-1}}\rangle.$
\end{proof}

This Lemma shows that as $n \to n+2$ there are always only four new pairs $(a_i,a_j) \in I_{n+2}$ that were not in $I_n$, that are those shown in items $2,3,4,$ and $5$ from Lemma~\ref{lemma: update rule}. This implies that there are only four new inner products $\langle \psi_{a_i}|\psi_{a_j}\rangle $ in $\Tr(T_{n+2}{\psi_{\mathrm{sw}}}^{(n+2)})$ that were not present in $\Tr(T_{n}{\psi_{\mathrm{sw}}}^{(n)})$.   

\begin{proof}[Proof of Theorem~\ref{theorem: main theorem appendix}]

We will prove this result by induction. It is trivial to see that the result holds for small values of $n$ such as $3,5,7$ etc. We take $n=5$ as our base case, so we can use Lemma~\ref{lemma: update rule}.

We now consider the case for $n \rightarrow n+2$, assuming that the result holds for any odd integer $n \geq 5$. The initial input state reads $$\ket{{\psi_{\mathrm{sw}}}^{(n+2)}} = \ket{\psi_{a_1}, \psi_{a_2}, \dots, \psi_{a_{n}},\psi_{a_{n+1}},\psi_{a_{n+2}}} \equiv \vert a_1,a_2,\dots,a_{n},a_{n+1},a_{n+2}\rangle,$$ where, from now on, we will just write the labels for clarity and simplicity of the calculation. Also, there is one extra SWAP gate for each of the operations $A_{n+2}, B_{n+2}$ with respect to $A_n, B_n$.

\begin{figure}
    \centering
    \includegraphics[width=0.75\linewidth]{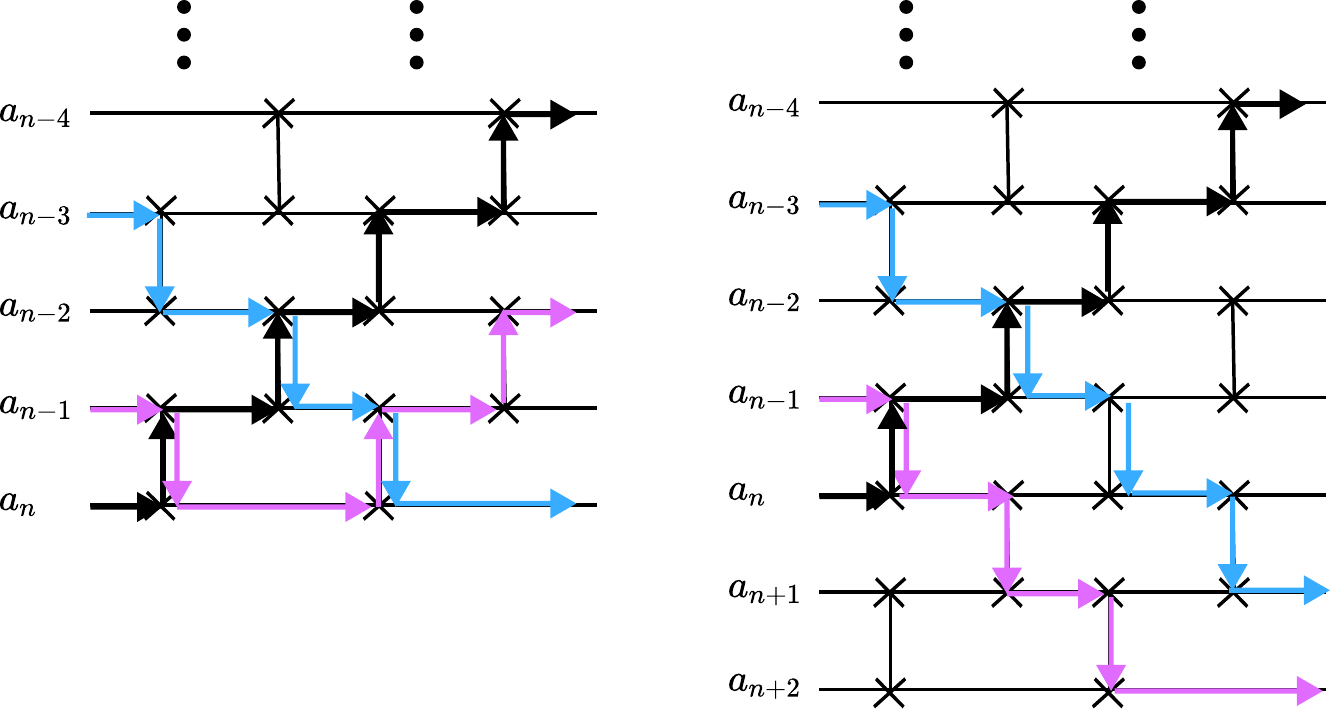}
    \caption{\textbf{From $T_n$ to $T_{n+2}$.} Note that if we change from $n \to n+2$ we have that the label permutation $a_n \to a_{n-4}$ continues to hold (shown by the black arrows). This effectively implies that $\langle \psi_{a_{n}}|\psi_{a_{n-4}}\rangle$ (for all $n \geq 5$) is present in both  $\Tr(T_n {\psi_{\mathrm{sw}}}^{(n)})$ and $\Tr(T_{n+2} {\psi_{\mathrm{sw}}}^{(n+2)})$. Similarly to all other inner products not affected by the change $T_n \to T_{n+2}$ since these are not interchanged with the last system $a_n$. If they \emph{are} interchanged with system $a_n$, such as is the case with $a_{n-1}$ and $a_{n-3}$, one needs to update them, and it is simple to see that from the form of $T_n$ this is as  described in Lemma~\ref{lemma: update rule}. }
    \label{fig: circuit inductive step}
\end{figure}

Picturing the quantum circuit with the added states at the bottom, we can follow their path through the SWAP gates to see where they end up (see Fig.~\ref{fig: circuit inductive step}). The state $\vert a_{n+1}\rangle$ goes first through a SWAP gate in the first layer (going down), then another on the third (going up) and then on the fourth layer (going up), such that it ends up in the slot $n$ in the output state. Following the state $\vert a_{n+2}\rangle $, which is the last one, we see that it does not skip any of the layers and always goes up, ending up in the slot $n-2$ of the output state.

Since we assume that $\Tr(T_n {\psi_{\mathrm{sw}}}^{(n)}) = \Tr(\psi_1\psi_2\cdots\psi_n)$, in order to show that the same holds for $n \to n+2$ it suffices to show, due to Lemma~\ref{lemma: update rule}, that $\vert a_i\rangle \stackrel{T_n}{\mapsto}\vert \tilde{a}_i\rangle = \vert a_{\permT^{-1}(i)}\rangle = \vert a_i+1\rangle$ for the new inner products in $\Tr(T_{n+2} {\psi_{\mathrm{sw}}}^{(n+2)}),$ given according to Lemma~\ref{lemma: update rule}, namely 
\begin{align}
    \alpha_1 &= \langle a_{n-2}|a_{n+2}\rangle,\\ \alpha_2 &= \langle a_{n}|a_{n+1}\rangle,\\ \alpha_3 &= \langle a_{n+1}|a_{n-3}\rangle,\\ \alpha_4 &= \langle a_{n+2}|a_{n-1}\rangle. 
\end{align}

To do so, we note that from the permutation rule described in Eq.~\eqref{eq: blockruleAPP}, and from the fact that $n+2$ is odd, we have that $a_{n+2}$ must be either in slots $a_{4s+1}$ or $a_{4s+3}$. We conclude the proof by explicitly calculating each such case for all inner products $\alpha_1,\alpha_2,\alpha_3,\alpha_4$ shown above. For the first case, we have that $n+2= 4s+1 = 2k'+1$ and, from that, $k'=2s$. Then, following the coefficients from Eq.~\eqref{eq: blockruleAPP}, the inner products are given by
\begin{align*}
    \alpha_1 &= \braket{a_{n-2}}{a_{n+2}} = \braket{a_{4(s-1)+1}}{a_{4s+1}} = \braket{(s-1)+1}{s+1} = \braket{s}{s+1} \\
    \alpha_2 &= \braket{a_{n}}{a_{n+1}} = \braket{a_{4(s-1)+3}}{a_{4(s-1)+4}} = \braket{k'+(s-1)+2}{2k'+1-(s-1)} = \braket{3s+1}{3s+2} \\
    \alpha_3 &=\braket{a_{n+1}}{a_{n-3}} = \braket{a_{4(s-1)+4}}{a_{4(s-2)+4}} = \braket{2k'+1-(s-1)}{2k'+1-(s-2)} = \braket{2k'-s+2}{2k'-s+3} \\
    \alpha_4 &=\braket{a_{n+2}}{a_{n-1}} = \braket{a_{4s+1}}{a_{4(s-1)+2}} = \braket{s+1}{k'-s+2} = \braket{s+1}{s+2}.
\end{align*}

And for the second case, where $n'=4s+3$ and $k'=2s+1$, we find
\begin{align*}
    \alpha_1 &=\braket{a_{n-2}}{a_{n+2}} = \braket{a_{4(s-1)+3}}{a_{4s+3}} = \braket{k'+(s-1)+2}{k'+s+2} \\
    \alpha_2 &=\braket{a_{n}}{a_{n+1}} = \braket{a_{4(s-1)+3}}{a_{4(s-1)+4}} = \braket{s+1}{k'+1-s} = \braket{s+1}{s+2} \\
    \alpha_3 &=\braket{a_{n+1}}{a_{n-3}} = \braket{a_{4(s-1)+4}}{a_{4(s-2)+4}} = \braket{k'-s+1}{k'-s+2} \\
    \alpha_4 &=\braket{a_{n+2}}{a_{n-1}} = \braket{a_{4s+1}}{a_{4(s-1)+2}} = \braket{k'+s+2}{2k'+1-s+2} = \braket{3s+3}{3s+4},
\end{align*}
where we can see that for both cases the condition $\braket{a_i}{a_{\permT^{-1}(i)}} = \braket{a_i}{a_i +1}$ is fulfilled for all the changed inner products.

In conclusion, by showing that the coefficients~\eqref{eq: blockruleAPP} work for $n=3$ and for every $n'=n+2$, by writing the initial state of the system based on it we will find all Bargmann invariants with order $n \geq 3$.
\end{proof}

\section{Other unitary operations for estimating Bargmann invariants with the quantum switch} \label{appendix: proof first eq: ansatz}

There are various different choices of unitary operations $A_n$ and $B_n$ that can be used as inputs of the quantum switch, and that are useful for measuring Bargmann invariants. For another example, let us focus on odd $n=2k+1$. One possibility is to choose permutations $\permAn$ and $\permBn$ defined (differently than those considered in the main text) as follows:
\begin{alignb}\label{eq: ansatz}
    \permAn &= \cycle[\quad]{m, m+1, ..., n-1, n},  \\
    \permBn &= \cycle[\quad]{1,m,n} \cycle[\quad]{2, n-1} \cycle[\quad]{3, n-2}\; ...\; \cycle[\quad]{m-2, m+2} \cycle[\quad]{m-1, m+1},
\end{alignb}
Above we take $m = (n+1)/2$. This solution exhibits a straightforward circuit pattern for all odd $n$. In Fig.~\ref{fig: ex5e7} we show two instances (the cases $n=5$ and $n=7$) of the unitary representations of the permutations just described using SWAP decompositions and we illustrate how the protocol works by working out explicitly the case $n=3$ for these two unitaries. 

To find the real part of the third-order pure state Bargmann invariants from the quantum switch operations $A_3$ and $B_3$ given by Eq.~\eqref{eq: ansatz}, we input the quantum switch with a pair of unitary channels given by $A_3=\text{SWAP}_{2,3}$ and $B_3=\text{SWAP}_{1,2}\; \text{SWAP}_{2,3}$. The unitary operation of the quantum switch becomes then $S_{A_3, B_3}$ given by Eq.~\eqref{eq: switcheq}. We then prepare the quantum state $S_{A_3, B_3}\vert +\rangle \vert \psi_1\psi_2\psi_3\rangle $, and after we apply a Hadamard unitary on the control qubit we end up with the following quantum state~\cite{procopio2015experimental}
\begin{align*}
  \frac{1}{2}\vert 0\rangle \{B_3,A_3\}\vert \psi_1\psi_2\psi_3\rangle + \frac{1}{2}\vert 1\rangle [B_3,A_3]\vert \psi_1\psi_2\psi_3\rangle. 
\end{align*}
If we now open the expression and apply the unitaries $A_3, B_3$ chosen we have that
\begin{alignb}    \ket{\psi_f^{\mathrm{SWITCH}}} &= \frac{1}{2} \Big[ \ket{0} \otimes (B_3A_3+A_3B_3) \ket{\psi_1 \psi_2 \psi_3} \\
    &\quad\;\; + \ket{1} \otimes (B_3A_3-A_3B_3) \ket{\psi_1 \psi_2 \psi_3} \Big] \\
    &= \frac{1}{2} \Big[ \ket{0} \otimes \big( \ket{\psi_2 \psi_1 \psi_3} + \ket{\psi_3 \psi_2 \psi_1} \big) \\
    &\quad\;\; + \ket{1} \otimes \big( \ket{\psi_2 \psi_1 \psi_3} - \ket{\psi_3 \psi_2 \psi_1} \big) \Big].
\end{alignb}
The probability to obtain $\vert 1\rangle \langle 1 \vert $ when measuring the control qubit with the $Z$ basis (equivalently to obtaining $\vert -\rangle \langle - \vert $ since we have already applied the Hadamard on the control qubit) is then given by $p_1 = (1 - \Re[\Delta_3])/2$, where $\Delta_3 = \braket{\psi_1}{\psi_2} \braket{\psi_2}{\psi_3} \braket{\psi_3}{\psi_1}$ is the third order Bargmann invariant.

Define the \emph{involution of words} as the operation 
\begin{equation}\label{eq: involution of words}
    (1, \, 2,\,  3, \dots,\,  n-1, \, n)^\dagger = (n,\,  n-1,\,  \dots,\,  3,\,  2,\,  1). 
\end{equation}
Without loss of generality, we will use the equivalent cyclic notation $(1, 2, \dots, n) \equiv (1 \, 2 \, \dots \, n)$ whenever convenient. With these, we are now ready to show the following.

\begin{proposition}
    Let $n\geq 1$ be any odd integer. Let $\permAn$ and $\permBn$ be a class of permutations defined as
\begin{align*}
    \permAn &= \cycle[\quad]{m, m+1, ..., n-1, n},  \\
    \permBn &= \cycle[\quad]{1,m,n} \cycle[\quad]{2, n-1} \cycle[\quad]{3, n-2}\; ...\; \cycle[\quad]{m-2, m+2} \cycle[\quad]{m-1, m+1},
\end{align*}
    where $n=2m-1$. Then, for all $n$, we have that 
    \begin{equation*}
        \permCn = \permAn\permBn\permAn^{-1} \permBn^{-1} = (1 \,\,\, 2 \,\,\, 3 \,\,\, \dots\,\,\, n).
    \end{equation*}
\end{proposition}

\begin{proof}

\begin{figure}[t]
	\centering
	\includegraphics[scale=.6]{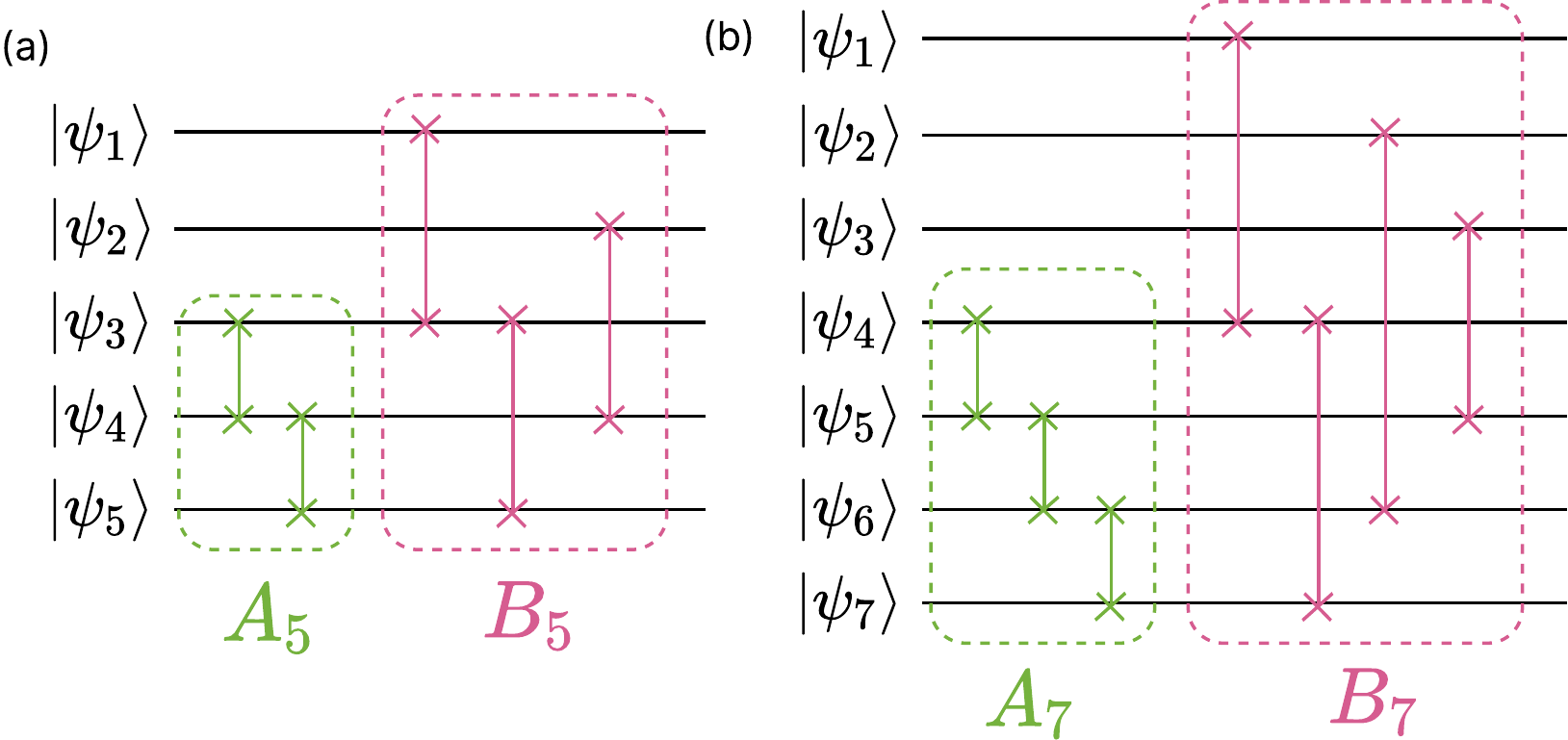}
	\caption{Unitary operation $B_nA_n$ applied to a target product state with (a) $n = 5$ and (b) $n = 7$ systems. Operation $A_n$ is a unitary associated to a $3$-cycle permutation, while $B_n$ swaps the first with the middle and then the final qubit, followed by iterative swaps between the outermost qubits up until the neighbors of the middle qubit.}
	\label{fig: ex5e7}
\end{figure}
We start with the product $\permAn\permBn$, which can be re-written as:
\begin{align*}
    \permAn\permBn &= \cycle[\quad]{m, m+1, ..., 2m-1} \cycle[\quad]{1,m,2m-1} \cycle[\quad]{2, 2m-2} \cycle[\quad]{3, 2m-3}\; ...\; \cycle[\quad]{m-1, m+1} \\
    &= \cycle[\quad]{m+1, ..., 2m-1, m} \cycle[\quad]{m, 2m-1, 1} \cycle[\quad]{2, 2m-2} \cycle[\quad]{3, 2m-3}\; ...\; \cycle[\quad]{m-1, m+1},
\end{align*}
where we have used $n=2m-1$, and used that $(a_1  \, a_2 \, \dots \, a_n) = (a_2 \, \dots \, a_n \, a_1)$ in cycle notation. Knowing that $\cycle[\;]{a,b}=\cycle[\;]{b,a} = \cycle[\;]{a,b}^{-1} = \cycle[\;]{a,b}^{^\dagger}$, we can simplify the above expression using the following relation
\begin{alignb}\label{procedure}
    \cycle[\quad]{..., 2m-1, m}\cycle[\quad]{m, 2m-1, 1} &= \cycle[\quad]{..., 2m-1} \cycle[\quad]{2m-1, m} \cycle[\quad]{m, 2m-1} \cycle[\quad]{2m-1, 1} \\
    &= \cycle[\quad]{..., 2m-1, 1}.
\end{alignb}
Where we have also merged the two cycles remaining, since when we have two cycles with the same element beside each other (as is $2m-1$), we can combine these two cycles into one following the property $   \cycle[\;]{a, b, c} \cycle[\;]{c, d, e} = \cycle[\;]{a, b, c, d, e}$. Identifying these occurrences is the only step required to proceed with the proof.

Moving forward, we are left with
\begin{align*}
    \permAn\permBn &= \cycle[\quad]{m+1, ..., 2m-1, 1} \cycle[\quad]{2, 2m-2} \cycle[\quad]{3, 2m-3}\; ...\; \cycle[\quad]{m-1, m+1} \\
    &= \cycle[\quad]{2m-1, 1, m+1, ..., 2m-2} \cycle[\quad]{2m-2, 2} \cycle[\quad]{3, 2m-3}\; ...\; \cycle[\quad]{m-1, m+1} \\
    &= \cycle[\quad]{2m-1, 1, m+1, ..., 2m-2, 2} \cycle[\quad]{3, 2m-3}\; ...\; \cycle[\quad]{m-1, m+1}.
\end{align*}

By iteratively doing this procedure, we find a single cycle permutation that represents the product $\permAn \permBn$
\begin{alignb}\label{eq: proofAB}
    \permAn \permBn = \cycle[\quad]{m+1, m-1, m+2, m-2, ..., 2m-1, 1}.
\end{alignb}
This result marks the first part of the proof that $\permAn\permBn \permAn^{-1} \permBn^{-1} = \permCn$. Next, we will find $(\permBn \permAn)^{-1}$, so that later we may combine them.

Therefore, we write the product $\permBn \permAn$ as
\begin{align*}
    \permBn \permAn &= \cycle[\quad]{1,m,2m-1} \cycle[\quad]{2, 2m-2} \cycle[\quad]{3, 2m-3}\; ...\; \cycle[\quad]{m-1, m+1} \cycle[\quad]{m, m+1, ..., 2m-1},
\end{align*}
and by doing the same simplifications as for $\permAn \permBn$, we end up with its cycle permutation
\begin{align*}
    \permBn \permAn &= \cycle[\quad]{1, m, m-1, m+1, ..., 3, 2m-3, 2, 2m-2}.
\end{align*}
To find the inverse, we just rewrite it from right to left:
\begin{alignb}\label{eq: proofBA}
    (\permBn \permAn)^{-1} = \cycle[\quad]{2m-2, 2, 2m-3, 3, ..., m+1, m-1, m, 1}.
\end{alignb}

Now the only step left is to find $A_nB_n(B_nA_n)^{-1}$. Multiplying equations~\eqref{eq: proofAB} and~\eqref{eq: proofBA}, we have
\begin{align*}
    \permAn\permBn(\permBn\permAn)^{-1} &= \cycle[\quad]{m+1, m-1, m+2, m-2, ..., 2m-1, 1} \cycle[\quad]{2m-2, 2, 2m-3, 3, ..., m, 1}.
\end{align*}

By doing the same procedure as in Eq.~\eqref{procedure},
\begin{align*}
    \permAn\permBn(\permBn \permAn)^{-1} &= \cycle[\quad]{2m-1, 1, m+1, m-1, ..., 2m-2} \cycle[\quad]{2, 2m-3, 3, ..., m, 1} \\
    &= \cycle[\quad]{2m-2, 2m-1, 1, m+1, m-1, ..., 2m-3, 3} \cycle[\quad]{2m-3, 3, ..., m, 1, 2} \\
    &= \cycle[\quad]{2m-2, 2m-1, 1, m+1, m-1, ..., 2m-3} \cycle[\quad]{3, ..., m, 1, 2},
\end{align*}
we see that the higher terms remain in the left cycle and the lower terms in the right one. If we keep simplifying until the only common term between these two cycle permutations is the number $1$, we are left with
\begin{align*}
    \permAn\permBn(\permBn\permAn)^{-1} &= \cycle[\quad]{1, m+1, m+2, ..., 2m-1} \cycle[\quad]{m, 1, 2, ..., m-2, m-1} \\
    &= \cycle[\quad]{m+1, m+2, ..., 2m-1, 1} \cycle[\quad]{1, 2, ..., m-2, m-1, m} \\
    &= \cycle[\quad]{m+1, m+2, ..., 2m-1, 1, 2, ..., m-2, m-1, m}.
\end{align*}
And if we rearrange it, $\permAn \permBn(\permBn \permAn)^{-1} = \cycle[\quad]{1, 2, ..., m-2, m-1, m, m+1, m+2, ..., 2m-1} = \permCn$, with $n=2m-1$.
    
\end{proof}

This result is interesting because it allows us to see that by choosing $A_n$ and $B_n$ in this way (different from the choice of the main text) we have that it is trivial to see that the Hadamard test shown in Fig.~\ref{fig: simulgeral} can simulate the quantum switch operation $S_{A_n,B_n}$ since we have that $A_nB_nA_n^\dagger B_n^\dagger = C_n$ is the cyclic unitary operator. 

\end{document}